\documentclass[pra,twocolumn,preprintnumbers,amsmath,amssymb,superscriptaddress,showpacs,longbibliography]{revtex4-1}
\usepackage{graphicx}
\usepackage{latexsym}
\usepackage{amsmath}
\usepackage{graphics}
\usepackage{amssymb}
\usepackage{layout}
\usepackage{verbatim}
\usepackage{amsfonts,epsfig}
\usepackage{natbib}
\usepackage{hyperref}

\usepackage{graphicx}
\usepackage{epsfig}

\newcommand{\ket}[1]{\left|#1\right>}
\newcommand{\bra}[1]{\left< #1 \right|}
\newcommand{\beq}{\begin{equation}}
\newcommand{\eeq}{\end{equation}}
\newcommand{\bea}{\begin{eqnarray}}
\newcommand{\eea}{\end{eqnarray}}

\newcommand{\mean}[1]{\langle{#1}\rangle{}}

\begin{document}

\title{Environmental noise spectroscopy with qubits subjected to dynamical decoupling}

\author{Piotr Sza\'{n}kowski}
\affiliation{Institute of Physics, Polish Academy of Sciences, Aleja Lotnikow 32/46, PL-02668 Warsaw, Poland}
\author{Guy Ramon}
\affiliation{Department of Physics, Santa Clara University, Santa Clara, California 95053, USA}
\author{Jan Krzywda}
\affiliation{Institute of Theoretical Physics, Faculty of Physics, University of Warsaw,
ulica Pasteura 5, PL-02093 Warsaw, Poland}
\author{Damian Kwiatkowski}
\affiliation{Institute of Physics, Polish Academy of Sciences, Aleja Lotnikow 32/46, PL-02668 Warsaw, Poland}
\author{{\L}ukasz Cywi{\'n}ski}
\email{lcyw@ifpan.edu.pl}
\affiliation{Institute of Physics, Polish Academy of Sciences, Aleja Lotnikow 32/46, PL-02668 Warsaw, Poland}

\date{\today}

\begin{abstract}
A qubit subjected to pure dephasing due to classical Gaussian noise can be turned into a spectrometer of this noise by utilizing its readout under properly chosen dynamical decoupling (DD) sequences to reconstruct the power spectral density of the noise. We review the theory behind this DD-based noise spectroscopy technique, paying special attention to issues that arise when the environmental noise is non-Gaussian and/or it has truly quantum properties. While we focus on the theoretical basis of the method, we connect the discussed concepts with specific experiments and provide an overview of environmental noise models relevant for solid-state based qubits, including quantum-dot based spin qubits, superconducting qubits, and NV centers in diamond.
\end{abstract}
\maketitle

\section{Introduction}
Pure quantum states of a small system (e.g.~a single qubit or a multi-qubit register) become mixed due to unavoidable interaction with the ``rest of the world'' --- this is the process of decoherence \cite{Zurek_RMP03}. Even for a quantum system perfectly isolated from all the unwanted influences, initialization, control (i.e.~forcing the quantum state to evolve in a prescribed way), and readout require the presence of an apparatus (usually referred to as a laboratory), and it is rather unrealistic to assume that the system-apparatus coupling can be switched at will between a sizable one (required for system manipulation) and a negligible one (required for very long-time coherence stability). In the vast majority of experimentally investigated quantum systems, especially the ones found in condensed matter environment, it is however not the control apparatus, but the ``rest of the world'' that is the main source of decoherence.
While it is hard to say how large a part of the ``rest of the world'' has to be considered in order to describe asymptotic decay of coherence (see e.g.~\cite{Blencowe_PRL13} and references therein for discussion of decoherence due to gravitational waves), we are most often interested in coherence on a timescale on which it is still sizable (i.e.~for times comparable to characteristic decoherence time $T_{D}$), or at least measurable with a precision determined by the readout procedure. For quantum systems with sizable coupling to their surroundings, such as condensed-matter based qubits that are embedded in a vibrating crystal lattice that also hosts paramagnetic spins, delocalized electrons, etc., in order to describe coherence dynamics on timescales comparable to $T_{D}$ it is most often enough to consider their nearest environment, the size of which can vary from nanometer to micrometer scales.

It is rather obvious that the time dependence of coherence decay of the quantum system ($S$) of interest is determined by the character of system-environment coupling and of the environment itself. It is less obvious how much (and what kind) of information about the environment ($E$) can be gathered from measurement of the decoherence of $S$. It is expected that with sufficient knowledge about $E$, one could take steps (such as modifying the qubit control scheme, or simply modifying the preparation of the whole physical system hosting $S+E$) leading to enhancement of the coherence time \cite{Biercuk_Nature09,Uys_PRL09,Ajoy_PRA11,Gordon_PRL08,Pasini_PRA10,Chen_PRA10,Wang_PRA13}.
The latter outcome is clearly desired if we want to use multiple copies of $S$ as qubits in a quantum computer --- large-scale quantum computation requires qubits with low enough error rates, determined roughly by the ratio of $T_{D}$ to the time of shortest unitary operation corresponding to a logical gate \cite{Nielsen_Chuang,Devitt_RPP13}. The characterization of $E$ can be viewed however as a goal in itself. As mentioned above, solid-state based qubits are effectively sensitive to a small volume of the surrounding material, so in principle the measurement of their decoherence could bring information about excitations and characteristic timescales of dynamics of the relevant degrees of freedom present in this volume. Furthermore, certain kinds of qubits can be tightly localized in a nanostructure that can be brought into a close contact with another condensed matter system. The most prominent example is the nitrogen-vacancy (NV) center in diamond \cite{Wrachtrup_JMR16}, which is a spin qubit (thus sensitive to fluctuations of external magnetic fields) that can be created a few nanometers from the surface of a diamond nanocrystal  that can then be brought into a vicinity of a source of magnetic fields \cite{Maletinsky_NN12}. In such a setup, the qubit becomes a probe that could be used to characterize various environments \cite{Staudacher_Science13,DeVience_NN15,Haberle_NN15,vanderSar_NC15,Wrachtrup_JMR16,Degen_arXiv16,Lovchinsky_Science16}.

From the above general description of the ``quantum system as an environment sensor'' paradigm it should be clear that it is going to be useful if
\begin{enumerate}
\item There is a description of possible influence of $E$ on $S$ that is amenable to experimentally practical procedure of reconstruction of properites of $E$ from the measurement of dynamics of $S$ initialized in some state, and possibly subjected to a prescribed control and readout protocol,
\item The above description is relevant for a sufficiently broad class of $S+E$ systems.
\end{enumerate}
For a \emph{given} environment, described by a microscopic model --- a quantum or a classical Hamiltonian with a density matrix or a classical probability distribution of initial states --- fulfilling the first condition should be in principle possible. Such a description however requires a large amount of a priori knowledge about the most relevant dynamical degrees of freedom of $E$. In fact, the whole field of high-resolution Nuclear Magnetic Resonance \cite{Haeberlen} can be viewed as an example: for a system of nuclei having a Hamiltonian of known general form (containing nucleus-specific Zeeman splittings, chemical shifts and quadrupolar splittings, and inter-nuclear  dipolar and  electron-mediated interactions), one can reconstruct the parameters of the Hamiltonian from the decay of the resonance signal of a sample subjected to an appropriately designed control protocol\footnote{The dynamical decoupling protocols considered in this review are in fact closely related, and often practically the same, as the multi-pulse protocols used in high resolution NMR.}.
The true usefulness of the paradigm arises when the second condition is fulfilled, and we can use $S$ to characterize a truly \emph{unknown} $E$.

In recent years it became clear that both of the above conditions are fulfilled if we assume that $S$ is subjected to pure dephasing (i.e.~we can neglect exchange of energy between $S$ and $E$), and $E$ is simply modeled as a source of classical Gaussian noise, fully characterized by its power spectrum $S(\omega)$ \cite{Goodman,Kogan}. When $S$ is a qubit, application of a sequence of short pulses, rotating its Bloch vector by angle $\pi$ about axes perpendicular to the quantization axis, not only leads to decoherence suppression due to so-called dynamical decoupling (DD)  \cite{Viola_PRA98,Viola_PRL99,Viola_JMO04,Kofman_PRL04,Khodjasteh_PRA07,Uhrig_PRL07,Gordon_PRL08,Cywinski_PRB08,Biercuk_JPB11,Yang_FP11,Khodjasteh_NC13,Suter_RMP16} of the qubit from its environment, but also establishes a simple relation between the decoherence and the spectrum $S(\omega)$ \cite{Alvarez_PRL11,Bylander_NP11,Almog_JPB11,Yuge_PRL11}. This relation follows from the observation that a qubit subjected to DD from Gaussian phase noise, effectively experiences the noise modified by a frequency-domain filter function \cite{Martinis_PRB03,deSousa_TAP09,Cywinski_PRB08,Biercuk_JPB11} determined by the pulse sequence, and that for a sequence in which pulses are applied periodically (with period $\tau$), the filter is sharply peaked at a frequency $f\! = \! 1/2\tau$ (or angular frequency $\omega_{p} \! =\! \pi/\tau$) and its odd harmonics.

This method of dynamical-decoupling-based noise spectroscopy (DDNS) has been employed for reconstruction of the environmental noise intrinsic to various qubit designs, including those based on trapped ions \cite{Biercuk_Nature09,Kotler_Nature11}, superconducting circuits \cite{Bylander_NP11}, semiconductor quantum dots \cite{Medford_PRL12,Dial_PRL13},  ultracold atoms \cite{Almog_JPB11}, phosphorous donors in silicon \cite{Muhonen_NN14}, and NV centers in diamond \cite{BarGill_NC12,Staudacher_Science13,Romach_PRL15}. It has also become the basis for many applications of spin qubits (specifically NV centers) for nanoscale-resolution sensing of environmental fluctuations \cite{Staudacher_Science13,DeVience_NN15,Haberle_NN15,Degen_arXiv16}, allowing for sensing of signals from single molecules \cite{Lovchinsky_Science16}.
In this review we will explain the theoretical basis of this technique and discuss its applicability to real systems that are not necessarily sources of classical Gaussian noise. We will give only a brief overview of experimental applications (mentioning example experimental papers when we discuss a particular theoretical issue), and a reader interested in a thorough and up-to-date discussion of experiments on environmental spectroscopy with dynamically decoupled qubits is referred to \cite{Degen_arXiv16}.

Let us stress that our focus here is solely on environmental noise spectroscopy methods employing dynamical decoupling of qubits experiencing phase noise. There are numerous other methods of characterization of $E$ based on observation of decoherence (or, more generally, dynamics) of a small quantum system. Transition rates between the two energy eigenstates of the qubit depend on the quantum noise spectrum \cite{Clerk_RMP10} at a frequency corresponding to qubit energy splitting \cite{Schoelkopf_spectrometer}, allowing for spectroscopy based on measurement of the qubit energy relaxation \cite{Astafiev_PRL04}. Spin-locking based spectroscopy of continuously driven qubit \cite{Loretz_PRL13,Yan_NC13}, that decays due to noise at the locking frequency, was also used to reconstruct noise spectrum in a broad range of frequencies. Measurement of the temporal profile of the decay of qubit's Rabi oscillations can also be related to the shape of the noise spectrum by varying the driving field amplitude \cite{Ithier_PRB05,Yoshihara_PRB14}. Finally, noise at frequencies comparable and lower than the frequency with which the qubit energy splitting is estimated from multiple measurements, can be reconstructed using real-time traces of slowly fluctuating qubit splitting \cite{Bialczak_PRL07,Sank_PRL12,Yan_PRB12,Reilly_PRL08,Malinowski_arXiv17}. A recently emerging method of optical spin noise spectroscopy can be viewed as an ensemble counterpart of this single-qubit technique \cite{Sinitsyn_RPP16}.

The structure of the review is the following. In Section \ref{sec:GaussianDDNS} we describe the DD-based noise spectroscopy method for a qubit experiencing pure dephasing due to its longitudinal coupling to Gaussian noise (either classical or quantum). Then in Section \ref{sec:noises} we give an overview of the most important types of environmental noises (not necessarily classical and Gaussian) affecting popular types of solid-state-based qubits.
In Section \ref{sec:GNG} we discuss the conditions under which the method for spectroscopy of Gaussian noise can be used in a reliable way to characterize the first spectrum \cite{Kogan} of a non-Gaussian noise, and we illustrate the general discussion with two examples relevant for solid-state based qubits: the coupling to a charge fluctuator being a source of Random Telegraph Noise (in Sec.~\ref{sec:RTN}) and the coupling to a precessing magnetic moment consisting with a small number of spins (in Sec.~\ref{sec:mu}). Finally, Section \ref{sec:extensions} contains a discussion of theoretically proposed extensions of the DD-based noise spectroscopy method. These include the case of nonlinear coupling between the qubit and the noise (that amounts to a specific instance of spectroscopy of non-Gaussian noise) in Sec.~\ref{sec:OWP}, a recently proposed method for reconstruction of higher-order spectra (so-called polyspectra) characterizing the non-Gaussian noises \cite{Norris_PRL16} in Sec.~\ref{sec:nonGaussian}, and using multiple qubits to reconstruct the cross-spectrum of noises affecting distinct qubits (i.e.~charaterizing the cross-correlations of two or more environmental noises) in Sec.~\ref{sec:multiple}.

\section{Spectroscopy of Gaussian phase noise using dynamical decoupling}  \label{sec:GaussianDDNS}
\subsection{Pure dephasing of the qubit}  \label{sec:PD}
The general Hamiltonian of qubit-environment coupling is
\beq
\hat{V}_{\rm QE} = \frac{1}{2}\mathbf{\hat{b}}(t)\cdot \hat{\mbox{\boldmath $\sigma$}} \,\, ,
\eeq
where $\mathbf{\hat{b}}$ is a vector of three environmental operators, two of them couple to the transverse ($x$ and $y$) operators of the qubit, and the third, $\hat{b}_{z}$, couples longitudinally (the $z$ axis is singled out since we assume finite energy splitting of the qubit's levels, given by $\hat{H}_{\rm Q}\! =\! \Omega\hat{\sigma}_{z}/2$, where $\Omega$ is the energy splitting).

A key approximation that we shall consider is to neglect the influence of transverse couplings. They are responsible for changes in the expectation value of the qubit's $\hat{\sigma}_{z}$ operator that have to be accompanied by exchange of energy between $Q$ and $E$ (for the nonzero $\Omega$ assumed here). However, in many practical situations,  the coherence of the qubit is limited by {\em pure dephasing}, due to the presence of $\hat{b}_{z}$, and not by energy relaxation. A simple explanation  is the following. For finite $\Omega$, efficient energy relaxation of the qubit requires the presence of appreciable number of excitations of $E$ having such an energy (and appreciable coupling to such excitations is necessary). On the other hand, pure dephasing does not involve any energy transfer, and environmental dynamics at all frequencies (excitations of $E$ having all possible energies) contribute to dephasing. Furthermore, condensed matter environments are typically rich in systems having low-energy excitations (e.g.~slowly switching two-level fluctuators responsible for $1/f$ noise, ubiquitous in solid state structures \cite{Paladino_RMP14}). As a result, most solid-state based qubits have their coherence times limited by dephasing, not by relaxation. Using the popular terminology, $T_{1}$ times of most qubits are orders of magnitude longer than their $T_{2}$ times.

We focus then on pure dephasing coupling,
\beq
\hat{V}_{\rm QE, PD} = \frac{1}{2} b_{z}(t) \hat{\sigma}_{z} = \frac{1}{2}v\xi(t) \hat{\sigma}_{z} \,\, , \label{eq:Hpd}
\eeq
where $b_{z}(t)$ is either a classical stochastic function, or a quantum operator $\hat{b}_{z}$ (we will omit the operator hat symbol whenever the quantumness of $b_{z}$ is not of special importance), and
in the second expression we have written $b_{z}(t)$ in terms of a coupling constant $v$ and a dimensionless variable/operator $\xi(t)$. 
Note that in the quantum noise case, the time dependence of $\hat{b}_{z}$ is governed by the self-Hamiltonian of the environment $\hat{H}_{\rm E}$, i.e.~$\hat{b}_{z}(t) \! = \! e^{i\hat{H}_{\rm E}t}\hat{b}_{z}e^{-i\hat{H}_{\rm E}t}$.

The powerful method of noise spectroscopy, which is the main topic of this review, crucially relies on this approximation. This follows not only from formal simplification of the theoretical treatment in the pure dephasing case (although simplicity of the theory makes it easier to connect it with experiment). In this setting, the treatment of the qubit as a passive (or non-invasive) sensor of the environmental dynamics, rather than an object that modifies the environmental fluctuations, is easier to justify than in the case of general $Q$-$E$ coupling.

\subsection{Dynamical decoupling with short pulses}  \label{sec:control}
We assume that the qubit can be initialized in a superposition of its energy eigenstates, subsequently evolving under the influence of $E$ while being subjected to a sequence of rotations by $\pi$ about its $x$ or $y$ axis, and finally that its state is read out after total evolution time $T$.

Application of a sequence of $\pi$ rotations is known to lead to {\em dynamical decoupling} (DD) of the qubit from its environment \cite{Viola_PRA98,Viola_PRL99,Kofman_PRL04,Cywinski_PRB08,Biercuk_JPB11,Yang_FP11,Suter_RMP16}. The basic principle, known since the seminal work of Erwin Hahn on spin echo, in which a single $\pi$ pulse is applied at $T/2$ \cite{Hahn_PR50}, is especially clear in the pure dephasing context. $\pi$ pulses interchange the amplitudes of the two qubit states, which is formally equivalent to flipping the sign of the $Q$-$E$ coupling. The influence of any noise that is quasi-static (i.e.~static on timescale of $T$) is completely removed at the readout time $T$ if the periods of time corresponding to the two signs of the coupling are equal. Generalizations of Hahn's echo to sequences of multiple pulses also date back to the 1950s, starting with the work of Carr and Purcell \cite{Carr_Purcell}.

In this review we will assume that the rotations of the qubit can be well approximated by a sequence of very short (essentially instantaneous) $\pi$ pulses, so that the qubit evolution consists of periods of free evolution, $\hat U_{F}(t_{k+1},t_{k})$, interspersed with $-i\hat{\sigma}_{x}$ or $-i\hat{\sigma}_{y}$ operators acting on the qubit at pulse times $t_{k}$. Of course, realistic pulses have finite time, but this, in principle, should not be an issue as long as the pulse duration is shorter than other relevant timescales in the problem. More importantly, they are imperfect. Reviewing the large body of literature devoted to the design of robust pulses is beyond the scope of this article. We only mention one important concept. When the pulse errors are systematic (i.e.~all the rotations are imperfect in the same way during time $T$), it is possible to design a sequence which is largely resilient to accumulation of errors. This can be achieved by choosing the pulse spacing (see, e.g., the so-called concatenated DD \cite{Khodjasteh_PRA07}), or by choosing a pattern of $\pi_{x}$ and $\pi_{y}$ pulses for a given pulse spacing. The latter approach was first demonstrated by the modification of the Carr-Purcell sequence by Meiboom and Gill \cite{Meiboom_Gill,Borneman_JMR10}, which avoids error accumulation when the qubit is initialized parallel to the axis of rotations. Subsequent modifications of patterns of $\pi_{x}$ and $\pi_{y}$ pulses in the Carr-Purcell sequence suppress error accumulation for all the initial qubit states \cite{Gullion_JMR90,deLange_Science10,Staudacher_Science13}, allowing us to safely  assume ideal pulses when describing experimental control sequences.

\subsection{Qubit dephasing under dynamical decoupling}  \label{sec:WPD}
Phase noise affects only the off-diagonal elements of the qubit state -- its coherence. The figure of merit is $W(T)$, the ratio of the qubit coherence at time $T$ and its initial value:
\beq
W(T) \equiv \frac{\langle {+}|\hat\rho_{Q}(T)|{-}\rangle}{\langle{+}|\hat\rho_{Q}(0)|{-}\rangle }
\,\, , \label{eq:W_def}
\eeq
where $|{\pm}\rangle$ are the eigenstates of $\hat\sigma_z$ qubit operator, and $\hat{\rho}_{Q}(T)$ is the qubit's density matrix at the readout time.  Note that for freely evolving qubit the LHS above should be multiplied by $e^{-i\Omega t}$ in order to cancel this trivial phase (for evolution under DD this phase is absent).
In the case of classical noise, $\hat\rho_Q(T)$ represents the evolution of the qubit averaged over the realizations of the noise. In the case of quantum noise, $\hat{\rho}_Q(T)$ should be interpreted as the qubit's reduced density matrix, $\hat{\rho}_{Q}(T) \! = \! \mathrm{Tr}_{E}[\hat{\rho}_{QE}(T)]$, where $\mathrm{Tr}_{E}(...)$ is the partial trace over the environmental degrees of freedom, and $\hat{\rho}_{QE}$ is the density matrix of the whole system.

We focus now on the evolution under a DD sequence of ideal $\pi$ pulses, taken here to be about the $x$ axis,  without any loss of generality of all the final formulas. Note that for definition (\ref{eq:W_def}) to be useful in the case of sequences with odd number of pulses, for which $\bra{+}\hat{\rho}_{Q}(T)\ket{-} \! \propto  \bra{-}\hat{\rho}_{Q}(0)\ket{+}$, such sequences should be augmented by adding one more $\pi$ pulse at the readout time $T$. With each DD sequence we can associate the so-called time-domain filter function, $f_{T}(t)$, which takes values of $\pm 1$ for $t \in [0, T]$ (and is equal to zero for other times), switching its sign at pulse times $t_{k}$:
\begin{equation}
f_T(t) = \sum_{k=0}^n(-1)^k \Theta\left(t-t_k\right)\Theta\left(t_{k+1}-t\right)\,\, ,
\end{equation}
where $\Theta (t)$ is the Heaviside step function, $n$ is the number of pulses in the sequence, and we have used the convention that $t_0 =0$ and $t_{n+1}=T$.

In what follows we assume that $b_{z}(t)$ represents a classical stochastic process, but the final expression connecting $W(T)$ with the spectrum of a Gaussian noise is the same in the quantum case, see \ref{app:quantum} for a derivation. In the absence of control over the qubit, its free evolution is governed by $\hat U_\mathrm{F}(T,0) = \exp\left[ -\frac{i}{2}\hat\sigma_z\,\int_0^T b_z(t)\mathrm{d}t \right]$.
The instantaneous ideal $\pi$ pulses cause qubit flips that interrupt normal phase evolution in such a way, that the evolution after the pulse picks immediately from the evolution before the pulse ended, hence
\begin{eqnarray}
&&\hat U(t_{n+1}=T,t_0=0) = \hat{U}_\mathrm{F}(T,t_{n})(-i\hat\sigma_x)\hat{U}_\mathrm{F}(t_{n},t_{n-1}) \ldots\nonumber\\
&&\times\hat U_\mathrm{F}(t_3,t_2)(-i\hat\sigma_x)\hat U_\mathrm{F}(t_2,t_1)(-i\hat\sigma_x)\hat U_\mathrm{F}(t_1,0)\nonumber\\
&&=(-i)^n\big( \ldots e^{-i\frac{1}{2}\hat\sigma_z \int_{t_2}^{t_3}b_z(t)\mathrm{d}t}\nonumber\\
&&\times e^{-i\frac{1}{2}(-\hat\sigma_z)\int_{t_1}^{t_2}b_z(t)\mathrm{d}t} e^{-i \frac{1}{2}\hat\sigma_z \int_0^{t_1} b_z(t)\mathrm{d}t} \big)\nonumber\\
&&=(-i)^n e^{-i \frac{1}{2}\hat\sigma_z \int_0^T f_T(t)b_z(t) \mathrm{d}t}\label{eq:evo_op}
\end{eqnarray}
where we used the relation $\hat\sigma_x \exp[ -i\alpha \hat\sigma_z ] \hat\sigma_x = \exp[ -i\alpha(\hat\sigma_x \hat\sigma_z\hat\sigma_x) ] = \exp[+i\alpha \hat\sigma_z ]$, and the consequent piecewise sign alteration of $\hat\sigma_z$ is now accounted for by the filter function $f_T(t)$. Note that adding a $\pi$ pulse at the final time $T$, or even replacing some (or all) of the $\pi_{x}$ pulses by $\pi_{y}$ rotations about the $y$ axis, gives a result for $\hat{U}(T,0)$ that differs from the one above only by a global phase. The latter cancels out from the final expression for $W(T)$, which reads \cite{Klauder_PR62,Martinis_PRB03,deSousa_TAP09,Cywinski_PRB08,Biercuk_JPB11}
\beq
W(T) = \left\langle\exp\left[-i\int_0^T f_T(t)b_z(t) \mathrm{d}t \right]\right\rangle  \equiv \left\langle e^{-i\phi(T)} \right\rangle \, ,\label{eq:W_av}
\eeq
where $\mean{\ldots}$ denotes averaging over realizations of the stochastic process $b_{z}(t)$.

\subsection{Environmental noise} \label{sec:stochastic}

Performing the average in Eq.~(\ref{eq:W_av}) is, in general a formidable task. It becomes however easy if we can assume that the noise $b_z(t)$ is {\it Gaussian}, i.e.~its properties are fully determined by its average $\mean{b_{z}(t)}$ (which we assume to be zero for simplicity) and its autocorrelation function, given in the classical ($\mathrm{cl}$) and quantum ($\mathrm{qm}$) noise cases by
\begin{eqnarray}
C^{\mathrm{cl}}(t_{1}-t_{2}) & = & \mean{b_{z}(t_{1})b_{z}(t_{2})} \,\, , \label{eq:AC} \\
C^{\mathrm{qm}}(t_{1}-t_{2}) & = & \frac{1}{2} \mathrm{Tr}_{E}\left( \hat\rho_E\{b_{z}(t_{1}),b_{z}(t_{2})\} \right ) \,\, , \label{eq:AQ}
\end{eqnarray}
where $\{\ .\ ,\ .\ \}$ is the anticommutator. Note that we have also assumed here that the noise is stationary, i.e.~its autocorrelation function depends only on the time difference $t_{1}-t_{2}$ (note that for quantum noise this is true when $[\hat{H}_{E},\hat{\rho}_{E}]=0$). The use of a symmetrized correlation function in the quantum case is not an outcome of our choice -- it is in fact dictated by the assumption of pure dephasing, as shown in \ref{app:quantum}.

In more technical terms, Gaussian character of the noise means that all the higher-order moments,
\begin{eqnarray}
&&M_k^{\mathrm{cl}}(t_1\ldots t_k) =\langle b_z(t_1)b_z(t_2)\ldots b_z(t_k)\rangle\,,\\
&&M_k^\mathrm{qm}(t_1\ldots t_k) =\frac{1}{2}\sum_{p\in \mathcal{S}_k}\prod_{i=1}^{k-1}\Theta(t_{p(i+1)}-t_{p(i)})\nonumber\\
&&\mathrm{Tr}_E(\hat\rho_E \{ \hat b_z(t_{p(k)}) , \{ \hat b_z(t_{p(k-1)}) , \ldots \{ \hat b_z(t_{p(2)}),\hat b_z(t_{p(1)})\}\ldots\}\})\,,\nonumber\\
\end{eqnarray}
(here $\mathcal{S}_k$ are all permutations of $\{1,\ldots k\}$) factorize into products of $C(t_i-t_j)$. From this point we shall suppress $\mathrm{cl}/\mathrm{qm}$ superscript when the nature of noise is not important. For example, the four-point moment is given by
\begin{eqnarray}
M_{4}(t_1,t_2,t_3,t_4) &&= C(t_1-t_2)C(t_3-t_4) \nonumber\\
&& +C(t_1-t_3)C(t_2-t_4) \nonumber\\
&& +C(t_1-t_4)C(t_2-t_3) \,\, . \label{eq:A4}
\end{eqnarray}
The non-Gaussian case, in which non-trivial multi-point correlation functions of $b_z(t)$ have to be taken into account
will be discussed at some length in Section \ref{sec:nonGaussian}. For now, let us assert that the Gaussian approximation for $b_z(t)$ is realistic in many experimentally relevant cases, and in fact the application of DD sequences used for noise spectroscopy can improve the accuracy of this approximation. This will be further discussed in Section \ref{sec:GNG}.

An important quantity characterizing a noise is the range of its correlation function, i.e. such $\tau_c$ that $C(\tau)\underset{|\tau|\gg \tau_c}{\longrightarrow} 0$. For classical noise, the correlation function can be thought of as a measure of the predictive power of the observer: on average, how reliably one can infer the value of the noise at $t_1$ given its value at $t_2$ (or {\it vice versa}). For example, the correlation time of deterministic signal is infinite, because given the initial condition in a form of, e.g. $\xi(t_2)$, it is in principle possible to determine with certainty the value of noise at any other (even very distant) time. The introduction of randomness involves a loss of information and it diminishes this level of confidence. As a result the correlation time is shortened. Obviously, such an interpretation is not compatible with quantum noise, since it does not make sense to consider definite values of operator quantities. Instead, the decay of quantum correlation function indicates ``dense'' structure of environmental energy levels. W can illustrate this point with a help of bath's Hamiltonian eigenstates, $\hat H_E |i\rangle = \epsilon_i |i\rangle$, by writing $C^{\mathrm{qm}}(t)=\mathrm{Re}\sum_{i}\langle i|\hat\rho_E|i\rangle \sum_j \langle i |\hat \xi|j\rangle e^{-i(\epsilon_i-\epsilon_j)t}\langle j|\hat \xi|i\rangle$. For $\hat\xi$ that couples only isolated levels, $C^{\mathrm{qm}}$ has mostly oscillatory character. When the levels are more dense, the ``trajectories'' of virtual excitations induced by $\hat\xi$ start to interfere with each other. This results in the correlation function decay on a time scale set by the spread of energy level differences (i.e.~the bandwidth of the environment), $\epsilon_i-\epsilon_j$. Naturally, this time scale defines the correlation time of the noise.

A noise for which $\tau_c\to 0$, so that $C(t)\propto \delta(t)$, is called {\it white}, while a noise with non-zero correlation time is referred to as a {\it colored}. Note, that the fact that a noise is colored {\it does not} necessary mean it is ``non-Markovian'', at least in a sense used in context of stochastic process theory \cite{VanKampen,Wang_RMP45} (and, furthermore, white noise is not necessarily a Markovian process \cite{Fulinski_PRE94}).
For classical noise, the memoryless, or Markovian, $\xi(t)$ is defined as such a stochastic process for which its probability distribution evolve according to $P_t(\xi) =\int \mathrm{d}\xi'\, W_{t,t'}(\xi,\xi')P_{t'}(\xi')$ for $t>t'$, with the propagator $W_{t,t'}(\xi,\xi')$ that satisfy the Chapman-Kolmogorov-Smoluchowski composition rule, $W_{t,t'}(\xi,\xi')=\int \mathrm{d}\xi''\, W_{t,t''}(\xi,\xi'')W_{t'',t'}(\xi'',\xi')$ for $t>t''>t'$. An extreme example of Markovian process is a deterministic process describing, e.g. trajectory of particle which motion is governed by Newtonian mechanics.
The Markovianity of noise plays an important role in most of the technical aspects of stochastic process theory, mostly because memoryless processes can be described by a fairly simple and unified formalism.
However, from the point of view of noise spectroscopy, the Markovian property has very little relevance. For example, there is only one Markovian process that is also Gaussian and stationary: the Ornstein-Uhlenbeck process \cite{Wang_RMP45}. It is distinguished among all other Gaussian noises merely by the shape of its correlation function, which is $\propto e^{-|t|/\tau_c}$. Interestingly, a prominent example of a non-Gaussian process, the Random Telegraph Noise (which will be discussed in the context of noise spectroscopy in Sec.~\ref{sec:RTN}), has the same form of two-point correlation function (but, being non-Gaussian, it is also characterized by infinite number of higher-order correlation functions that do not factorize as in Eq~(\ref{eq:A4})).

As a side note, let us mention that white noise $b_z(t)$ leads to exponential decay of coherence (see the next Section), which can be described using a time-local Master equation (involving time-independent Lindblad operators) for dynamics of qubit's density matrix. The process of decay of coherence is thus ``Markovian''. For discussion of non-Markovian features of dynamics of a {\it system} driven by noise, but not of the {\it noise} to which the system is subjected, see e.g.~\cite{Rivas_RPP14}.

\subsection{Dynamical decoupling as noise filtering}  \label{sec:filtering}

An equivalent, more convenient definition of the Gaussian property is that for $\phi(T)$ being a linear functional of $b_z(t)$ we have
\beq
W(T) =\mean{e^{-i\phi(T)}} = e^{-\frac{1}{2}\mean{\phi^2(T)}} \equiv e^{-\chi(T)} \, . \label{eq:Wchi}
\eeq
Here, the {\em attenuation factor} $\chi(T)$, is given by
\begin{eqnarray}
\chi(T) & = & \frac{1}{2}\langle \phi^2(T) \rangle \nonumber\\
& = & \frac{1}{2}\int_{0}^{T} \mathrm{d}t_{1} \int_{0}^{T} \mathrm{d}t_{2}\, f_{T}(t_{1})f_T(t_2)C(t_1-t_2) \,\, \nonumber \\
& = & \frac{1}{2}\int_{-\infty}^{\infty} \frac{\mathrm{d}\omega}{2\pi} |\tilde{f}_{T}(\omega)|^{2} S(\omega)  \,\, , \label{eq:chi}
\end{eqnarray}
where $\tilde{f}_{T}(\omega) = \int \mathrm{d}t e^{-i\omega t}f_{T}(t) $ is the Fourier transform of the time-domain filter function, and $S(\omega)$ is the {\it spectrum} (or power spectral density) of noise, given  by
\beq
S(\omega) = \int_{-\infty}^{\infty} \mathrm{d}t e^{-i \omega t} C(t) \, . \label{eq:S}
\eeq
The Fourier transform of time-domain filter function has a characteristic form, conveniently parametrized by the set of ratios $\delta_k = t_k /T$:
\begin{eqnarray}
|\tilde{f}_{T}(\omega)|^2 & = & T^2 \left|\sum_{k=0}^n (-1)^k e^{-i\omega T\left(\frac{\delta_{k+1}+\delta_{k}}{2}\right)}\right.\nonumber\\
		&&\times\left.(\delta_{k+1}-\delta_k)\,\mathrm{sinc}\!\left(\frac{\omega T(\delta_{k+1}-\delta_{k})}{2}\right)\right|^2\nonumber \\
	&=& \frac{\left | \sum_{k=0}^{n} (-1)^{k} ( e^{-i\omega T \delta_{k+1}} -  e^{-i\omega T \delta_{k}}) \right |^2}{\omega^2}\nonumber\\
	&\equiv& \frac{2 F(\omega T)}{\omega^2}\,,\label{eq:F}
\end{eqnarray}
which lead to a widely used form form \cite{Cywinski_PRB08} of Eq.~(\ref{eq:chi})
\beq
\chi(T) \! =\!\! \int_{-\infty}^\infty \!\!\frac{\mathrm{d}\omega}{2\pi} \frac{F(\omega T)}{\omega^2}S(\omega) \! =\!
	 \int_{0}^{\infty} \!\! \frac{\mathrm{d}\omega}{\pi} \frac{F(\omega T)}{\omega^2} S(\omega), \label{eq:chiF}
\eeq
where we used the relation $S(-\omega) = S(\omega)^* = S(\omega)$ for spectral density of a symmetric correlation function $C(t)$ for both classical and quantum noises, see Eqs.~(\ref{eq:AC},\ref{eq:AQ}). Depending on one's preferences, either $|\tilde{f}_{T}(\omega)|^2$ or $F(\omega T)$ can be called the frequency domain filter function.

The above results show that pure dephasing due to Gaussian noise can be naturally viewed as occurring due to the influence of noise filtered by a function defined by the pulse sequence applied to the qubit \cite{Martinis_PRB03,Uhrig_PRL07,Cywinski_PRB08,Biercuk_JPB11}. This is in fact an example of a more general observation that the decoherence, calculated up to the second order in coupling for a general form of qubit driving (possibly other than a sequence of short pulses), can be related to an appropriately defined filtered noise spectrum \cite{Kofman_Nature00,Kofman_PRL01,Kofman_PRL04,Gordon_JPB11}, see Sec.~\ref{sec:shorttimes}.

\begin{figure}[tb]
	\epsfxsize=1.0\columnwidth
	\vspace*{-0.0 cm}
	\centerline{\epsffile{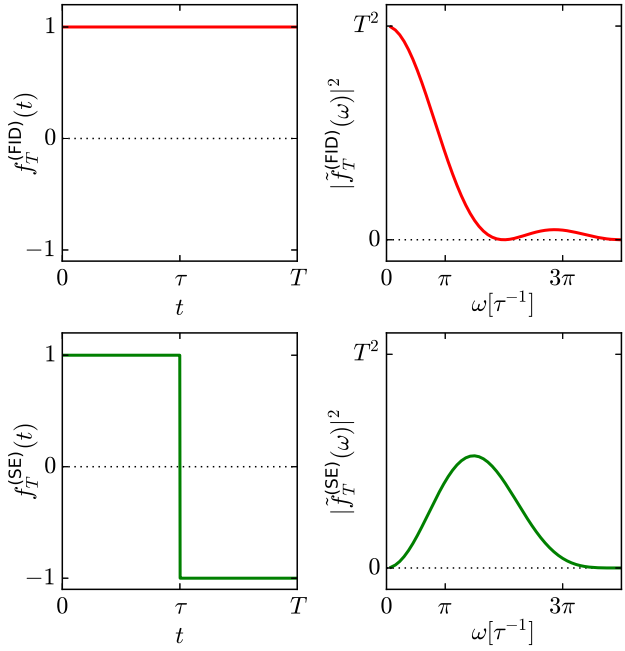}}
	\vspace*{-0 cm}
	\caption{Time- and frequency-domain filter functions for freely evolving qubit (upper panels), and a qubit subjected to an echo sequence (lower panels). }
	\label{fig:FID_ECHO}
\end{figure}

In the simplest case of no pulses, i.e.~for a freely evolving qubit or, using terminology borrowed from NMR, qubit experiencing free induction decay (FID) we have \cite{Ithier_PRB05,deSousa_TAP09}
\beq
|\tilde{f}_T^{(\mathrm{FID})}(\omega)|^2 = \frac{2 F^{(\rm FID)}(\omega T)}{\omega^2} = T^2 \mathrm{sinc}^2 \left(\frac{\omega T}{2}\right) \,\, ,\label{eq:FID_filter}
\eeq
see Fig.~(\ref{fig:FID_ECHO}), and the attenuation function reads
\beq
\chi^{(\mathrm{FID})}(T) = \frac{1}{2}\int_{-\infty}^{\infty}  \frac{\mathrm{d}\omega}{2\pi}  T^2\mathrm{sinc}^2\left(\frac{\omega T}{2}\right) S(\omega)
\eeq
In the long-time limit, $(T/2\pi) \mathrm{sinc}^2(\omega T/2)$ approaches a Dirac delta function, and if $S(\omega)$ is flat flat in frequency range $|\omega| \! <\!  2\pi/T$ for realistically long $T$, we obtain an asymptotically exponential decay of coherence with $\chi(T) \rightarrow TS(0)/2$ \cite{Bergli_NJP09}.
Such a decay occurs for {\it any} pulse sequence when the noise is white, i.e.~$S(\omega)\! =\! {\rm const}$ up to a large frequency $\omega_{c}$ -  to derive this one has to use $\int f^{2}_{T}(t){\rm d}t = \int |\tilde{f}_{T}(\omega)|^2 {\rm d}\omega/2\pi = T$ in Eq.~(\ref{eq:chi}).
Exponential decay of a freely evolving qubit in fact rarely happens for qubits embedded in a solid-state environment, in which we most often encounter many low-energy excitations, and thus large noise power at low frequencies. If most of the total noise power $\sigma^2$ is concentrated at frequencies much lower than typical $1/T$ used in experiments, so that
\beq
\sigma^2 \equiv \int_{-\infty}^{\infty} \frac{\mathrm{d}\omega}{2\pi} S(\omega)   \approx  \int_{-\frac{1}{T}}^{\frac{1}{T}} \frac{\mathrm{d}\omega}{2\pi} S(\omega),
\eeq
we should rather make the following low frequency (LF) approximation:
\begin{eqnarray}
\chi^{(\mathrm{FID,LF})}(T) &\approx& \frac{1}{2}\int_{-\frac{1}{T}}^{\frac{1}{T}}  \frac{\mathrm{d}\omega}{2\pi} |\tilde{f}_T^{(\rm FID)}(\omega)|^2 S(\omega)\nonumber\\
	 &\approx& \frac{T^2}{2}\mathrm{sinc}^2 (0)\int_{-\frac{1}{T}}^{\frac{1}{T}}S(\omega)\frac{\mathrm{d} \omega}{2\pi} =  \frac{\sigma^2  T^2}{2}
\end{eqnarray}
which leads to Gaussian decay of coherence:
\beq
W^{(\mathrm{FID,LF})}(T) = e^{-(T/T_{2}^{*})^2} \,\, ,\label{eq:Winhnbroad}
\eeq
where $T_{2}^{*} \! \equiv \! \sqrt{2}/\sigma$ is the decay time due to inhomogeneous broadening. The decay here occurs due to fluctuations with frequencies $|\omega| < 2\pi/T$, i.e.~due to random shifts of the qubit energy splitting that are constant during each evolution of duration $T$, but which vary from one repetition of qubit evolution to another. This is the definition of a {\em quasi-static} environment dominating observed decoherence.

An even more interesting case is that of $1/f^{\beta}$ type noise, common in solid state setting \cite{Paladino_RMP14}, for which $S(\omega) \propto 1/\omega^{\beta}$ down to unobservably small (or nonexistent, see e.g.~\cite{Niemann_PRL13}) low-frequency cutoff. The observed $T_{2}^{*}$ time depends then on the value of this cutoff, that is typically given by the inverse of the total data acquisition time, $T_{M}$, e.g.~for $\beta\! =\! 1$, we have $T_{2}^{*} \propto \sqrt{1/\ln (T_{M}/T)}$ \cite{Schriefl_NJP06}. Note that dependence of $T_{2}^{*}$ on $T_{M}$ was observed for a quantum dot-based qubit exposed to $1/f^\beta$ charge noise \cite{Dial_PRL13}.

Application of a {\em balanced} sequence of pulses, defined by $\tilde{f}_T(0)=\int f_{T}(t) \rm{d}t \! =\! 0$, obviously removes the influence of zero-frequency noise, but the degree of suppression of low-frequency noise depends on the pulse pattern. For the simplest case of spin echo (SE), in which $\delta_1 = 1/2$, we have \cite{Ithier_PRB05,deSousa_TAP09}
\beq
|\tilde{f}^{(\rm{SE})}_{T}(\omega)|^2 = \frac{16 \sin^4 \frac{\omega T}{4}}{\omega^2} \,\, , \label{eq:FSE}
\eeq
see Fig.~(\ref{fig:FID_ECHO}), so that the LF contribution to the attenuation factor is
\beq
\chi^{(\rm{SE,LF})}(T) \approx \frac{T^4}{32} \int_{0}^{\frac{2\pi}{T}} \frac{\mathrm{d}\omega}{\pi} \omega^2 S(\omega) \,.
\eeq
Therefore, the LF part of the spectral density is being suppressed by $\omega^2$. Sequences with more pulses offer even stronger LF suppression, in the sense that the low-frequency expansion of the respective filter function starts with higher powers of $\omega$. For example, the periodic (PDD) sequence composed of $n \in \mathrm{odd}$ equidistant pulses (not that PDD with even $n$ is not balanced), defined by $\delta_k \! = \! k/(n+1)$, gives a filter
\begin{equation}
|f^{(\rm PDD)}_{T}(\omega)|^2\! =\! \frac{4}{\omega^2}\tan^2\!\! \frac{\omega T}{2(n+1)}\sin^2\!\frac{\omega T}{2}
	\label{eq:FPDD} \,\, ,
\end{equation}
so that the LF noise is suppressed by $\omega^2$ factor (as in the SE case)
while for the Carr-Purcell (CP) sequence, defined by $\delta_{k} \! =\! (k-\frac{1}{2})T/n$, we have for even $n$
\begin{equation}
|f^{(\rm CP)}_{T}(\omega)|^2 = \frac{16}{\omega^2} \frac{\sin^4 \frac{\omega T}{4n}}{\cos^2 \frac{\omega T}{2n}} \sin^2 \frac{\omega T}{2}
\,\, .  \label{eq:FCP}
\end{equation}
so that the LF noise is suppressed by $\omega^4$ factor, and if the decoherence is dominated by the LF contribution, we have $\chi(T) \! \propto\! T^6$.

One can ask ``what is the DD sequence that suppresses the low-frequency noise most effectively?'' Despite the fact that the design of pulse sequences has been an active area of research in NMR for many years, an elegant answer was given only in 2007 \cite{Uhrig_PRL07}. It was shown that if we define the ``best high-pass filter'' as the one for which, for a given number of pulses $n$, the leading term in the Taylor expansion of $F(\omega T)$ around zero has a maximal power of $\omega T$, then the pulse spacing has to be given by:
\beq
\delta_{k} = \sin^2 \left(\frac{\pi k}{2(n+1)}\right)\, ,
\eeq
which leads to $F(\omega T) \propto (\omega T)^{2n+2}$ for $\omega T \ll 1$. This condition for maximal suppression of environmental influence for short $T$ holds also in purely quantum setting, in which one also obtains $\chi(T) \! \propto \! T^{2n+2}$ \cite{Yang_PRL08,Lee_PRL08,Uhrig_NJP08}, justifying the name of the universal DD (UDD) sequence. Note however that UDD is more effective than other DD sequences in decoherence suppression when the classical noise spectrum has a hard cutoff at high frequency $\omega_{\rm HF}$ (so that $S(\omega) \! \sim \! \exp(-\omega/\omega_{\rm HF})$ for $\omega \! > \! \omega_{\rm HF}$), and when $\omega_{\rm HF}T \! \ll \! 1$ is fulfilled. When this is not the case, e.g.~when the noise spectrum has a power-law tail (or when the bandwidth of quantum environment is larger than $1/T$ \cite{Lee_PRL08,Uhrig_NJP08}), CP sequence leads to stronger decoherence suppression \cite{Cywinski_PRB08,Pasini_PRA10,Wang_PRA13}.

It should be clear now that an accurate measurement of $W(T)$ (and thus $\chi(T)$) is a source of information on the noise spectrum, but due to the integral form of $\chi(T)$ it is not possible, in general, to reconstruct a completely unknown $S(\omega)$ from a signal measured under a given DD sequence.
Some progress can be made if we have some {\em a priori} information about the spectrum. For example, if we expect the spectrum to have a hard cutoff at high frequency $\omega_{\rm HF}$ so that $S(\omega) \! \sim \! \exp(-\omega/\omega_{\rm HF})$ for $\omega \! > \! \omega_{\rm HF}$, then applying an $n$-pulse UDD sequence of total length $T<2\pi/\omega_{\rm HF}$ results in \cite{Cywinski_PRB08}
\beq
\chi^{(\rm UDD)}(T) \!\approx \! \frac{16(n+1)^2}{[(n+1)!]^2} \!\left( \frac{T}{4}\right)^{\!\! 2n+2}\!\!\!\!\int_0^{\omega_{\mathrm{HF}}}\!\! \frac{\mathrm{d}\omega}{2\pi} \omega^{2n}S(\omega).
\eeq
In this setting, measurements of the attenuation factor give access to even moments of $S(\omega)$, which allows for its reconstruction (note that the spectrum is an even, non-negative function).

Another example is an often-encountered noise in condensed matter setting with $S(\omega) \! =\! A^{\beta+1}/\omega^{\beta}$ (where $A$ has the units of frequency). For this noise, a change of variables in Eq.~(\ref{eq:chiF}) gives
\beq
\chi^{(\beta)}(T) = (AT)^{\beta+1} \int_{0}^{\infty} \frac{\mathrm{d}x}{\pi} \frac{F(x)}{x^{2+\beta}}  \,\, ,
\label{eq:chibetan0}
\eeq
so that a fit of the super-exponential function $e^{-(T/T_{2})^\alpha}$ to the observed coherence decay allows one to extract the exponent $\beta \! =\! \alpha-1$ \cite{Cywinski_PRB08}.

The above examples show how decoherence due to a low-frequency noise is suppressed by dynamical decoupling sequences that lead to high-pass filtering in $|\tilde{f}_T(\omega)|^2$ (or $F(\omega T)$). Is it however crucial to also pay attention to the high-frequency features of the filter functions. For periodic pulse sequences, in which the $\pi$ pulses are evenly spaced, or a block of pulses is repeated many times, the corresponding filters also exhibit pass-band and stop-band (or notch) behavior \cite{Biercuk_JPB11}, i.e.~noise in specific narrow frequency ranges is allowed to pass freely, while it is very strongly suppressed in other ranges. This is not obvious from inspection of the filter functions in Eqs.~(\ref{eq:FPDD}) and (\ref{eq:FCP}), but becomes apparent after plotting them, see Fig.~\ref{fig:CP_PDD_om}. A transparent analytical analysis of the pass-band behavior will be give in the next Section. Here let us only note that these filters consist of periodically spaced narrow peaks at $\omega T\! =\! (k-\frac{1}{2})2\pi(n+1)$ for PDD and $\omega T = (k-\frac{1}{2}) 2\pi n$ for CP \cite{Cywinski_PRB08}. This leads to an approximate solution to Eq.~(\ref{eq:chibetan0}) that reads
\beq
\chi^{(\beta)}(T) \approx \frac{(AT)^{\beta+1}}{\pi^2(2\pi n)^\beta} \sum_{k=1}^{\infty} \frac{1}{(k-\frac{1}{2})^{2+\beta}} \equiv \left(\frac{T}{T_{2}}\right)^{\beta+1} , \label{eq:chibetan}
\eeq
for CP and similarly for PDD where that factor $(2\pi n)^{\beta}$ in the denominator is replaced by $(2\pi(n+1))^\beta$ (note that in the above equation we have corrected a typo in \cite{Cywinski_PRB08}). The resulting scaling, $T_{2} \propto n^\gamma$ with $\gamma \! =\! \beta/(\beta+1)$ was used to estimate $\beta$ in several experiments \cite{Nadj_Perge_Nature10,deLange_Science10,Medford_PRL12}.
Note that in \cite{Medford_PRL12}, the value of $\beta\! = \! 2.6$ obtained from fitting of $T_{2}$ vs $n$ dependence for large $n$ was shown to be consistent with an even-odd effect observed for small $n$, for which distinct low-frequency behavior of even/odd $n$ CP sequences \cite{Cywinski_PRB08} visibly affects the coherence time.

The frequency-comb nature of the CP/PDD filters \cite{Ajoy_PRA11} leads to much more general consequences.
The pass-band behavior of periodic pulse sequences is the basis of the most direct method of noise spectroscopy with dynamical decoupling, and the following Section is devoted to the description of this method.

\subsection{Noise spectroscopy with periodic sequences}  \label{sec:periodic_ddns}
Here we review the key theoretical results, showing how a qubit subjected to dynamical decoupling can be used as a spectrometer of environmental noise. The basic idea is the following: a periodic DD sequence, with $f_{T}(t)$ characterized by period $T_{p}$ and the corresponding characteristic angular frequency $\omega_p = 2\pi/T_p$, modulates the noise experienced by the qubit. Noise frequencies $\omega \! \ll \! \omega_p$ are strongly suppressed (``echoed away''), while $\omega \! \gg \! \omega_p$ self-average to zero -- the mechanism is basically that of motional narrowing. The frequencies that pass through the filter are $\omega_p$ and its harmonics, and for a DD sequence consisting of many pulses, only they contribute to the observed decoherence. This mechanism was described in a seminal work on using qubits as sensors of ac fields \cite{Taylor_NP08}, compared to  creation of ``optical interference from diffraction  grating'' in \cite{Ajoy_PRA11}, termed ``quantum lock-in'' in \cite{Kotler_Nature11}, and mentioned in the context of sensing nuclear spins with a NV center in \cite{Zhao_NN11}.
It was translated into easy-to-use spectroscopic formulae for general noise spectra in \cite{Alvarez_PRL11,Bylander_NP11,Yuge_PRL11,Almog_JPB11}.

Let us write the time-domain filter function as
\beq
f_T(t) =\Theta\left(T-t\right)\Theta\left(t\right) \sum_m c_{m\omega_p} e^{i m\omega_p t} \, \, ,\label{eq:fFseries}
\eeq
where
\beq
c_{\omega} = \frac{1}{T}\int_0^T e^{-i \omega t}f_T(t) \mathrm{d} t \label{eq:c}
\eeq
is the Fourier series coefficient, and the characteristic frequency of the sequence, $\omega_p$ is the smallest multiple of $2\pi/T$ present in the expansion. For example, for a sequence of odd number, $n$, of equidistant pulses (the PDD sequence), $\omega_p = 2\pi/(2\tau)=\pi(n+1)/T$, where $\tau$ is the delay between consecutive pulses and $T=(n+1)\tau$ is the total duration of the sequence. For a general pulse sequence with no apparent internal periodicity, the smallest frequency possible in the expansion of $f_{T}(t)$ is the fundamental frequency defined by the duration of the sequence, $2\pi/T$.

\begin{figure}[tb]
	\epsfxsize=1.0\columnwidth
	\vspace*{-0.0 cm}
	\centerline{\epsffile{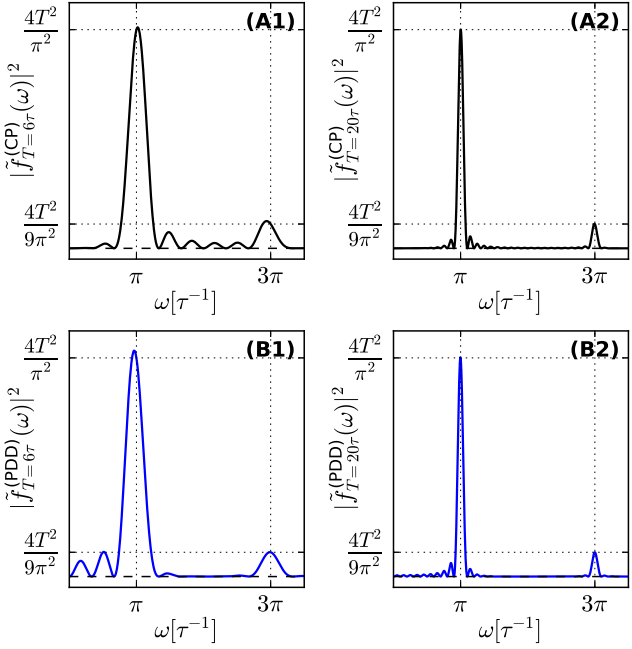}}
	\vspace*{-0 cm}
	\caption{Frequency-domain filter functions, $|\tilde{f}_{T}(\omega)|^2$ for two pulse sequences: the upper panels depict CP sequence with $n=6$ (A1) and $n=20$ (A2) pulses, while the lower panels show PDD sequence with $n=5$ (B1) and $n=19$ (B2) pulses. The peaks of the filters of both sequences are located at $\omega \! =\! \omega_{p} \! =\! \pi/\tau$ and its odd harmonics, with widths $\propto 1/n$, and their heights follow from $|c_{m\omega_p}|^2 \propto 1/m^2$, see text.}
	\label{fig:CP_PDD_om}
\end{figure}

Now we write the Fourier transform of $f_{T}(t)$ in the following form:
\begin{eqnarray}
&&\tilde{f}_T(\omega) =  \int_{-\infty}^\infty \!\!\! \mathrm{d} t\,e^{-i \omega t}f_T(t)\nonumber\\
&& =  \int_0^T \!\!  \mathrm{d} t\,e^{-i\omega t}\left(\sum_m c_{m\omega_p}e^{i m\omega_p t}\right)\nonumber\\
&& = \sum_m c_{m\omega_p} e^{-i\frac{T(\omega-m\omega_p)}{2}}T \mathrm{sinc}\left( \frac{T(\omega-m\omega_p)}{2}\right)\,,
\end{eqnarray}
and then taking its modulus square one gets
\begin{eqnarray}
&& |\tilde{f}_T(\omega)|^2 =  T\sum_{m}| c_{m\omega_p}|^2\,T\,\mathrm{sinc}^2\left(\frac{T(\omega-m\omega_p)}{2}\right)\nonumber\\
&&+\sum_{m_1\neq m_2} c_{m_1\omega_p}c^*_{m_2\omega_p} e^{i T \omega_p\left(\frac{m_1-m_2}{2}\right)}\nonumber\\
&&\times T \,\mathrm{sinc}\left(\frac{T(\omega-m_1\omega_p)}{2}\right)T\, \mathrm{sinc}\left(\frac{T(\omega-m_2\omega_p)}{2}\right).\label{eq:comb_approx}
\end{eqnarray}
As $T\to\infty$, both $T\mathrm{sinc}( \omega T/2 )$ and $T\mathrm{sinc}^2( \omega T/2 )$ approach $2\pi\delta(\omega)$. Consequently, the second term in Eq.~(\ref{eq:comb_approx}) becomes negligible compared to the first one, and the frequency space filter function can be approximated by a series of $\delta$-functions at multiples of the characteristic frequency $\omega_p$, with weights given by squared moduli of the Fourier coefficients $|c_{m\omega_p}|^2$. This is illustrated in Fig.~\ref{fig:CP_PDD_om} for CP and PDD sequences. The peaks at $\omega_{p}\ =\! \pi/\tau$ and its odd harmonics are clearly visible, and they become narrower and sharper with increasing $n$ (equivalently, $T$).

If the spectral density can be treated as approximately constant in a narrow frequency range corresponding to the width of the $\mathrm{sinc}^2$ functions regularizing the $\delta$-functions, Eq.~(\ref{eq:chi}) can be approximated as
\beq
\chi(T) \approx T \sum_{m>0} |c_{m\omega_p}|^2 S(m\omega_p),\label{eq:spectro_formula}
\eeq
where we assumed no zero frequency component (balanced sequence with $\int f_{T}(t)\mathrm{d}t \! = \! 0$). This is the quantitative relation between the decoherence rate and the noise spectrum evaluated at a set of discrete frequencies.
It shows that the sequence acts as a frequency comb: it picks out the values of the noise spectrum only at the multiples of the characteristic frequency $\omega_p$ and outputs the weighted average of the spectrum, with weights given by the Fourier coefficients of the filter function.

The accuracy of Eq.~(\ref{eq:spectro_formula}) is discussed in \cite{Piotr_unpublished}. Here we mention that replacing $|\tilde{f}_T(\omega)|^2$ with a series of delta-like functions results in a relative error that roughly scales with the total duration of the sequence as $\sim (1-e^{-T/\tau_c})\tau_c/T$ for Lorentzian spectrum, and $\sim(1-e^{-\frac{1}{2}T^2/\tau_c^2})\tau_c/T-(1-\sqrt{2/\pi}\int_0^{T/\tau_c}  e^{-\frac{1}{2}t^2}\mathrm{d}t)$ for Gaussian-shaped spectrum. The error goes to zero for $T$ much longer then $\tau_c$ -- the correlation time of the noise. In the case of very narrow-bandwidth noise, it is possible that at finite $T$, the width of the nascent delta functions of the filter is actually larger than the bandwidth of $S(\omega)$. The delta-function approximation for $\tilde{f}_{T}(\omega)$ cannot be used then. For an example, see Section \ref{sec:mu}.

\begin{figure}[tb]
	\epsfxsize=1.0\columnwidth
	\vspace*{-0.0 cm}
	\centerline{\epsffile{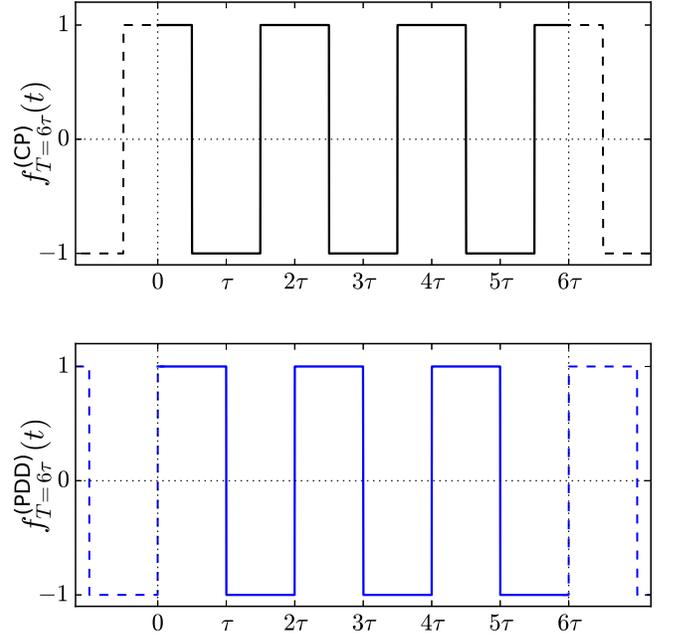}}
	\vspace*{-0 cm}
	\caption{Time-domain filter function $f_{T}(t)$ for $n=6$ pulse CP sequence (upper panel) and $n=5$ pulse PDD sequence (lower panel). The interpulse time is $\tau$, and the total sequence time is $T\! = \! 6\tau$. Dashed lines show the periodic extension of the filter function to $t$ outside of $[0,T]$ interval. The two extended filters are related through time-shift by $t_{0} \! =\! \tau/2$. This means that the Fourier series coefficients from Eq.~(\ref{eq:c}) differ only by phase between the two sequences. }
	\label{fig:CP_PDD_t}
\end{figure}

Since the Fourier series expansion uniquely determines a function defined on interval $[0,T]$, it follows that the only sequences with well defined characteristic frequencies higher than $2\pi/T$ ($\pi/T$) for even (odd) number of pulses (i.e.~with  $c_\omega$ nonzero only for $\omega= m\omega_p$ with integer $m$), are sequences of equidistant pulses and all sequences phase-shifted with respect to them.  By {\it phase-shifted} we mean here sequences with corresponding filter functions $f_T^{(1)}(t)$ and $f_T^{(2)}(t)$ (that are nonzero only for $0<t<T$) such that their periodically extended versions $f^{(i)}_{T\mathrm{-period}}(t) \equiv \sum_{k}\Theta((k+1) T-t)\Theta(t-k T)f_T^{(i)}(t-k T)$ can be transformed one into another by a simple translation, i.e. there exists $t_0$ such that $f_{T\mathrm{-period}}^{(1)}(t) = f_{T\mathrm{-period}}^{(2)}(t-t_0)$, see Fig.~\ref{fig:CP_PDD_t} for example involving CP and PDD sequences. The Fourier series coefficients of phase-shifted sequences differ only by a phase factor $c_{m\omega_p}^{(1)} =e^{i m\omega_p t_0}c_{m\omega_p}^{(2)} $. For example, a CP sequence of $n+1$ ($n\in\mathrm{odd}$) pulses applied at $\frac{1}{2}\tau,\frac{3}{2}\tau,\frac{5}{2}\tau,\ldots ,\frac{n+1}{2}\tau$ is phase-shifted with respect to a PDD sequence with pulses applied at $\tau,2\tau,\ldots,n\tau$ and $c_{m\omega_p}^{(\mathrm{CP})} = e^{i m \frac{\pi}{2}} c_{m\omega_p}^{(\mathrm{PDD})}$. This phase difference is irrelevant for the spectroscopic formula, Eq.~(\ref{eq:spectro_formula}). This is illustrated in Fig.~\ref{fig:CP_PDD_om} where one can see that for large $n$ the $|\tilde{f}_{T}(\omega)|^2$ filters for CP and PDD sequences that are related by a phase-shift are indistinguishable in the high-frequency region in which their peaks are the most important features.

More generally, the frequency comb structure can also be achieved with sequences without well defined (non-trivial) characteristic frequency. As mentioned above, the spectroscopic formula, Eq.~(\ref{eq:spectro_formula}), is a good approximation for a long sequence duration, $T$. In the case of phase-shifted CP sequences, $T$ can be easily extended by adding more pulses. Such modification does not alter the Fourier decomposition of the sequence, hence the only result is the improvement of the quality of the approximation. For more complicated sequences with pulses that are not equidistant, the long $T$ limit is obtained by repeating a base sequence of duration $T_B$ multiple ($M$) times, so that $T=MT_B$ is effectively extended without changing the structure of the filter in Fourier space:
\begin{eqnarray}
& & c_\omega^{(\mathrm{repeated})} = \frac{1}{MT_{B}}\int_{0}^{MT_B} f^{(\mathrm{repeated})}_{MT_B}(t) e^{i\omega t} \mathrm{d}t \, , \nonumber\\
& & = \frac{1}{T_B}\int_{0}^{T_B} f^{(\mathrm{base})}_{T_B}(t) e^{i\omega t} \mathrm{d}t  = c_\omega^{(\mathrm{base})}\, .
\end{eqnarray}

\subsection{Spectral density reconstruction}  \label{sec:spectrum_reconstruction}

For a given sequence with characteristic frequency and a set of Fourier coefficients, $\{c_{m\omega_p}\}$, it is only possible to measure the average $\chi(T)/T =\sum_{m>0}|c_{m\omega_p}|^2 S(m \omega_p)$. In general this does not provide enough information to reconstruct the spectral density of the noise. However, assuming that the spectrum and/or the Fourier coefficients of the sequence become negligibly small at high frequencies, the average can be approximated by a finite number of terms: $\sum_{m>0}|c_{m\omega_p}|^2 S(m\omega_p) \approx \sum_{0<m\leq m_c}|c_{m\omega_p}|^2 S(m \omega_p)$. In particular, when it is justified to take $m_c=1$, the method gives a direct access to the noise spectrum at $\omega_p$. For example, for a CP sequence with $\omega_p = \pi/\tau$, one gets
\begin{equation}
\chi(T) \approx T|c_{\omega_p}^{(\mathrm{CP})}|^2 S(\omega_p) = \frac{4 T}{\pi^2}S(\omega_p) \,\, . \label{eq:singledelta}
\end{equation}
Using this simple equation is the most straightforward method of noise spectroscopy: one measures coherence for sequence durations $T\! =\! n\tau$ with increasing $n$, and for large enough $n$ (in the ``spectroscopic limit'') the results should fall on $e^{-\Gamma n\tau}$ curve \cite{Alvarez_PRL11}, and if Eq.~(\ref{eq:singledelta}) is a good approximation, the decay rate is given by $\Gamma \! =\! 4S(\omega_{p})/\pi^2$. Then, by adjusting $\tau$ one can scan over the whole spectrum.

Generally, when $m_c \! > \!1$, the spectrum reconstruction becomes a problem of solving the linear equations with known coefficients $|c_{m\omega_p}|^2$ ($m\leq m_c$) for the set of unknowns $S(m\omega_p)$. This was addressed in the work of \'{A}lvarez and Suter \cite{Alvarez_PRL11} that put the noise spectroscopy technique described here on firm footing. In order to successfully determine the spectrum, one must generate a system of equations by performing a set of measurements with different pulse sequences $\{c_{m\omega_i}^{(i)}\}$. The sequences must be chosen in an appropriate way, so that: (i) they do not yield dependent equations, (ii) there is an overlap in filtered frequencies (i.e. there is a set of integers $m,m'$ such that $m \omega_i=m'\omega_j$ for $i\neq j$), so that the equations share the same unknowns, thereby forming a system, (iii) the error from cutting-off high frequencies is acceptable.

As an example, for a CP sequence with Fourier coefficients $c_{m\omega_p}^{(\rm CP)} = 2e^{i m\frac{\pi}{2}}/(i\pi m)$ for $m\in \mathrm{odd}$ and zero otherwise, the system of equations can be constructed in the following way. We start with a certain $\omega_1=\pi/\tau_1 = \pi n_1/T_1$, where $n_1$ and $T_1$ are the number of pulses and total duration of the first sequence. The first measurement involves power spectrum values at odd multiples of $\omega_1$: $\sum_{m}|c^{(1)}_{m\omega_1}|^2 S(m\omega_1) = \frac{4}{\pi}\left( S(\omega_1) + \frac{1}{3^2}S(3\omega_1)+ \frac{1}{5^2}S(5\omega_1)+\ldots\right)$. The second equation is obtained by applying another CP sequence with adjusted number of pulses $n_2$ and total time $T_2$ so that $\omega_2 = \pi n_2/T_2 = 3 \omega_1$, then the measured averaged spectrum is $\sum_{m}|c^{(2)}_{m\omega_2}|^2 S(m\omega_2) = \frac{4}{\pi}\left( S(3\omega_1) + \frac{1}{3^2}S(9\omega_1)+\frac{1}{5^2}S(15\omega_1)+\ldots\right)$. The next equation is obtained by applying a sequence with frequency matching the next harmonic present in the original sequence, namely $\omega_3 = 5 \omega_1$, and so on, until the characteristic frequency of the last sequence reaches the chosen cut-off frequency present in the first sequence. Because of the cut-off, this method produces a finite and consistent system of equations.

The method described above can be conveniently summarized in matrix notation. Denote by $\chi^{(s)}(T_s)$ the attenuation factor measured in an experimental run with applied pulse sequence $s$, described by the filter function $f_{T_s}^{(s)}(t)$. Then the relation between measured data and the spectral density can be cast onto a linear transformation $U$ \cite{Alvarez_PRL11} with matrix elements derived from the coefficients $c^{(s)}_{m\omega_s}$:
\begin{equation}
\frac{\chi^{(s)}(T_s)}{T_s}\approx\sum_{m<m_c}|c^{(s)}_{m\omega_s}|^2 S(m\omega_s) \equiv \sum_{m}U_{s, m} S_{m}\,.\label{eq:spec_recon}
\end{equation}
In the case of the example described above, only values of spectrum at odd multiples of $\omega_1=\pi/\tau_1$ are present, hence we have $S_{m} = S[(2m-1)\omega_1]$. The matrix $U$ is a sparse upper triangle and its elements can be written in a compact way as
\beq
U_{s, m} = \sum_{n=1}^\infty \frac{4}{\pi^2 n^2} \delta_{(2m-1),(2s-1)n} \,\, ,
\eeq
with the values of indices $s,m = 0,1\ldots (m_c-1)/2$ truncated by the assumed cut-off. Thus, the reconstruction of the spectral density has been reduced to inverting the matrix $U$, as first described by \'{A}lvarez and Suter in \cite{Alvarez_PRL11}. The restrictions set upon the pulse sequences translate into unambiguous and easily verifiable properties of $U$, e.g., it has to be a well-conditioned invertible matrix.

The reconstruction procedure can become more straightforward for repeated base sequences without a well-defined characteristic frequency. As explained in previous Section, for large number $M$ of repetitions of a base sequence having duration $T_{B}$, the frequency comb picks out frequencies being multiples of $2\pi/T_B$. Forming consistent systems of independent equations for such sequences (with corresponding non-singular $U$-matrices) requires less careful consideration in comparison to phase-shifted CP sequences. For example, using base sequences with pulses applies at random times it is almost guaranteed that each sequence generated in such a way will have non-zero Fourier coefficients for every multiply of fundamental frequency $2\pi/T_{B}$, and it is almost impossible to obtain two sequences with exactly the same distributions. Consequently the resultant matrix $U$ is almost surely invertible.
In \cite{Norris_PRL16} using blocks consisting of a few concatenated DD sequences of various orders stringed together was advocated, and it was shown that using such a family of sequences possessing incommensurate periodicities, one can extend the spectrum reconstruction to higher frequencies, compared to using the \'{A}lvarez-Suter reconstruction with CP sequences obeying the same constraints for pulse spacing (e.g.~for the precision with which a realistic pulse can be applied at required time).
The downside is that the matrix is also dense, which might lead to unfavorably high condition number and overall drop in the accuracy of the reconstruction, especially due to systematic error caused by the cut-off.

\begin{figure}[tb]
\epsfxsize=0.95\columnwidth
\vspace*{-0.0 cm}
\centerline{\epsffile{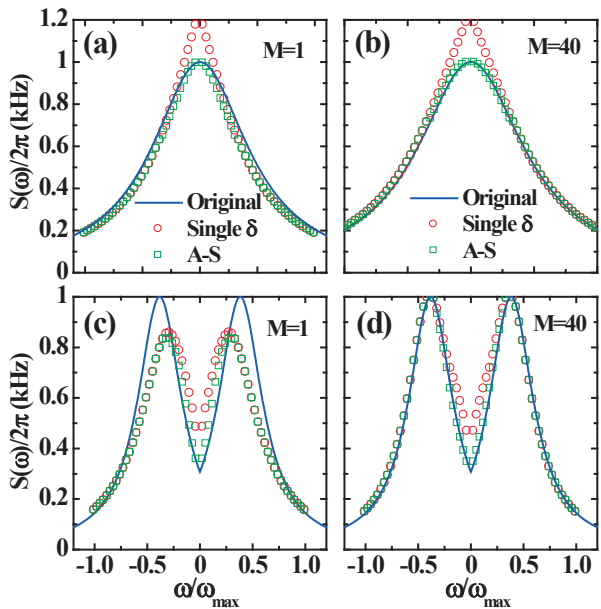}}
\vspace*{-0 cm}
\caption{Original (solid lines) and reconstructed (symbols) Gaussian noise spectrum defined in Eq.~(\ref{eq:LorentzianS}). Open circles (squares) correspond to the single-peak approximation from Eq.~(\ref{eq:singledelta}) (\'{A}lvarez-Suter protocol described in the text). The Lorentzian weight is $S_0=2\pi$ kHz, and its center and width are $\omega_0=0$ and $\tau_c^{-1}=4$ kHz in Figures (a) and (b) and $\omega_0=3$ kHz and $\tau_c^{-1}=2$ kHz in Figures (c) and (d). In all cases we take the minimal interpulse time to be $\tau_{32}=4\times 10^{-4}$ sec, setting the frequency cut-off $\omega_{\rm max} =\pi/\tau_{32}$. In (a) and (c) we use two-pulse CP sequences, while in (b) and (d) these base sequences are repeated 40 times.}
\label{FigLor}
\end{figure}

Finally, phase-shifted CP sequences do not include a zero frequency component in their filter functions. Therefore, it is impossible to reconstruct the power spectrum at exactly $\omega=0$ with such sequences. One can only approach zero frequency asymptotically by, e.g. extending the duration while keeping the number of pulses fixed. Alternatively, the exact zero-frequency component can be accessed by introducing an unbalanced sequence, $c_0 \propto \int f_T(t)\mathrm{d}t \neq 0$, into the protocol, as it was done in \cite{Norris_PRL16}. Note that for $1/f$ type noise the qubit exposed to low frequency noise decoheres very strongly, and consequently both of these methods have to rely on measurements of very small values of coherence, that might not be measurable in the presence of realistic amounts of noise.

We conclude this section with an example that illustrates the method and some of the considerations mentioned above, in generating a reliable and robust noise spectrum reconstruction. Consider a Gaussian noise, characterized by a Lorentzian spectrum:
\beq
S(\omega)=\frac{S_0}{1+\tau_c^2\left(|\omega|-\omega_0\right)^2},
\label{eq:LorentzianS}
\eeq
and depicted by the solid lines in Figure \ref{FigLor}, where $\omega_0=0$ in Figs.~(a) and (b) and $\omega_0=3$ kHz in Figs.~(c) and (d). We have performed noise reconstruction using $32$ CP sequences with variable interpulse delays: $\tau_1, \tau_{2}=\tau_1/2, \tau_3 = \tau_1/3, \ldots, \tau_{32} = \tau_1/32$.
The frequency range in the reconstructed spectra is bound by $\omega_{\rm min}=\pi/{\tau_1}$ and $\omega_{\rm max}=32\omega_\mathrm{min}$ (i.e. $m_c=32$). Open circles in Fig.~\ref{FigLor} correspond to the single-peak approximation, Eq.~(\ref{eq:singledelta}), whereas open squares depict the results of the full \'{A}lvarez-Suter protocol, Eq.~(\ref{eq:spec_recon}). In both cases, the measured data, $\chi^{(s)}(T_s)$, used in the protocol, have been simulated by exact integration over the filtered original noise spectrum.  In Figs.~(a) and (c) we have used two-pulse sequences ($n\! =\! 2$) with durations $T_s\! = \! 2\tau_s$ (i.e.~short total duration), compromising the spectroscopic approximation, whereas in Figs.~(b) and (d) we have used $n\! =\! 80$ pulses (equivalently, the base 2-pulse sequences from (a) and (c) are repeated $M\! = \! 40$ times). The resulting reconstruction with these extended sequence durations is considerably improved as the spectroscopic formula, Eq.~(\ref{eq:spectro_formula}), becomes more accurate.

\section{Condensed matter environment of a qubit as a source of Gaussian and non-Gaussian noise} \label{sec:noises}
In this Section we will not attempt to give a full overview of decoherence of solid-state based qubits. Our goal here is to show that treating the environment as a source of noise (either classical Gaussian/non-Gaussian or quantum Gaussian) is, perhaps surprisingly, a good approximation in the case of qubits experiencing pure dephasing. In  Section \ref{sec:decoh_sources} we overview the dominant decoherence sources for a wide variety of solid-state based qubits, followed by a more detailed discussion of charge and spin noise in the two subsequent sections.

\subsection{Dominant decoherence sources for solid-state based qubits}  \label{sec:decoh_sources}
Qubits embedded in a solid-state environment are subjected to a wide array of perturbations associated with a plethora of possible excitations of the electronic and nuclear degrees of freedom of the host crystal. Metallic components, e.g.~gates defining the quantum dots, are sources of voltage noise associated with low-energy electron-hole excitations of the Fermi sea of electrons. Insulating elements of nanostructures, particularly the strongly disordered glassy ones, such as insulating oxide layers, contain multiple two-level systems (TLSs) having nonzero dipole moment, that are sources of charge noise of Random Telegraph character --- see Section \ref{sec:RTN} below for more discussion.
Furthermore, both the localized unpaired electrons (from dangling bonds, or localized on donors/defects) and the spinful nuclei of the atoms, from which a given crystal is built, are sources of magnetic field noise. Finally, lattice vibrations (phonons), apart from being a source/sink of energy leading to qubit's energy relaxation, also perturb the energy levels of qubits, e.g.~by creating local fluctuating electric fields in polar crystals.

Solid-state based qubits can be roughly divided into the ones possessing sizable electric or magnetic moment, i.e.~into so-called {\em charge} and {\em spin} qubits. The presence of such a moment allows for coupling to external electric or magnetic fields, and thus for qubit manipulation. It also determines the type of environmental excitations that are dominating the decoherence of the qubit. Let us give a few examples.

\subsubsection{Superconducting qubits.}
All superconducting qubits are based on the Josephson tunnel junction that provides nonlinearity and very low dissipation at low temperatures --- critical properties for an operational qubit. At its simplest form, the device consists of a small superconducting island coupled to both a superconducting lead with flux-dependent coupling energy $E_J$ and to a gate voltage $V_g$. The basic Hamiltonian reads:
\beq
\hat H=4E_C (\hat{N}-N_g)^2-E_J \cos \varphi,
\eeq
where $E_C=e^2/2C$ is the charging energy associated with the combined capacitance of the junction and gate, $C$, $\hat{N}\equiv \sum_m m \ket{m}\bra{m}$ is the number operator for the Cooper-pairs accumulating on the island, $N_g$ is the offset charge induced by the gate electrode, and $\varphi$ is the superconducting phase difference across the tunnel layer, which is a conjugate variable to $\hat{N}$.

Note that the number operator $\hat{N}$ has integer eigenvalues, whereas $N_g$ is a continuous variable with values affected by uncontrolled interactions with the environment. Various circuits have been designed primarily to suppress the qubit susceptibility to the fluctuations of this offset charge. Different strategies are aimed at reducing sensitivity to charge noise or flux noise, by operating at different $E_J/E_C$ ratios. These approaches lead to various types of Josephson qubits, some of which are briefly described below.

The ground state energy of the island depends on the number of electrons on it through the charging energy and the parity term, which equals either zero for even number of electrons (forming a BCS ground state) or the superconducting gap for odd number of electrons (where an unpaired electron occupies a quasiparticle state).
In these devices the charging energy is always smaller than the superconducting gap, such that the ground state holds only even number of electrons. Moreover, for certain gate voltage values, there is a degeneracy of ground states differing by one Cooper pair. As a result the ground state at these degeneracy points is formed from an equal mixture of the states $\ket{N}$ and $\ket{N+1}$.

Josephson charge qubits typically operate in the regime where $E_J \lesssim E_C$ \cite{Nakamura_PRL02}. Close to the degeneracy point, the symmetric and antisymmetric superpositions, $\ket{N} \pm \ket{N+1}$, form an avoided crossing, where $E_J$ is the tunnel coupling splitting the two states. The resulting effective Hamiltonian is \cite{Shnirman_PS02}
\beq
\hat H_{\rm eff}=-\frac{1}{2} \Delta E_C \hat\sigma_z- \frac{E_J}{2} \hat\sigma_x,
\label{eq:Hcharge}
\eeq
where $\Delta E_C=4E_C (1-2N_g)$ is the difference in the charging energies of the two states.
The charge qubit is controlled by the gate voltage in the $z$ axis and the applied flux through the junction in the $x$ axis. These controls are susceptible to charge and flux noises, respectively, and while low temperature ($k_B T \ll E_C$) strongly suppresses charge noise, it is still the dominant noise source at this regime. We defer discussing specific proposed models for charge noise sources to section \ref{sec:chargeSC}, pointing here only that the measured coherence in these systems is limited by low-frequency fluctuations of background charges in the substrate. To first order we can neglect flux fluctuations and consider only longitudinal noise with respect to the qubit control field, where the effective Hamiltonian assumes the form of Eq.~(\ref{eq:Hpd}). In this picture, charge fluctuators are capacitively coupled to the island and induce qubit dephasing.

The most straightforward approach to reduce the sensitivity to charge noise is to operate the qubit at the degeneracy point, also known as the optimal working point (OWP). At this point the noise is transverse to the qubit control field, and thus couples to it only quadratically \cite{Ithier_PRB05} (see Sec.~\ref{sec:spin} for further discussion of decoherence at the OWP, and Sec.~\ref{sec:OWP} for a description of noise spectroscopy at such a point). Several designs of charge qubits have been employed to further enhance their charge noise immunity. Among these, the Transmon utilizes a transmission line to increase the gate capacitor, such that $E_C$ is reduced. The increased ratio $E_J/E_C$ greatly flattens the energy bands of the device, thereby reducing its sensitivity to charge noise at an extended range of working positions \cite{Koch_PRA07,Schreier_PRB08}.

A flux qubit consists of a loop with several Josephson junctions, each with energy $E_J^{(i)}(1-\cos \delta_i )$, where $\delta_i$ is the phase drop on the $i$th junction \cite{Mooij_Science99, Friedman_Nature00}. Within the loop, these phase drops are related through $\sum_i \delta_i +2\pi (\Phi/\Phi_0)=2\pi n$, where $\Phi$ is the magnetic flux within the loop, $\Phi_0=h/2e$ is the flux quantum, and $n$ is an integer. Flux qubits operate at the regime where $E_C \ll E_J$. As an example consider three junctions where $E_J^{(1)}=E_J^{(2)}\equiv E_J$ and $E_J^{(3)} =\alpha E_J$, with $\alpha <1$ \cite{Mooij_Science99}. Neglecting charging energies of the junctions and magnetic energy associated with the loop inductance, the effective Hamiltonian for the device is a periodic function of $(\delta_1,\delta_2)$:
\bea
H_{\rm eff}(\delta_1,\delta_2) &=&E_J \left(2-\cos \delta_1 -\cos \delta_2 \right)+ \nonumber \\ && \alpha E_J \left[ 1-\cos\left(2\pi\Phi/\Phi_0+\delta_1 +\delta_2\right)\right].
\eea
When $0.5< \alpha < 1$, each cell includes a double minimum potential, which becomes symmetric when the loop encloses exactly half a flux quantum. The Hamiltonian can then be written in the basis of the eigenstates of these minima, taking a form similar to Eq.~(\ref{eq:Hcharge}). Here, the $z$-axis control is obtained through the loop flux, $\Phi$, and the tunnel-coupling between the minima ($x$-axis control) can be achieved by replacing the third junction with two Josephson junctions in a secondary loop, whose combined Josephson energy is tunable by the magnetic flux through it. Flux qubits are thus fully controlled by tunable magnetic fluxes and the Josephson energies, making them susceptible predominantly to flux noise and critical current noise.

Lastly, phase qubits consist of a larger ($\sim 10 \mu$m), current-biased Josephson junction, making the charging to Josephson energy ratio even smaller than that of flux qubits, and as a result providing better immunity to charge noise \cite{Martinis_PRL02}.

\subsubsection{Spin qubits.}  \label{sec:spin}
These are the qubits based on spin degree of freedom of carriers localized in a semiconductor - when the dynamics of spatial degrees of freedom of an electron or a hole is quenched by localization (due to a binding potential of an impurity, or a conifinig potential of a quantum dot), the spin remains the only degree of freedom.
Although the name suggests that these qubits should couple exclusively to magnetic field fluctuations, the reality is more complicated. Qubits based on a spin of an electron \cite{Loss_PRA98,Hanson_RMP07,DeGreve_RPP13,Liu_AP10} or a hole \cite{Warburton_NM13} in a quantum dot, an electron bound to a donor \cite{Pla_Nature12,Zwanenburg_RMP13}, or an electron complex associated with a deep impurity level such as nitrogen-vacancy (NV) center in diamond \cite{Dobrovitski_ARCMP13}, are obviously very sensitive to fluctuations of other nearby magnetic moments. The majority of such qubits are in fact decohering due to the hyperfine interaction with the nuclear spins of the host material atoms \cite{Chekhovich_NM13}.
The nuclear bath is certainly the most relevant decoherence source for quantum dots based on III-V compounds such as GaAs \cite{Koppens_Science05,Johnson_Nature05,Petta_Science05,Koppens_PRL08} and InGaAs \cite{Greilich_Science06,Bechtold_NP15}, as all the isotopes of Ga, As, and In are spinful, but this holds also for qubits residing in silicon \cite{Tyryshkin_JPC06,Maune_Nature12} and carbon \cite{Childress_Science06,Reinhard_PRL12}, having a natural concentration of spinful $^{29}$Si and $^{13}$C isotopes, respectively (isotopic purification diminishes of course the influence of the nuclear bath). The influence of nuclear spins can however be overwhelmed by fluctuations of magnetic fields due to paramagnetic impurities (localized electronic spins) \cite{Hanson_Science08,deLange_Science10} or simply due to other spin qubits existing in the same sample \cite{Witzel_PRL10,Tyryshkin_NM12}.
In both cases the relevant qubit-environment interaction is
\beq
\hat{H}_{QE} = \sum_{k}\sum_{a,b=x,y,z} \hat{S}^{a} A^{ab}_{k} \hat{I}^{b}_{k} \equiv  \mathbf{\hat{S}}\cdot \mathbf{\hat{h}} \,\, , \label{eq:hf}
\eeq
where $A^{ab}_{k}$ are the components of a tensor of coupling to the $k$-th bath spin, $\mathbf{I}_{k}$ is the operator of the $k$-th bath spin, and $\mathbf{\hat{h}}$ is the operator of the effective field exerted by the bath on the qubit (i.e.~the so-called Overhauser field, in the case of nuclear bath).

However, even such single-spin qubits can be affected by charge noise. This can happen due to the presence of spin-orbit coupling, or due to the presence of magnetic field gradients \cite{Pioro_NP08,Kawakami_PNAS16}. While both allow for driving electron spin resonance with applied electric fields, they also make the spin susceptible to unwanted electric field fluctuations. For electrons, the spin-orbit coupling is much stronger in III-V compounds as compared with Si, and it is possible that the dephasing of certain types of quantum dot single-spin qubits is in fact due to charge noise from a nearby gate \cite{Press_NP10,Bechtold_NP15}. Charge noise can also couple to the spin via hyperfine interaction playing an intermediary role: electric shifts of the carrier wavefunction translate into fluctuations of the nuclear Overhauser field experienced by the carrier spin -- the possible existence of such a coupling has been invoked in a few experimental works \cite{Bluhm_NP10,Malinowski_arXiv17}.

Even more importantly, there is a vigorously researched family of qubits based on two \cite{Petta_Science05,Bluhm_NP10,Malinowski_NN17,Botzem_NC16,Kim_Nature14} or three \cite{Medford_PRL13,Medford_NN13,Hung_PRB14,Russ_Burkard_16} quantum dots, whose logical states reside in a subspace of a larger Hilbert space. For example, in a two-electron, singlet-triplet qubit, localized in a double dot, the singlet $\ket{S}$ and unpolarized triplet  $\ket{T_{0}}$ states of two electrons form the relevant subspace (with the polarized triplet states being separated in energy by applied magnetic field much larger than the typical Overhauser fields in the two dots \cite{Coish_PRB05,Hung_PRB13}). When tunneling of the electrons between the two dots is allowed, the exchange splitting $J$ between $S$ and $T_{0}$ states is nonzero. The Hamiltonian of the qubit, in basis of the $\ket{S}$ and $\ket{T_{0}}$ states, is given by \cite{Taylor_PRB07}
\beq
{\hat H}_Q =(J \hat\sigma_z+\delta h \hat\sigma_x)/2 \,\, ,
\eeq
where $\delta h$ is the inter-dot difference of Overhauser field components parallel to the external magnetic field. While tuning $\delta h$ by dynamical nuclear polarization is a challenging and time-consuming task \cite{Bluhm_PRL10,Foletti_NP09}, $J$ is  highly sensitive to the bias between the dots, providing accessible means to control the qubit working point. For $J\! = \! 0$ a state initialized as $\ket{S}$ (which is a superposition of eigenstates of $\hat\sigma_x$ in the Hamiltonian above) undergoes dephasing due to fluctuations of the Overhauser fields \cite{Petta_Science05,Bluhm_NP10,Botzem_NC16,Malinowski_NN17,Malinowski_arXiv17}.
On the other hand, at large interdot bias we have $J \gg \delta h$.  Since $J$ is of electrostatic origin, it is  susceptible to charge noise \cite{Coish_PRB05,Hu_PRL06,Culcer_APL09,Ramon_PRB10}, and a state initialized as a superposition of $\ket{S}$ and $\ket{T_{0}}$ undergoes dephasing due to this noise \cite{Dial_PRL13}.
Note that in both cases the Hamiltonian is $\approx [{\rm const}+b_{i}(t)]\hat{\sigma}_{i}$ with $i\! =\! x$ $(z)$ and $b_{i}\! =\! \delta h$ $(J)$,  and we are dealing with pure dephasing due to linear coupling to $b_{i}(t)$ noise. However, if we assume a fixed $\delta h \! \gg \! J(t)$, and consider dephasing of superpositions of eigenstates of $\hat\sigma_x$, then to the lowest order in $J/\delta h$ we have
\begin{equation}
\hat{H} \approx \left[ \delta h + \frac{J^{2}(t)}{2\delta h}\right] \hat\sigma_x \,  . \label{eq:HOWP}
\end{equation}
Quadratic dependence of the qubit energy splitting on the noisy parameter means that we are effectively at the OWP (more precisely the OWP corresponds to $\arctan{\delta h/J}\! =\! \pi/2$ \cite{Bergli_PRB07,Ramon_PRB12}).

\subsection{Charge noise: from Random Telegraph Noise to $1/f$ noise}

Charges jumping between two states in the vicinity of a qubit play a key role in limiting coherence times of many qubit systems \cite{Paladino_RMP14}. These sources of random telegraph noise (RTN) have various system-dependent origins, but in many cases, they can be treated as classical two-level systems. This approach, commonly referred to as the spin-fluctuator model, is typically justified when the fluctuators couple more strongly to their own environment than to the qubit (over damped fluctuators) \cite{Paladino_PRL02, Galperin_PRL06}. Analysis of quantum charge fluctuators was performed in \cite{Abel_PRB08}, and several other studies examined the applicability of the classical RTN model by considering the ratio between the qubit-fluctuator coupling strength and the fluctuator decoherence rate \cite{Wold_PRB12}, or by quantifying the classical to quantum transition using several nonclassicality criteria \cite{Trapani_PRA15}.

When one or few charge fluctuators are strongly coupled to the qubit, its coherence factor exhibits non-Gaussian signatures in the form of plateaus \cite{Galperin_03}, as was demonstrated early on in Josephson charge qubits \cite{Nakamura_PRL02}. In contrast, it has been known for a long time that a collection of statistically-independent weakly-coupled RTN sources with a wide distribution of switching rates, $\gamma_i$, leads to a characteristic $1/f$ noise \cite{Schriefl_NJP06}. More precisely, $1/f$ noise is obtained for a set of fluctuators with uniform distribution of $\log \gamma_i$. This distribution occurs naturally in Anderson-type models that are commonly used to account for the bistable fluctuators \cite{Anderson_72}. In these models, the charge is trapped in a double-well potential and tunnels between the two wells. Since the tunneling matrix element depends exponentially on geometrical parameters of the trap, which tend to be uniformly distributed, a log-uniform distribution of the switching rates follows \cite{Bergli_NJP09}. Simulations of charge fluctuator ensembles, with parameters drawn randomly from wide distributions of switching rates and coupling strengths, have shown a crossover between non-Gaussian RTN to Gaussian $1/f$ noise, for ensemble sizes of $\sim 40-50$ fluctuators. Moreover, as the number of fluctuators increased, coherence times became dominated by a majority of weakly-coupled fluctuators, and their average over many runs of randomly drawn distributions have shown considerable narrowing in their standard deviations \cite{Ramon_unpublished}.

Despite the long history of their study, the physical origins of charge fluctuations in solid state devices are controversial, and various microscopic models have been proposed to account for them. Below we review briefly some of these models, as they pertain to the systems that were surveyed in section \ref{sec:decoh_sources}.

\subsubsection{Charge  noise models in superconducting qubits.}
\label{sec:chargeSC}

The coherence of superconducting devices is limited by either charge or flux noise, depending on the $E_J/E_C$ ratio. In addition, all devices are affected by fluctuations in the Josephson coupling energy (or equivalently in the critical current of the tunnel junction). Ultimately, though arising from different microscopic mechanisms, all of these noises involve in one form or another low-frequency fluctuating charges, making it difficult to single them out.

One of the first models for charge noise suggested electron tunneling between a localized state in the insulator and a metallic gate \cite{Paladino_PRL02, Abel_PRB08}. This model was criticized based on relaxation rate measurements \cite{Astafiev_PRL04}, and the experimental setup that suggested that all gates and leads in the qubit vicinity should be in a superconducting state \cite{Nakamura_PRL02}. In a different model, a Cooper pair is split and the two electrons tunnel separately to localized states in the insulator \cite{Faoro_PRL05,Lutchyn_PRB08}. While a constant density of these localized states leads to the observed dependence of relaxation rates on the energy splitting of the qubit states, the model seemed to require an inordinately high concentration of localized states \cite{Faoro_PRL06}. Quantum treatment of the qubit-fluctuator coupling was shown to relax the above requirement \cite{Abel_PRB08}, so at present there is no apparent agreement on the correct microscopic picture behind charge noise in Josephson devices.

All Josephson qubits decohere also due to dielectric losses. These losses result from spurious tunneling TLSs that reside in the amorphous dielectric covering the
circuits, or in the dielectric forming the tunneling barrier in the Josephson junction. The interaction of these systems with their environment causes them to switch their states, resulting in low-frequency noise. This noise source was argued to be central to the operability of Josephson qubits, and was modeled by resonant absorption of two-level defects, successfully accounting for microwave qubit measurements \cite{Martinis_PRL05}.

Finally, let us mention a recently emerging point of view that in order to understand $1/f$ noise generated by many TLSs affecting a superconducting device, interactions between the TLSs have to be taken into account \cite{Burin_PRB15,Muller_PRB15,Lisenfeld_NC15,Burnett_NC14}, and the standard tunneling model \cite{Anderson_72} of TLS fluctuations in glasses has to be revised. DD-based noise spectroscopy methods described here were not used in these works, but they are in fact an example of how using qubits as probes of environmental fluctuations can shed light on long-researched problems in condensed matter.

\subsubsection{Charge-noise models in semiconductor quantum-dot devices.}

Charge noise in quantum-dot based qubits has been relatively less studied as compared with nuclear-induced noise \cite{Culcer_APL09, Ramon_PRB10}. Generally speaking, background charge noise in the sample and electrical noise in the gate voltages are the main sources for charge-induced fluctuations in lateral gate-defined devices. Various mechanisms were suggested, including gate leakage currents via localized states near the electrodes, charge traps near the quantum point contacts (QPCs), donor centers near the gate surface, Johnson noise from the gate electrodes, and switching events in the doping layer, typically located at an interface 100 nm below the surface \cite{Petta_Science05, Pioro_PRB05}. Similarly to the case of Josephson qubits, the main difficulty in interpreting noise measurements is insufficient information that would allow to distinguish between these microscopic mechanisms. In addition, noise levels differ considerably across samples, even when they originate from a single batch. While all the above mechanisms fall into either an Anderson-type model or a tunneling TLS model, more study is needed to determine if their specific characteristics may lead to distinct qubit dynamics. Promising directions may be to explore distinct sensitivities to the qubit's working position, or to use more than a single qubit to probe the charge environment.

Measurements of the background charge fluctuations in GaAs quantum dots have shown a linear temperature dependence characteristic of $1/f$ noise, suggesting a uniformly distributed trap activation energy \cite{Jung_APL04}. Random telegraph noise in GaAs lateral gated structures was measured and characterized by Pioro-Ladri\`{e}re {\it et al.} \cite{Pioro_PRB05}. The noise was attributed to electrons that tunnel through the Schottky barrier under the gate into the conduction band, becoming trapped near the QPC, and causing fluctuations in the conductance with typical frequency of ~ 1 Hz. Applying a positive gate bias during the device cooldown significantly lowered the noise by reducing the density of ionized donors near the surface, thereby suppressing the electron tunneling \cite{Pioro_PRB05}. Telegraph noise induced by the QPCs was also measured in double and triple coupled QDs \cite{Taubert_PRL08}.
Finally, we mention a recent study of microscopic models of fluctuator dynamics, where fluctuators were assumed to be independent, and in turn coupled to independent thermal baths \cite{Beaudoin_PRB15}. Closed-form expressions provided distinct temperature dependence of the coherence time and the stretching parameter, allowing, in principle, to verify experimentally prevailing noise sources.

\subsection{Flux noise and critical current noise affecting superconducting qubits}
Magnetic flux noise in SQUIDs has been studied for over 20 years with no conclusive answers. The noise magnitude has been shown to be largely insensitive to device area or to the type of substrate, suggesting that the noise origin is local. One model has attributed flux noise to electrons jumping between traps, where their spins are fixed in random orientations \cite{Koch_PRL07}. To account for the observed noise magnitude, a large density of unpaired spins was required. A second model proposed spin flips of paramagnetic dangling bonds due to interactions with tunneling two-level-systems and phonons \cite{deSousa_PRB07}. A different mechanism, based on electron spin diffusion through localized states along the surface of the superconductor, was also proposed \cite{Faoro_PRL08}, and later extended to show that in a typical configuration, a fraction of the spins can form strongly-coupled pairs, behaving as two-level fluctuators, and resulting in $1/f$ noise in the magnetic properties of the circuit \cite{Faoro_PRB12}. Finally, local paramagnetic moments trapped by potential disorder at the metal-insulator interface, were also proposed as a potential source of fluctuating spins \cite{Choi_PRL09}.

As with charge and flux noises, the microscopic source of fluctuations in the critical current in Josephson junctions has been a long-standing open problem. Initially these fluctuations were attributed to charges tunneling or hopping between localized states in the barrier forming glass-like TLSs. The trapped charges induce Coulomb repulsion that reduces tunneling through the junction, modulating the junction effective area with each trapping and un-trapping process. Each of these charge fluctuators serves as an RTN source in the critical current, producing Lorentzian peaks in the measured noise spectrum. Typically, these traps are local and noninteracting, so that collectively they produce a distribution of Lorentzians that superimpose to create a $1/f$-like noise spectrum. A detailed experimental work that compared this dephasing channel with flux noise contributions provided supporting evidence to the general picture of a collection of RTN signals from trapped charges but the microscopic origin of these signals remained unclear \cite{VanHarlingen_PRB04}. Furthermore, a $(k_B T)^2$-dependence of the noise spectrum was observed, in disagreement with the assumption of constant TLS density of states \cite{Wellstood_APL04}. A TLS density proportional to their level spacing was proposed \cite{Shnirman_PRL05} and supported by a microscopic model \cite{Faoro_PRL05} to account for this temperature dependence. Later, a model that corroborated with these results and with similar noise measurements in normal-state junctions, suggested that the critical current noise is due to electrons trapped in subgap states that may be forming at the superconductor-insulator interface \cite{Faoro_PRB07} --- a mechanism similar to the one proposed for charge noise \cite{Faoro_PRL06}.

\subsection{Spin noise from a spin bath}
Decoherence of spin qubits due to their interaction with a bath of other spins (either nuclear or electronic) has been a subject of intense research during last 15 years, for reviews see e.g.~\cite{Yang_RPP17,Cywinski_APPA11,Fischer_SSC09}. We stress that the treating the influence of a spin bath as being due to effectively classical noise is far from obvious, and in fact we do not expect such a treatment to be correct in general (i.e., for all regimes of qubit-bath coupling, all protocols of qubit driving, and at all timescales). However, it is currently clear that employing this point of view often gives insight into experimental observations, especially when the spin qubit is subjected to echo or other DD sequences. Let us mention some results of this kind, without delving deeply into open theoretical issues related to the precise formulation of mapping spin baths on classical noise sources.

In the case of an electronic spin (specifically the NV center in diamond) interacting with a bath of other electronic spins, so that the central spin-bath spin coupling is the same as intra-bath couplings, comparison between calculations and various experiments \cite{Hanson_Science08,Dobrovitski_PRL09,deLange_Science10} has shown that treating the spin bath as a source of classical Ornstein-Uhlenbeck noise is a very good approximation. Heuristic arguments for such a mapping  were given in the supplementary materials of \cite{deLange_Science10}. The accuracy of this mapping was later examined in several other works, in which various approximate schemes were employed to calculate the quantum bath dynamics \cite{Witzel_PRB12,Witzel_PRB14,Wang_PRB13}.

The arguments presented in \cite{deLange_Science10} do not apply to III-V based quantum dot spin qubits and certain kinds of Si-based qubits, where the dominant noise source originates from  hyperfine coupling with the nuclear spins. In this case, the intra-bath coupling (either intrinsic, via dipolar interactions, or extrinsic, via electron-mediated interactions \cite{Yao_PRB06,Cywinski_PRB09}) is much weaker than the qubit-bath couplings. In a seminal study devoted to central spin decoherence due to intrinsic dynamics of nuclear bath, elementary excitations of the bath (flip-flops of nuclear pairs) were treated classically, as sources of RTN \cite{deSousa_PRB03}. This calculation was in semi-quantitative agreement with echo observations on phosphorous donors in Si \cite{Tyryshkin_PRB03}. Quantum treatments of intrinsic dynamics of nuclear pairs \cite{Yao_PRB06,Liu_NJP07} and of larger clusters of nuclei \cite{Witzel_PRB05,Witzel_PRB06,Saikin_PRB07,Yang_CCE_PRB08,Yang_CCE_PRB09} soon followed, showing excellent agreement with echo measurements in Si:P \cite{Tyryshkin_JPC06,Abe_PRB10}, and offering predictions for decoherence under influence of various DD sequences \cite{Yao_PRL07,Witzel_PRL07,Lee_PRL08}. The main observation here is that many of these results (obtained with explicitly quantum-mechanical calculations) could be explained using classical noise-based reasoning. For example, CP decays calculated for GaAs in \cite{Witzel_PRL07}, and comparison of UDD and CP decays for GaAs and Si:P \cite{Lee_PRL08}, suggested that a nuclear bath in large GaAs dots can be treated as a source of noise with an effective high-frequency cutoff, while the nuclear bath of P donors in Si corresponds to a spectrum with no such cutoff in the relevant frequency range. The latter result was obtained using a classical treatment of stochastic dynamics of nuclear pairs in \cite{deSousa_TAP09}.

Regimes in which the nuclear bath dynamics can be treated (semi-)classically were discussed in a few other papers \cite{Reinhard_PRL12,Ma_PRB15} (see also \cite{Yang_RPP17} for more examples), using various criteria for distinguishing ``classical'' from ``quantum'' behavior of decoherence. Here our focus is on the applicability of the Gaussian (classical or quantum) approximation in the context of DD-based noise spectroscopy, and in Sec.~\ref{sec:mu} we provide a concrete example demonstrating this kind of consideration within a simple spin bath model.
An interesting idea for testing if the nuclear bath is a source of Gaussian noise was put forth in \cite{Zhao_PRL11}. For NV centers in diamond, corresponding to spin $1$, one can base the qubit on either $S^{z} = 0$, $1$, or $S^{z}=\pm 1$ states. For $b_{z}(t)$ in Eq.~(\ref{eq:Hpd}) being a Gaussian noise, the $W(T)$ function in the second case should be equal to the fourth power of $W(T)$ in the first case. The breaking of this relationship was predicted \cite{Zhao_PRL11} and experimentally confirmed \cite{Huang_NC11} for NV center subjected to specific DD sequences.

Another interesting question arises at low magnetic fields, at which the transverse hyperfine couplings in Eq.~(\ref{eq:hf}) have to be taken account. When treated in second order, they lead to an effective quadratic coupling to transverse components of the Overhauser field \cite{Cywinski_PRL09,Cywinski_PRB09,Bluhm_NP10,Neder_PRB11,Botzem_NC16,Malinowski_NN17,Malinowski_arXiv17}. Many key features of the dynamics of echo \cite{Cywinski_PRL09,Cywinski_PRB09,Bluhm_NP10,Neder_PRB11,Botzem_NC16} and of CP signals \cite{Malinowski_NN17,Malinowski_arXiv17} follow from treating the transverse fluctuations as Gaussian noise with a spectrum concentrated at a few discrete frequencies, corresponding to possible differences of the Zeeman energies of various nuclear isotopes.

The case of linear coupling of a spin qubit to a nuclear bath having its spectrum concentrated at the Larmor precession frequencies of the relevant nuclei is at the heart of recently-developed techniques of nanoscale NMR imaging with NV centers \cite{Staudacher_Science13,DeVience_NN15,Haberle_NN15,Degen_arXiv16,Lovchinsky_Science16}. We discuss this case further in Sec.~\ref{sec:mu}

\section{When can we use Gaussian approximation for environmental noise?} \label{sec:GNG}
From Section~\ref{sec:noises} it is clear that while Gaussian noise statistics is often encountered in solid state environment (or at least it is often reasonable to assume its presence until experiments prove otherwise), the possible presence of non-Gaussian features of noise cannot be ruled out. In this Section we will discuss a set of conditions {\em sufficient} for using Gaussian approximation when performing DD based noise spectroscopy. The latter caveat is important: we will not try to exhaustively  answer the question of applicability of the Gaussian approximation to decoherence calculation in general (a partial answer to this question is contained in Sections \ref{sec:shorttimes} and \ref{sec:CLT} below). Our focus is on limits of reliability of the above-described method of reconstruction of $S(\omega)$ (the so-called {\em first spectrum} in the case of non-Gaussian noise \cite{Kogan}) for noises that are not exactly Gaussian. We present two examples of such non-Gaussian noises encountered in solid state setting: the Random Telegraph Noise (relevant for superconducting and quantum-dot based qubits sensitive to charge noise) and noise due to magnetic moments (consisting of a few spins) precessing in an external magnetic field (relevant for sensing of single spins or clusters of spins with NV centers).

\subsection{Small decoherence limit} \label{sec:shorttimes}
Whenever the coherence function $W(T)$ is close to unity, it can be expanded to lowest (second) order in the qubit-noise coupling,
\beq
W(T) \approx 1 -\left( \frac{1}{2}\int \frac{\mathrm{d}\omega}{2\pi} |\tilde{f}_T(\omega)|^2 S(\omega) \right)\!= 1 - \chi_{2}(T) \, .\label{eq:weak_v}
\eeq
Here $\chi_{2}(T)$ is simply the attenuation function from Eq.~(\ref{eq:chi}), and the subscript indicates that it is a second order term in the so-called cumulant expansion that will be discussed in section \ref{sec:gaussianizationNS}. Consequently, in this low-decoherence regime, the spectroscopy method presented in Sections \ref{sec:periodic_ddns} and \ref{sec:spectrum_reconstruction} is applicable, both for classical and quantum noise of general statistics.

It is important to note here that the close and simple relation between the decoherence of the qubit, calculated to second order in coupling $v$, and the first spectrum of environmental noise, is much more general, i.e.~it holds also for decoherence of qubits subjected to a control protocol other than a sequence of short $\pi$ pulses. Formulas analogous to Eq.~(\ref{eq:weak_v}), showing that periodic driving of the qubit makes its decoherence depend only on a set of frequencies of environmental noise, first appeared in a seminal work, in which periodic modulation of the coupling between a single quantum state and a continuum was considered \cite{Kofman_PRL01}. Several other settings, in which the second-order calculation allows one to relate the decoherence to an overlap between a properly defined bath spectrum and a filter function resulting from external control of the system, were also considered \cite{Kofman_Nature00,Kofman_PRL01,Kofman_PRL04,Gordon_JPB11}, see \cite{Zwick_PRAPL16} for a recent discussion that puts these works in the context of most popular DD-based noise spectroscopy.

\subsection{`Gaussianization' via Central Limit Theorem} \label{sec:CLT}
The Central Limit Theorem (CLT) is often brought out as an argument for the Gaussian approximation of noise statistics. The exact formulation of CLT is as follows: assume that a phase noise has the form $b_z(t) = \sum_{i=1}^M (v/\sqrt M)\xi^{(i)}(t)$, and each $\xi^{(i)}(t)$ is an independent, not necessarily Gaussian noise with identical spectral densities, $S^{(1)}(\omega) = S^{(2)}(\omega)=\ldots=S^{(M)}(\omega)\equiv S(\omega)$. In the limit of large $M$, each qubit-noise coupling tends to zero, eventually reaching weak coupling regime where the second order expansion becomes justified. However, the cumulative contribution form infinite number of weak noises is indistinguishable from that of single noise with Gaussian statistics:
\begin{eqnarray}
W(T) &&= \prod_{i=1}^M \Big\langle e^{-i \frac{v}{\sqrt M}\int f_T(t) \xi^{(i)}(t)\mathrm{d}t} \Big\rangle \nonumber\\
&& \underset{v T\ll M}{\approx}\left( 1 - \frac{\chi_2(T)}{M} \right)^M\underset{M\to \infty}{\longrightarrow} e^{-\chi_2(T)}\,.
\end{eqnarray}
Evidently, the assumptions of CLT are very idealized, and as such, not likely to be strictly satisfied in realistic settings. Nevertheless, the mechanism of CLT is robust enough, so that it approximately works in a wider range of cases. The key requirement is that the phase noise can be decomposed into almost independent noises with similar spectra. One should bear in mind, that those constituent noises do not have to correspond to signals emitted by tangible sources, like e.g. nuclear magnetic moments of crystal lattice. The stochastic processes $\xi^{(i)}$ might as well represent, for example, signals of a whole complex of (possibly strongly) interacting sources, or even parts of the noise fluctuating on a well separated time scales (see, Sec. \ref{sec:OWP}). Then, the requirement of statistical independence is easier to satisfy. The scaling of couplings with increasing number of noises results naturally from conservation of energy, as the constituent noise can only carry a fraction of power of the total phase noise.

The approximate application of CLT can be illustrated as follows. We rewrite the second order contribution from each constituent noise as $\chi_2^{(i)}(T) = (\overline{\chi}_2(T)+\delta\chi_2^{(i)})/M$. This form takes into account the scaling of the couplings and it parameterizes the differences between spectra of the components with $\delta\chi^{(i)}_2$. Assuming that $\delta\chi_2^{(i)}\ll \overline{\chi}_2$ $\forall$ $i$, and $M\gg 1$, we get
\begin{eqnarray}
&&W(T) \approx \prod_{i=1}^M \left(1 - \frac{\overline{\chi}_2(T)+\delta\chi^{(i)}_2}{M}\right)\nonumber\\
&&\approx e^{-\overline{\chi}_2(T)}\left(1-\frac{1}{M}\sum_{i=1}^M \delta\chi_2^{(i)} \right)\approx e^{-\overline{\chi}_2(T)}\,,
\end{eqnarray}
where we neglected terms of order $(\delta\chi_2^{(i)})^2$ or higher and assumed that the deviations from $\overline{\chi}_2$ average out to zero.

\subsection{`Gaussianization'' in the context of noise spectroscopy} \label{sec:gaussianizationNS}
A formal solution for decoherence function defined in Eq.~(\ref{eq:W_av}), can be written
in the terms of the so-called {\it cumulant expansion},
\beq
W(T) = \exp\left[  \sum_{k=1}^{\infty} (-i)^{k} \chi_{k}(T) \right]  \label{eq:Wdefcum}
\eeq
where the $k$th order attenuation factor $\chi_{k}(T)$ is of order $k$ in the noise $b_z(t)$ \cite{Kubo_JPSJ62,VanKampen_P74_1,VanKampen_P74_2,Cywinski_PRA14,Norris_PRL16}, and it is related to the $k$-th cumulant \cite{Kubo_JPSJ62,VanKampen_P74_1,VanKampen_P74_2} of noise, $C_{k}(t_1,t_2\ldots,t_k)$, in the following way:
\begin{eqnarray}
&&\chi_k(T) = \frac{1}{k!}\int_0^T \mathrm{d}^k t_i \left(\prod_{i=1}^{k} f_T(t_i)\right) C_{k}(\mathbf{t})\nonumber\\
&&= \int_0^T \!\! \mathrm{d}t_1\int_0^{t_1}\!\! \mathrm{d}t_2\ldots\int_0^{t_{k-1}}\!\!\! \mathrm{d}t_{k} \left(\prod_{i=1}^{k} f_T(t_i)\right)C_{k}(\mathbf{t}) \, ,\label{eq:cum_k}
\end{eqnarray}
where we have used the vector shorthand notation $\mathbf{t} = (t_1,t_2\ldots,t_k)$. For example, assuming zero average $\langle b_z(t)\rangle =0$, the second cumulant coincides with correlation function, $C_2(t_1,t_2)= C(t_1,t_2)=\langle b_z(t_1)b_z(t_2)\rangle$, and the fourth cumulant is given by
\begin{eqnarray}
C_4(t_1,t_2,t_3,t_4) &=& \langle b_z(t_1)b_z(t_2)b_z(t_3)b_z(t_4) \rangle\nonumber\\
&&  - C(t_1,t_2)C(t_3,t_4)\nonumber\\
&&-C(t_1,t_3)C(t_2,t_4)\nonumber\\
&&-C(t_1,t_4)C(t_2,t_3)\,.\label{eq:C_4}
\end{eqnarray}
Overall, the noise cumulants, $C_{k}$, are characterized by a fundamental property stemming from their abstract definition as a sum of all connected diagrams of given order: $C_{k}(t_1\ldots t_{k}) \underset{|t_i-t_j|\gg\tau_c}{\longrightarrow}   0$, with $\tau_c$ being the correlation time of the noise. This can be illustrated using $C_{4}$ given above: suppose time arguments $t_3$ and $t_4$ are being moved away from $t_1$ and $t_2$. When $|t_2-t_3|\gg\tau_c$ while $|t_1-t_2|\sim|t_3-t_4|\ll\tau_c$. Then $b_z(t_1)$ is still correlated with $b_z(t_2)$, as well as $b_z(t_3)$ with $b_z(t_4)$, but each group is now statistically independent. Hence, the moment factorizes, $\langle b_z(t_1)b_z(t_2)b_z(t_3)b_z(t_4) \rangle = \langle b_z(t_1)b_z(t_2) \rangle \langle b_z(t_3)b_z(t_4) \rangle = C(t_1,t_2)C(t_3,t_4)$, and it cancels with one of the correlation function product. The two remaining products also factorize and vanish, thus making the cumulant go to zero.

For Gaussian noise only $C_{2}$ (and thus $\chi_{2}$) does not vanish for any set of values of its
arguments, and stationary Gaussian noise is fully characterized by its spectrum $S(\omega)$ given by Eq.~(\ref{eq:S}). For non-Gaussian noise the latter quantity is referred to as the {\it first spectrum} $S_{1}(\omega)$ \cite{Kogan}, but the noise is fully characterized by an infinite family of so-called {\it polyspectra}, see \cite{Norris_PRL16} and references therein.
For stationary noise the $(k-1)$th polyspectrum $S_{k-1}$ is given by the multidimensional Fourier transform of the noise cumulant over $k-1$ time differences $\tau_i = t_i - t_{i+1}$,
\begin{eqnarray}
S_{k-1}(\boldsymbol{\omega}) &=& S_{k-1}(\omega_1,\ldots,\omega_{k-1})\nonumber\\
&=& \int_{-\infty}^\infty \!\!\!\! \mathrm{d}^{k-1}\tau_i \,e^{-i \boldsymbol{\omega}\cdot\boldsymbol{\tau}}
C_k\left(\boldsymbol{\tau} \right)
\,,
\end{eqnarray}
with shorthand notation $\boldsymbol{\omega}\cdot\boldsymbol{\tau} = \sum_{j=1}^{k-1}\omega_j \tau_j$ (note, that due to symmetry of $C_{k}$ in the ordering of its arguments, the same result is obtained for any other choice of time differences). We will relate $\chi_{k}(T)$ for qubit subjected to DD sequence to $S_{k-1}(\boldsymbol{\omega})$ in Sec.~\ref{sec:nonGaussian}. For now only qualitative structure of the higher-order attenuation functions will be important.

In order to ascertain the importance of  the non-Gaussian features of the noise we extract the Gaussian part from the coherence function,
\beq
W(T) =e^{-\chi_{2}(T)} e^{\sum_{k>2}(-i)^k\chi_k(T)} \equiv W_{\rm G}(T) W_{\rm NG}(T) .\label{eq:W_G_W_NG_split}
\eeq
Now, $W_{\rm NG}(T)$ contains the non-Gaussian part of the cumulant series
\begin{equation}
\chi_\mathrm{NG} \equiv \sum_{k>2}(-i)^k\chi_{k}(T)\,.
\end{equation}
The most obvious condition for the Gaussian approximation to be valid is $W_{\rm NG}(T) \! \approx \! 1$, or equivalently
\begin{equation}
\chi_{\rm NG}(T) \! \ll \! 1\,. \label{eq:strong}
\end{equation}
We will refer to this as the ``strong'' condition for Gaussianization, that guarantees that the {\em relative} discrepancy between $W_{\rm G}(T)$ and the full $W(T)$ is small. We now focus on the rather common case in which $\chi_{k}$ with odd $k$ vanish -- when this is not fulfilled the discussion below pertains to $|W(T)|$ rather than $W(T)$.
The condition (\ref{eq:strong}) can then be converted to $\chi_{4}(T) \! \ll \! 1$ under the reasonable assumption that, since the whole non-Gaussian part of cumulant series converges to a small value, it is bounded by the first term of the sum, i.e. $|\chi_4(T)| \geq |\chi_\mathrm{NG}|$.

However, our goal here is not to obtain the accurate description of the coherence in itself, but rather to guarantee that the method of noise spectroscopy described in Sec.~\ref{sec:periodic_ddns} and \ref{sec:spectrum_reconstruction} allows for an accurate reconstruction of the first spectrum of noise, $S_{1}(\omega)$. Bearing this goal in mind, the natural way to quantify the quality of the Gaussian approximation is to examine the error of the reconstruction procedure, which assumes a concrete relation between the decoherence rate and the first spectrum. Therefore, we should inspect the non-Gaussian corrections to the spectroscopic formula (\ref{eq:spectro_formula}), restricted to settings where the characteristic frequency of the applied pulse sequence, $\omega_p$, is fixed and the other parameters approach the spectroscopic limit (e.g. for periodic DD sequences one should keep the inter-pulse interval constant and the number of pulses large).
For the procedure to give a small {\em relative} error in the reconstructed $S_{1}(\omega)$ we simply have to require that $\chi_{\rm NG}(T) \ll \chi_{2}(T)$. Note here that when $\chi_{2} \! \gg \! 1$ this condition does not preclude $\chi_{\rm NG} > 1$. In the latter case it is not obvious that $\chi_{\rm NG}(T) \approx \chi_{4}(T)$, but we will nevertheless assume that $\chi_{4}$ is at least of the same order of magnitude as the whole sum of all the non-Gaussian contributions. We arrive then at the following ``weak'' condition for effective ``Gaussianization'' of noise that is relevant for spectroscopy purposes:
\beq
\chi_{4}(T) \ll \chi_{2}(T) \,\,. \label{eq:weak}
\eeq

One often encounters a condition for approximate Gaussianity of the noise that reads $\chi_{4}/\chi^{2}_{2} \! \ll \! 1$ (see for example \cite{Cywinski_PRA14}). The appealing feature of this condition is that both numerator and denominator are of the same order in coupling $v$, so that this is a statement about the non-Gaussianity of the filtered noise {\em itself}. But when we treat the qubit as a probe of environmental dynamics, we are interested in the relation between the signal that we extract using this probe and the properties of noise. Qubit-noise coupling is then an important parameter that {\em should not} disappear from the discussion, especially since we can often vary $v$ in order to change the timescale of the coherence decay and thus to access distinct frequency ranges of fluctuations of $E$. Both the strong and weak Gaussianization conditions depend on $v$ since they characterize not the noise itself, but the noise {\em as it is experienced by the qubit}.

%
%
\subsection{Noise with a finite correlation time}
In this section we examine the weak condition for ``Gaussianization'' (\ref{eq:weak}) for the case of non-Gaussian noise with a well defined correlation time $\tau_c$, so that the spectroscopic limit, $T\gg \tau_c$, is attainable. For simplicity we assume below that a phase-shifted CP sequence with inter-pulse time $\tau_p$ is used, and a single-peak approximation of the spectroscopic formula (\ref{eq:singledelta}), is accurate enough for the reconstruction of $S(\omega)$. Our main goal is to investigate, under these settings, the significance of corrections due to non-Gaussian features of the noise.

For a general noise subjected to DD filtering it is difficult to make a statement about the behavior of $\chi_{4}(T)$ that turns Eq.~(\ref{eq:weak}) into an inequality that could be used with little {\it a priori} knowledge. Nevertheless, in the spectroscopic regime it is expected that for a {\em stationary} noise the $k$th order attenuation factor will scale as $\chi_{k}(T) \sim \! T \tau_c^{k-1}\propto n$, where $n$ is the number of pulses.
With this scaling we can provide a rough estimate of the fourth cumulant:
\beq
\chi_4 \sim T \tau_c^3 = \frac{\tau_c}{T} (T \tau_c)^2 \sim \frac{\tau_c}{T}\chi_2^2 \propto \frac{\chi_2^2}{n}\,.
\eeq
The weak Gaussianization condition from Eq.~(\ref{eq:weak}) now reads
\beq
\chi_4 \sim \frac{1}{n}\chi^2_2 \ll \chi_2 \,\, \Rightarrow \chi_2 \ll n
\eeq
and using the spectroscopic formula, Eq.~(\ref{eq:singledelta}), for $\chi_2$ we arrive at
\beq
\frac{4 \tau_p}{\pi^2}S\left(\frac{\pi}{\tau_p}\right) \ll 1\,,
\eeq
which is equivalent to
\begin{equation}
S_{1}(\omega_p) \ll \omega_p\,.
\label{Gcond}
\end{equation}
While this result follows from a few rather crude approximations, it shows that for $n\tau_p \! \gg \! \tau_c$, the boundaries of the regime in which the non-Gaussian corrections are irrelevant can be outlined based on the knowledge of $S(\omega)$. Strong breaking of the above inequality by $S(\omega_p)$, inferred from a spectroscopic procedure based on assumption of Gaussian noise, should be treated as a suggestion that the procedure can be unreliable, and its consistency should be checked (e.g.~by using the reconstructed $S(\omega)$ to predict coherence decay under a different DD sequence and then comparing this prediction with experiment). Of course, more accurate Gaussianity conditions can be derived if we have more a priori knowledge about the noise. We offer an example in the following Section.

\subsection{Random Telegraph Noise}  \label{sec:RTN}
Classical sources of random telegraph noise (RTN) are the quintessential example of non-Gaussian noise. They can be modelled by two-level charge fluctuators (TLFs), where the stochastic variable of the RTN is $\xi(t)=\pm 1$, and switching between these two possible values occurs with given rate $\gamma$ (for simplicity we consider here a symmetric TLF).  The coupling of the TLF with the qubit is $v$, so that $b_{z}(t) \! =\! v\xi(t)$. A simple calculation \cite{Machlup_JAP54} gives autocorrelation function of $\xi(t)$ of the $e^{-2\gamma t}$ form, and the resulting first spectrum is Lorentzian:
\beq
S(\omega)=\frac{4\gamma v^2}{(4\gamma^2+\omega^2)} \,\, .
\label{eq:SRTN}
\eeq
This noise is non-Gaussian, and examples of treatment of its multipoint correlation functions can be found in appendices of \cite{Cywinski_PRB08} and \cite{Ramon_PRB15}.
Many works have analyzed the effects of one or more RTN sources on a qubit at different working positions and under various DD protocols \cite{Laikhtman_PRB85,Paladino_PRL02,Cywinski_PRB08,Bergli_PRB07,Chen_PRA10,Ramon_PRB12,Schriefl_NJP06,Galperin_PRL06,Galperin_PRB07,Ramon_PRB15}. We confine our discussion below to pure dephasing, where exact analytical results for the qubit decoherence under various control sequences are known.

One feature of the RTN model which is of particular interest here, is the ability to tune the strength of non-Gaussian effects seen be the qubit by changing the coupling strength to switching rate ratio, $\eta \equiv v/\gamma$. According to the previous section, taking into account the spectrum from Eq.~(\ref{eq:SRTN}) we expect the Gaussian approximation to be accurate for weakly-coupled TLFs, $\eta \ll 1$, whereas a strongly-coupled TLF should induce pronounced non-Gaussian effects. The latter have been well known, exhibiting, e.g., plateaus in the qubit decay signal. We state here, as a concrete example, the exact result for the qubit decoherence function under $n$-pulse CP sequence \cite{Ramon_PRB15}:
\bea
W^{(\mathrm{CP})} (T) \!&=&\! \frac{e^{-\gamma T}}{2\mu^n} \left[  \frac{\cosh \gamma \mu \tau-\eta^2}{\mu \sqrt{\sinh^2 \gamma \mu \tau +\mu^2}} \left(\lambda_+^n -\lambda_-^n \right) + \right. \nonumber \\ && \left. \left( \lambda_+^n +\lambda_-^n \right) \right],
\label{depCPs}
\eea
where $\mu=\sqrt{1-\eta^2}$ and
\beq
\lambda_\pm = \sinh \gamma \mu \tau \pm \sqrt{\sinh^2 \gamma \mu \tau +\mu^2}.
\label{lsym}
\eeq
Eq.~(\ref{depCPs}) should be compared with the results of a cumulant expansion, Eq.~(\ref{eq:Wdefcum}), which in the RTN case includes only even orders. The first few terms can be calculated by direct evaluation of the time integrals over multi-point correlators \cite{Cywinski_PRB08,Ramon_PRB15}, where the leading (Gaussian) contribution is $W_G^{(\mathrm{CP})}(T)=\exp(-\chi^{(\mathrm{CP})}_2(T))$ with:
\bea
\chi_2^{(\mathrm{CP})}(T)\!& =&\! \frac{\eta^2}{4} \left[ 2n(\gamma \tau-\tanh \gamma \tau) - \right. \nonumber \\ && \left. \left( 1-\!(-1)^n e^{-2 \gamma T} \right) (1- {\rm sech} \gamma \tau )^2 \right]
\label{depCPG}
\eea
One can easily verify that the expansion of the exact result, Eq.~(\ref{depCPs}), in $\eta$ matches the cumulants to the corresponding order.

\begin{figure}[tb]
\epsfxsize=0.9\columnwidth
\vspace*{-0.0 cm}
\centerline{\epsffile{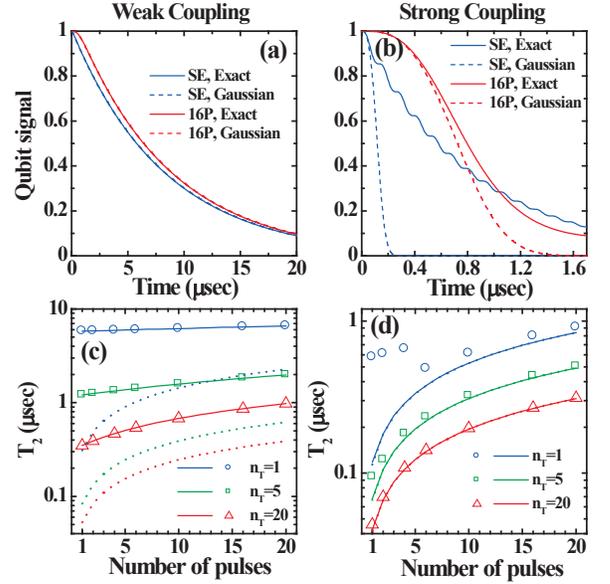}}
\vspace*{-0.1 cm}
\caption{(color online) Qubit coherence factor vs.~time for SE and 16-pulse CPMG at pure dephasing, calculated using the exact solution, Eq.~(\ref{depCPs}), (solid lines) and the Gaussian approximation, Eq.~(\ref{depCPG}) (dashed lines). (a) Single weakly-coupled TLF; (b) single strongly-coupled TLF. Figures (c) and (d) depict dephasing times vs.~number of control pulses for 1, 5, and 20 identical TLFs at weak and strong coupling, respectively. Symbols (solid lines) correspond to the exact solution (Gaussian approximation). The short time limit, Eq.~(\ref{short}), is also shown by dotted lines (notice that the latter completely coincides with the Gaussian approximation in the strong coupling regime). TLF parameters are $\gamma=0.1 \mu$eV, $v=0.01\mu$eV for Figs.~(a) and (c), and $\gamma=5 $neV, $v=0.2\mu$eV for Figs.~(b) and (d). Figure reprinted with permission from \cite{Ramon_PRB15}, copyright 2015 American Physical Society.}
\label{FigRTN1}
\end{figure}

Figs.~\ref{FigRTN1}(a) and (b) depict qubit dephasing due to a single TLF, calculated using the exact result, Eq.~(\ref{depCPs}), and the Gaussian approximation, Eq.~(\ref{depCPG}), for SE and 16-pulse CP. The Gaussian result holds well throughout the entire decay timescale for weak coupling [dashed and solid lines coincide in Fig.~\ref{FigRTN1}(a)], whereas pronounced non-Gaussian behavior develops in the strong coupling case, dominating the qubit signal in Fig.~\ref{FigRTN1}(b). As the number of control pulses increases, the deviations from Gaussian behavior are pushed to longer times, since increased $n$ within a fixed run time increases the control sequence filtering frequency, so that the sampled noise spectrum gets closer to the Gaussian limit, Eq.~(\ref{Gcond}).

Due to the linear qubit-TLF coupling at pure dephasing, extending the above results to any number of fluctuators, $n_T$, is straightforward, and the qubit's coherence factor is simply the product of contributions from all TLFs: $W (T)=\prod_{i=1}^{n_T} W_i (T)$. In the Gaussian limit this leads to a qubit attenuation factor that is a sum over $n_T$ factors, weighted by TLF parameter distribution \cite{Galperin_03}. Examining the asymptotic behavior of our exact result, adopted to $n_T$ TLFs, we have at short times:
\beq
\chi^{(\mathrm{CP})}(T) \underset{\gamma_i, v_i \ll T^{-1}}{\longrightarrow} \frac{T^3}{6n^2} \sum_{i = 1}^{n_T} {{\gamma _i}v_i^2},
\label{short}
\eeq
suggesting similar time- and $n$-dependence as that of a Gaussian noise with a soft ($\omega^2$) cutoff. Similarly, we find the weak- and strong-coupling asymptotic behaviors as:
\begin{equation}
\chi^{(\mathrm{CP})}(T) \!\!\! \underset{v_i, T^{-1} \ll \gamma_i}{\longrightarrow} \!\! \left\{ \!\! {\begin{array}{*{20}{l}}
{\frac{T}{2}\sum\limits_{i = 1}^{{n_T}} {\frac{{v_i^2}}{{{\gamma _i}}}} ,}&{\frac{{{\gamma _i}T}}{n} \gg 1}\\
{\frac{{{T^3}}}{{6{n^2}}}\sum\limits_{i = 1}^{{n_T}} {{\gamma _i}v_i^2} ,}&{\frac{{{\gamma _i}T}}{n} \ll 1}
\end{array}} \right.
\label{weak-asym}
\end{equation}
and
\begin{equation}
\chi^{(\mathrm{CP})}(T) \!\!\! \underset{\gamma_i, T^{-1} \ll v_i}{\longrightarrow} \!\!\! \left\{ \!\!\! {\begin{array}{*{20}{l}}
{\sum\limits_{i = 1}^{{n_T}} \!\left({{\gamma _i}T \!-\! \frac{{n{\gamma _i}}}{{{v_i}}}\sin \frac{{{v_i}T}}{n}} \right)\! ,}&\!\!\!\!\! {\frac{{{v_i}T}}{n} \gg 1}\\
{\frac{{{T^3}}}{{6{n^2}}}\sum\limits_{i = 1}^{{n_T}} {{\gamma_i}v_i^2} ,}&\!\!\!\!\! {\frac{{{v_i}T}}{n} \ll 1}
\end{array}} \right.
\label{strong-asym}
\end{equation}
respectively. These asymptotes elucidate the interplay between TLF parameters, ensemble size, and number of control pulses, in determining the qubit dephasing dynamics. First we observe that both the short- and long-time limits for the weak coupling case can be obtained directly from the Gaussian result, Eq.~(\ref{depCPG}), reaffirming the validity of the Gaussian approximation for weakly coupled TLFs. In contrast, in the strong-coupling case, only the short-time limit converges to the short-time Gaussian result, demonstrating the onset of non-Gaussian effects at longer times [see Fig.~\ref{FigRTN1}(b)].

Figs.~\ref{FigRTN1}(c) and (d) quantify noise Gaussianization with increased number of pulses, by depicting dephasing times for ensembles of identical TLFs. The Gaussian limit is reached when $\tau=T/n \ll \gamma^{-1}, v^{-1}$, corresponding to the short time asymptotic. In addition, we observe that the Gaussian limit is reached with fewer control pulses as the number of TLFs increases, reminiscent of $1/f$ (Gaussian) noise generated from large ensembles of TLFs with a uniform distribution of $\log \gamma_i$ \cite{Schriefl_NJP06, Bergli_NJP09}. For strong coupling shown in Fig.~\ref{FigRTN1}(d), Gaussianity is completely restored with 20 TLFs.
In the weak coupling regime, where the Gaussian result holds for any number of pulses and TLFs for the chosen parameters, the TLFs switch many times between control pulses and we are in the motional narrowing regime, where the long-time limit holds (compare with the short time limit result depicted by dotted lines). Here, increasing $n$ has little effect on the qubit coherence, up to unrealistic number of pulses [the weak-coupling long-time limit given in Eq.~(\ref{weak-asym}) is strictly independent of $n$, but subleading contributions have a mild $n$-dependence, as seen in Fig.~\ref{FigRTN1}(c)]. Alternatively, increased number of TLFs will induce a shorter timescale for qubit decay and a departure from the motional narrowing regime, resulting in a greater benefit from increasing $n$. For the chosen parameters, the short-time limit is reached for 200 TLFs with $n=20$.

\begin{figure}[t]
\epsfxsize=0.65\columnwidth
\vspace*{-0.0 cm}
\centerline{\epsffile{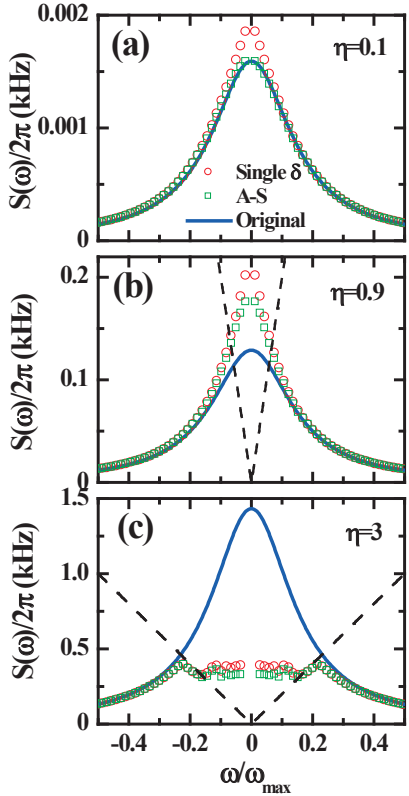}}
\vspace*{-0.25 cm}
\caption{(color online) Original (solid lines) and reconstructed (symbols) single RTN power spectrum. Open circles (squares) correspond to the single-peak approximation (\'{A}lvarez-Suter protocol). The original power spectrum is given by Eq.~(\ref{eq:SRTN}) with TLF switching rate of $\gamma=1$ kHz and coupling strengths of: (a) $v=0.1\gamma$, (b) $v=0.9\gamma$, and (c) $v=3\gamma$. The dashed lines in plots (b) and (c) correspond to $S(\omega)=|\omega|$. In all plots, $\tau_{32}=1/(4\gamma)$, setting the frequency cut-off $\omega_{\rm max}=\pi/\tau_{32}$, and the run time of the longest sequence is $T_1=32/\gamma$. The two-pulse CP base sequences are repeated 40 times.}
\label{FigRTN2}
\end{figure}
The above analysis provides useful guidance when attempting to reconstruct the RTN power spectrum from Eq.~(\ref{eq:SRTN}).
In figure \ref{FigRTN2} we demonstrate power spectrum reconstruction for a single RTN, using 32 two-pulse CP sequences. We examine three coupling strengths, comparing the single-peak approximation (open circles), with the \`{A}lvarez-Suter protocol (open squares), similarly to the Gaussian noise reconstruction presented in section \ref{sec:spectrum_reconstruction}. The control sequences are repeated 40 times to minimize errors due to the $\delta$-approximation, Eq.~(\ref{eq:spectro_formula}), allowing us to attribute poor reconstruction exclusively  to noise non-Gaussianity. Using the \'{A}lvarez-Suter protocol, we obtain a faithful spectrum reconstruction in the weak-coupling case, which deteriorates with increased coupling strength. The dashed lines in Figs.~{\ref{FigRTN2}(b) and (c) correspond to $S(\omega)=|\omega |$, so that all spectral peaks above these lines are in violation of the Gaussianity condition, Eq.~(\ref{Gcond}). In the weak-coupling limit depicted in Fig.~{\ref{FigRTN2}(a), the entire spectrum is below this line (not shown). We note that the error induced by invoking the single-peak approximation, Eq.~(\ref{eq:singledelta}), is largely insensitive to the coupling strength.

\subsection{Noisy magnetic moment precession}  \label{sec:mu}
One of the most actively pursued experimental applications of DD-based noise spectroscopy is using NV centers in diamond for detection of single nuclear spins or clusters of nuclei associated, for example, with single protein molecules in the vicinity of the qubit. In these nanoscale NMR experiments, the presence of specific nuclear species is sensed by matching peaks in the reconstructed noise spectrum with the precession frequency associated with them \cite{Staudacher_Science13,Wrachtrup_JMR16,Degen_arXiv16}. 
This method is based on the assumption that the nuclear noise is classical and Gaussian.
Since the volume from which an appreciable signal is gathered can be as small as $\sim 100$ nm$^3$, or equivalently, the signal comes from a single molecule containing up to hundreds of nuclei \cite{Lovchinsky_Science16}), it is not obvious {\em a priori} that the classicality and Gaussianity assumptions are reasonable. Below we compare a simple exact quantum calculation of DD decay due to coupling with $M$, non-interacting environmental spins with a calculation in which the magnetic field generated by these spins is represented by a Gaussian stochastic process (the classical noise approximation).

In the quantum model we take the Hamiltonian of longitudinal coupling between the qubit and the environmental $J\! =\! 1/2$ spins:
\begin{eqnarray}
&&\hat{V}_{QE} = \frac{1}{2}\hat{\sigma}_z\sum_{k=1}^{M} ( A_{\perp}^{(k)}\hat{J}_{x}^{(k)} + A_{\parallel}^{(k)}\hat{J}_{z}^{(k)}),\\
&&\hat H_E =  \sum_{k=1}^{M}\omega^{(k)}\hat{J}_{z}^{(k)}\, ,
\end{eqnarray}
where we eliminated the coupling to $\hat{J}_y^{(k)}$  by rotating the coordinate system. We restrict our discussion to the simplified case where all $\omega^{(k)}$ are equal to $\omega_0$, and all $A_{\perp/\parallel}^{(k)}$ are equal to $A_{\perp/\parallel}$.
We now define parameters that quantify the strength of the qubit-environmental spin coupling as compared with the Hamiltonian of $E$
\beq
\epsilon_{\perp/\parallel} \equiv \frac{A_{\perp/\parallel}}{\omega_0} \,\, .
\eeq
Here we focus on the weak-coupling regime, $\epsilon_{\perp/\parallel} \! \ll \! 1$, preserving the general setting of a qubit acting as a probe of the intrinsic dynamics of $E$.

The initial state of the environment is assumed to be described by a high-temperature density operator $\hat{\rho}_{ E}(0) \! =\! \hat{I}/2^{M} = \prod_{k=1}^{M}\hat{I}^{(k)}_{2\times 2}/2$. Since the Hamiltonian is a sum over $k$ commuting terms, and the density matrix of $E$ contains no correlations between the nuclear spins, the evolution of the qubit can be factorized into a product of $M$ terms, each describing the contribution to the decoherence function from one of the spins:
\beq
W^\mathrm{qm}(T) = \prod_{k=1}^{M} W^{(k)}(T) = \left[ W^{(1)}(T) \right]^{M} \,\, ,\label{eq:full_coh_spins}
\eeq
where in the last equality we have exploited the fact that all environmental spins are equivalent.

For a qubit probe controlled by an $n$-pulse PDD sequence with characteristic frequency matching the Zeeman splitting of the environmental spins, $\omega_p = \omega_0$ (with total duration $T=(n+1)\pi/\omega_0$), the single spin decoherence can be calculated exactly, see \ref{app:W_1}. To lowest order in $\epsilon_{\perp/\parallel}$ the result reads
\begin{equation}
W^{(1)}(T)\approx\cos\left(\frac{\omega_0 T}{2\pi}\,\phi\right) = \cos\left[\frac{(n+1)}{2}\,\phi\right]\,\,,  \label{eq:W1phi}
\end{equation}
where the angle $\phi= 2\epsilon_\perp + O(\epsilon^3)$, and for $n \! \ll \! \epsilon^{-3}$ we have \cite{Ma_PRAPL16}:
\begin{equation}
W^{(1)}(T)\approx\cos\left(\frac{\omega_0 T}{\pi}\epsilon_\perp\right)=\cos\left[(n+1)\,\epsilon_{\perp}\right]  \,\, . \label{eq:W1cos}
\end{equation}
The full decoherence function is simply
\beq
W^{\rm qm}\left(T=\frac{(n+1)\pi}{\omega_0}\right ) \approx \cos^M\left[ (n+1)\epsilon_{\perp}  \right ] \,\, . \label{eq:WmuQ}
\eeq

The above quantum expression should be compared with the classical result, which we derive by treating the collection of environmental spins as a classical precessing magnetic moment $\mbox{\boldmath ${\mu}$} \! =\! \sum_{k=1}^{M}\mathbf{J}^{(k)}$. The time dependence of its components are given by
\begin{eqnarray}
\mu_{x}(t) & = & \mu_{x}(0)\cos \omega_0 t - \mu_{y}(0)\sin \omega_0 t \,\, , \label{eq:mux} \\
\mu_{y}(t) & = & \mu_{y}(0)\cos \omega_0 t + \mu_{x}(0)\sin \omega_0 t  \,\, , \label{eq:muy}  \\
\mu_{z}(t) & = & \mu_{z}(0) \,\, ,
\end{eqnarray}
and the coherence of the qubit reads
\begin{eqnarray}
W^{\rm cl}(T) = \Big\langle \exp&&\Big [-i \int_{0}^{T} \!\!\mathrm{d}t  f^{(\mathrm{PDD})}_{T}(t)\big( A_\parallel \mu_z(t)\nonumber\\
&&+ A_\perp( \mu_{x}(t)+\mu_y(t)\big) \Big ]  \Big \rangle \,. \label{eq:WmuCdef}
\end{eqnarray}
Here $\mean{\ldots}$ denotes the average over Gaussian distribution of initial values of the moments, $P(\mu_{x,y}(0)) =\! e^{-\mu_{x,y}(0)^2/2\sigma_\perp^2}/\sqrt{2\pi}\sigma_\perp$ and $P(\mu_z(0)) =e^{-\mu_z(0)^2 /2\sigma_\parallel^2}/\sqrt{2\pi}\sigma_\parallel$ , with $\sigma_{\perp/\parallel}^2 = M A^2_{\perp/\parallel}$, appropriate for large $M$ and high-temperature approximation for the density matrix of $E$. A simple calculation gives the attenuation factor, $\chi(T)$, given in Eq.~(\ref{eq:chi}), with spectral density of the noise found as
\beq
S(\omega) = \pi A^2_{\perp} \sigma^2 [ \delta(\omega+\omega_0) +  \delta(\omega-\omega_0) ] \,\, . \label{eq:Smu}
\eeq
Note that $A_\parallel$ disappears from the classical result since the contribution from $\mu_z(t)=\mu_z(0)$ is eliminated by the balanced DD sequence. For such a sharply peaked spectrum we cannot use the spectroscopic formula for $\chi(T)$ given by Eq.~(\ref{eq:spectro_formula}). Performing  exact integration of $S(\omega)$ with the exact form of $\tilde{f}^{(\mathrm{PDD})}_{T}(\omega)$ given by Eq.~(\ref{eq:FPDD}), we find
\beq
W^{\rm cl}\left(T=\frac{(n+1)\pi}{\omega_0}\right ) = e^{-\frac{1}{2}M(n+1)^2\epsilon_\perp^2} \,\, .  \label{eq:WmuC}
\eeq

 \begin{figure}[tb]
\epsfxsize=0.95\columnwidth
\vspace*{-0.0 cm}
\centerline{\epsffile{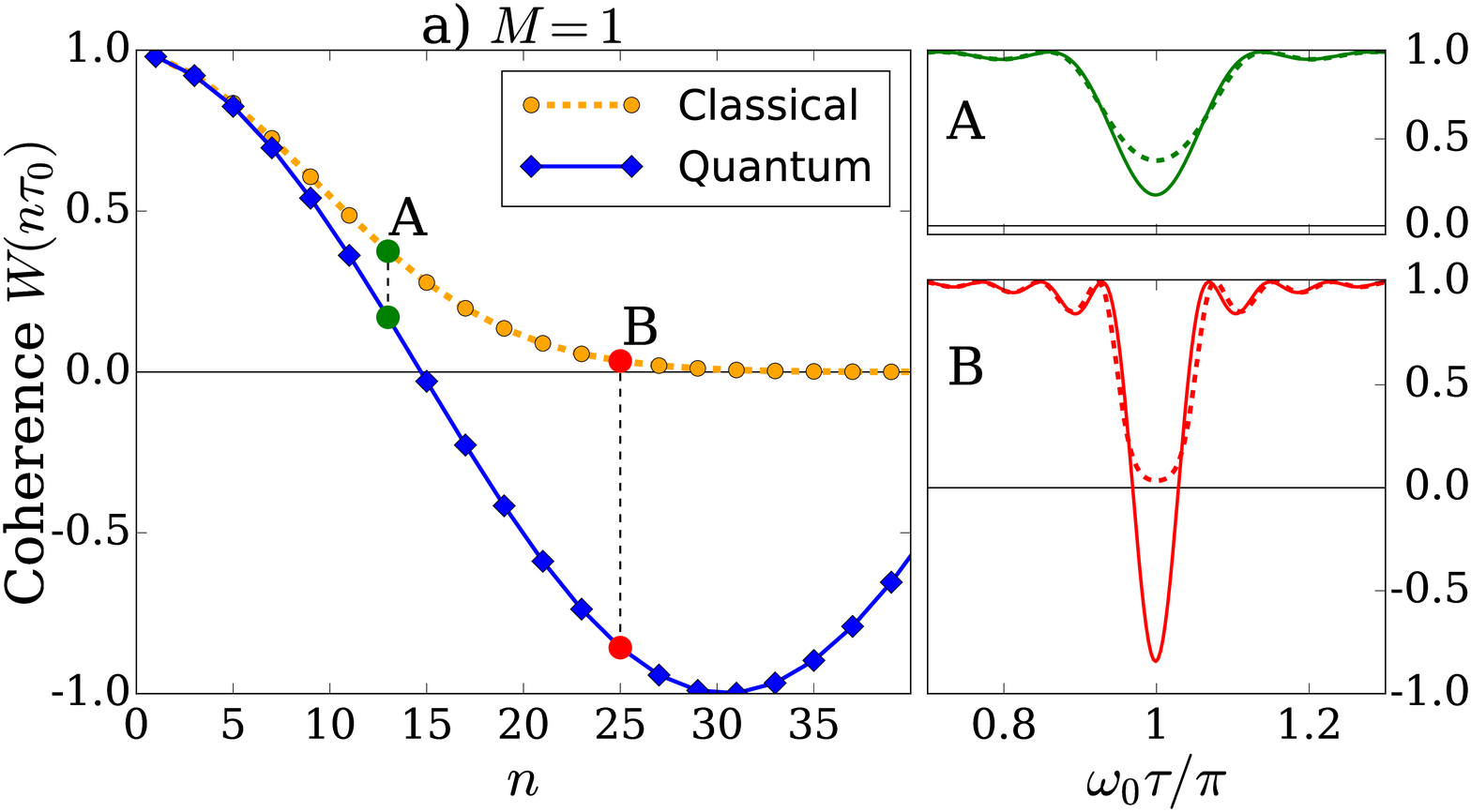}}
\vspace*{-0 cm}
\epsfxsize=0.95\columnwidth
\vspace*{-0.0 cm}
\centerline{\epsffile{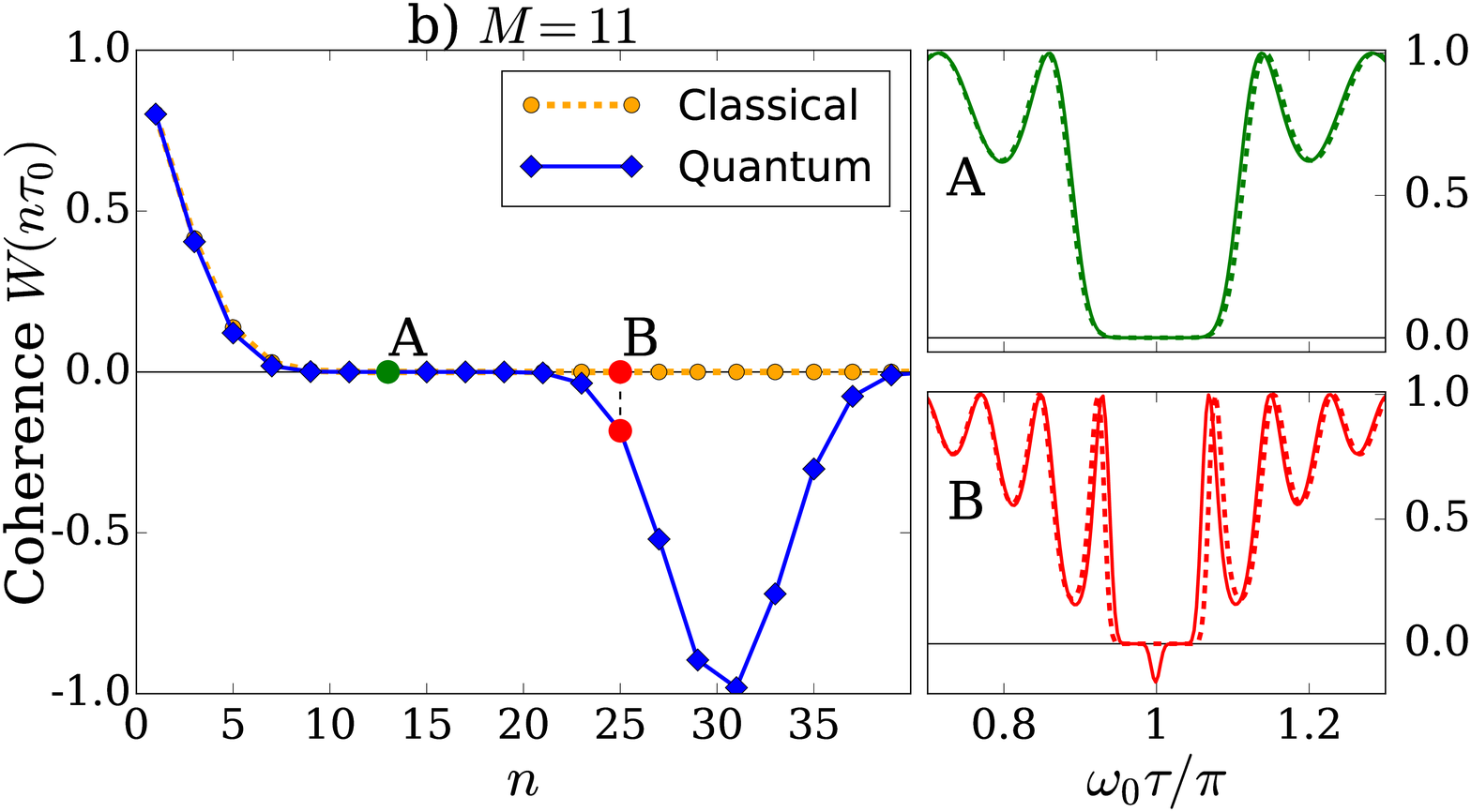}}
\vspace*{-0 cm}
\caption{The comparison of $W^\mathrm{cl}$ from Eq.~(\ref{eq:WmuC}) and $W^\mathrm{qm}$ from Eq.~(\ref{eq:WmuQ}) for a) $M=1$ and b) $M=11$ environmental spins. The left panels show quantum and classical coherences as functions of $T=n\tau_0$, where $\tau_0 = \pi/\omega_0$ -- the interpulse interval compatible with Zeeman splitting of environmental spins. The panels on right depict the coherences as a function of interpulse interval $\tau$ with PDD pulse sequences at $n=13$ in the upper, and $n=25$ in the bottom panel (corresponding to points $A$ and $B$, respectively, in the left panels).}
\label{fig:WclVSWqm}
\end{figure}

It is clear that the quantum expression, Eq.~(\ref{eq:WmuQ}), is qualitatively different from the classical and Gaussian one, Eq.~(\ref{eq:WmuC}), when $n\epsilon_\perp$ is not $\ll \! 1$. The exact result exhibits then large-amplitude oscillations about zero, while the classical result gives vanishing coherence.
Focusing on the $n\epsilon_\perp \! \ll \! 1$ regime, let us examine now under what conditions can the classical Eq.~(\ref{eq:WmuC}) be used for a {\em quantitative} analysis of the probed signal. It is important to take into account the following practical consideration: in a realistic situation, in which the coherence is measured with a reasonable accuracy, one could try to use the above formulas to extract information about the magnetic moment only if $W(T)$ is not too small. From Eq.~(\ref{eq:WmuC}) we see that this condition means that the number of pulses has to be $n \lesssim 1/\sqrt{M}\epsilon_{\perp}$, and for such $n$ the relative difference between the classical and the quantum expressions is
\beq
\frac{|W^{\rm qm}(T) - W^{\rm cl}(T)|}{W^{\rm cl}(T)} \approx \frac{M(n\epsilon_\perp)^4}{12} \lesssim \frac{1}{12 M}\underset{M\to\infty}{\longrightarrow} 0 \,.
\eeq
We see that in this ``experimentally practical'' regime, the error made by using the classical result, Eq.~(\ref{eq:WmuC}) is small, and it diminishes with increasing number of spins $M$. In other words, the larger the magnetic moment, the more classical the noise becomes (provided that $n\epsilon_\perp \! \ll \! 1$). This is an example of ``Gaussianization via the Central Limit Theorem'', but it should be noted that by restricting our considerations to the ``experimentally practical'' regime we obtained this result even without having to assume the scaling of couplings $A\propto 1/\sqrt M$. This is basically the same as the effect shown in Figs.~\ref{FigRTN1}(c) and  \ref{FigRTN1}(d), where one can see that for a large number of noise sources, the characteristic decoherence time $T_{2}$ is well-described by the Gaussian approximation.

The above discussion is illustrated in Fig.~\ref{fig:WclVSWqm}, in which we compare $W^{\rm cl}(T)$ and $W^{\rm qm}(T)$ for $\epsilon_\perp = 0.05$ and $M \! = \! 1$, $11$. The figures also demonstrate the dependence of the classical and quantum decoherence function on detuning of the sequence characteristic frequency, $\omega_p =\pi/\tau$, from $\omega_0=\pi/\tau_0$.

\section{Extensions of the dynamical-decoupling based methods}  \label{sec:extensions}
In this Section we describe some recent theoretical proposals for extending the DD-based noise spectroscopy method in two directions: quantitatively characterizing the properties of non-Gaussian noises (i.e.~reconstructing the higher-order noise polyspectra defined in Sec.~\ref{sec:gaussianizationNS}), and using multiple qubits, each affected by a noise source, to sense the presence of cross-correlations between various noises, and to reconstruct the corresponding cross-spectra.

\subsection{Spectroscopy at an Optimal Working Point}  \label{sec:OWP}
As discussed in Sec.~\ref{sec:noises}, one often encounters a situation in which the qubit is coupled to a square of noise, i.e.~we have the qubit-noise coupling given by $b_z(t)=v\xi^{2}(t)$, see Eq.~(\ref{eq:HOWP}) and preceding discussion. Such Optimal Working Points (also known as sweet spots or clock transitions) have been identified for   superconducting charge \cite{Vion_Science02,Ithier_PRB05} and flux \cite{Yoshihara_PRL06,Kakuyanagi_PRL07} qubits, semiconductor quantum dot (QD) charge qubits \cite{Petersson_PRL10}, mixed electronic-nuclear spin qubits based on electrons bound to bismuth donors in silicon \cite{Wolfowicz_NN13,Balian_PRB14},
and triple QD spin qubits \cite{Medford_PRL13,Malinowski_REQ_arXiv17}.
It is crucial to note that even if the $\xi(t)$ process is Gaussian, $\xi^2(t)$ is {\em not}.
The coherence function, $W(T)$, can then be written using the cumulant expansion as in Eq.~(\ref{eq:Wdefcum}). Assuming that $\xi(t)$ is Gaussian, the attenuation factors can be expressed in terms of the spectral density of $\xi(t)$, $S(\omega)$ \cite{Makhlin_PRL04,Cywinski_PRA14}
\begin{eqnarray}
\chi_{k}
& = & \frac{(2v)^{k}}{2 k!} \int \frac{\mathrm{d}^k\omega_{i}}{(2\pi)^k} \prod_{i=1}^{k}S(\omega_i)\nonumber\\
&&\times \tilde{f}_T(\omega_1-\omega_2)\tilde{f}_T(\omega_2-\omega_3)\ldots \tilde{f}_T(\omega_k-\omega_1) \, .\label{eq:chi_k_OWP}
\end{eqnarray}
This result appeared first in a seminal work that considered a qubit, quadratically coupled to a bath of noninteracting bosonic modes, generating Gaussian quantum noise, evolving freely with no pulse sequence applied \cite{Makhlin_PRL04}. A generalization to the case of qubit subjected to dynamical decoupling, constructed with noise spectroscopy as the goal, was given recently \cite{Cywinski_PRA14}. The two main results of that work are summarized below.

Firstly, for $\xi(t)$ with a finite correlation time $\tau_c$, it is possible to reach the spectroscopic regime, $T\gg \tau_c$ (see Sec. \ref{sec:periodic_ddns}), where the cumulant expansion can be truncated to $\chi_2$ (i.e. the Gaussian approximation to $W(T)$), and the spectroscopy method of Sec. \ref{sec:spectrum_reconstruction} becomes applicable
\begin{equation}
-\frac{\log W(T)}{T} \underset{T\gg \tau_c}{\longrightarrow} \chi_2 = \sum_{m>0}|c_{m\omega_s}|^2 v^2 S_{\xi^2}(m\omega_s)\,.
\end{equation}
Here $S_{\xi^2}(\omega)$ is the spectrum of square noise $\xi^2(t)$ and it is found to be given by the convolution of the spectra of $\xi(t)$,
\beq
S_{\xi^2}(\omega) = \int_{-\infty}^{\infty} \frac{\mathrm{d}\omega'}{\pi} S(\omega') S(\omega'-\omega) \,.
\eeq

Secondly, for low-frequency $\xi(t)$, where most of the noise power is concentrated at very low frequencies (e.g. noise with spectrum of form $1/|\omega|^\beta$, $\beta>1$), it is possible to approximate the {\it full} cumulant series with a closed expression. The physical picture behind this approximation is as follows. The noise $\xi(t)$ can be split into fast (high-frequency, $\mathrm{HF}$) part and slow, practically static part (low-frequency, $\mathrm{LF}$), $\xi(t) \approx \xi_\mathrm{LF}+\delta\xi_\mathrm{HF}(t)$. Each part is statistically independent, because of the vast difference between slow and fast time scale. Moreover, since we're dealing with low-frequency noise, the quasi-static part is much larger then the fast, dynamic part, hence
\beq
\xi^2(t) =  (\xi_\mathrm{LF}+\delta\xi_\mathrm{HF}(t))^2 \approx \xi^{2}_{\rm LF} + 2\xi_{\mathrm{LF}}\delta \xi_\mathrm{HF}(t)\, . \label{eq:xi_separation}
\eeq
The quasi-static shift of the qubit splitting, $\xi_\mathrm{LF}$, changes randomly between measurements, but not during each run. Therefore, for a qubit subjected to a balanced pulse sequence, the LF term, $\xi_\mathrm{LF}^2$, is removed completely, and the remaining dynamical term, $2\xi_\mathrm{LF}\delta\xi_\mathrm{HF}(t)$ is filtered by a frequency comb. Due to the assumed independence between the LF and HF parts, the averaging over the high- and low-frequency noise is carried out in steps: first, one have to evaluate a Gaussian average over the process $\delta \xi_\mathrm{HF}(t)$ for fixed $\xi_{\rm LF}$, and then average the result over a Gaussian distribution of $\xi_{\rm LF}$.
This physical picture of a slow noise renormalizing the qubit coupling to a fast noise was first elucidated in works devoted to dynamical decoupling from $1/f$ noise originating from many two-level fluctuators, each of them being a source of non-Gaussian Random Telegraph Noise \cite{Bergli_PRB06,Ramon_PRB15}. The slowest (and more numerous) fluctuators shift the position of the qubit's ``working point'' from the optimal one, activating the linear coupling to fast fluctuators.

Using the above approximation one arrives at $\chi_{2k} \! \approx (-2)^{k+1}\chi^k_2 / k$ (only even orders are considered because the odd ones can be eliminated by application of sequence with odd number of pulses), and the resummation of the cumulant expansion gives
\beq
W(T) \approx \frac{1}{\sqrt{1+2\chi_2(T)}}  \,\, , \label{eq:WOWP}
\eeq
where $\chi_2(T)$ is given by Eq.~(\ref{eq:chi_k_OWP}) with $k=2$. In the spectroscopic limit ($T$ much larger then the correlation time of the HF part) we can use the spectroscopic formulas from Sections \ref{sec:periodic_ddns} and \ref{sec:spectrum_reconstruction} to approximate $\chi_2(T)$, allowing us to carry out noise spectroscopy for $S_{\xi^2}(\omega)$.

We note that for a quadratic coupling to low-frequency Gaussian noise, we obtain a power-law decay: at long $T\! =\! n\tau$ we have $W(n\tau) \propto 1/\sqrt{n}$ while for linear coupling we have $W(n\tau)\! \propto \! \exp(-n)$. This is a rather general feature of the decoherence due to low-frequency quadratic noise (or low-frequency transverse noise in the presence of large longitudinal splitting of the qubit) that appears also for free evolution \cite{Cucchietti_PRA05,Cywinski_PRL09,Cywinski_PRB09,Cywinski_APPA11,Hung_PRB13,Szankowski_QIP15} and Rabi oscillations decay \cite{Koppens_PRL07,Dobrovitski_PRL09}. In addition, similarly to the case of free evolution discussed in Sec.~\ref{sec:filtering}, and in contrast to the case of DD for {\it linear} coupling to noise, $\chi_{2}(T)$ in Eq.~(\ref{eq:WOWP}) depends on the low-frequency cutoff of the noise (i.e.~the total data acquisition time),
see \cite{Cywinski_PRA14} for details (the fact that $\chi_{2}$ depends now on the total power of LF noise should be clear from the preceding discussion). Consequently, special care should be taken when interpreting DD data for the qubit at Optimal Working Point subjected to low-frequency noise (for examples of recent experiments see e.g.~\cite{Medford_PRL13,Malinowski_REQ_arXiv17}.

\subsection{Spectroscopy of polyspectra of general non-Gaussian noises} \label{sec:nonGaussian}
In the previous section we discussed an approximate resummation of the cumulant expansion that is appropriate for the special case of quadratic coupling to a Gaussian noise. When dealing with a general non-Gaussian noise, one needs to address the cumulant expansion, Eq.~(\ref{eq:Wdefcum}), on a term-by-term basis. Gaussian noise is fully described by its spectral density. The problem of characterizing a non-Gaussian noise can become vastly more complex, as one needs to address the whole series of cumulants of Eq.~(\ref{eq:Wdefcum}). In particular, Marcinkiewicz's theorem suggests that Gaussian noise is the only noise with finite number of non-zero cumulants (namely one if the average is assumed to vanish) \cite{Marcinkiewicz_MZ39}. One approach would be to split the series into the Gaussian part $\chi_2$ (i.e. only the second order attenuation factor) and $\chi_\mathrm{NG} = \sum_{k>2}(-i)^k\chi_k$, and to use this remainder, as a whole, to characterize the non-Gaussian noise (see Eq. (\ref{eq:W_G_W_NG_split})).

An alternative approach for the case of stationary noise was proposed in \cite{Norris_PRL16}. It assumes that, although infinite, the cumulant series can be truncated at some point. If that is the case, then the cumulants, or rather their multi-dimensional Fourier transforms termed {\it polyspectra} (see Sec.~\ref{sec:gaussianizationNS}), can be reconstructed term-by-term using a properly extended DD method. We point that this assumption is expected to be valid up to a certain timescale, see e.g.~Fig.~3 in Ref.~\cite{Cywinski_PRA14}, in which the contribution of the 4th cumulant to decoherence caused by quadratic coupling to Ornstein-Uhlenbeck noise was shown to be a small correction up to a certain $T$ while keeping $n$ fixed.
On the other hand, at long $T$ (in the spectroscopic limit of $T\gg \tau_c$), the decoherence due to strongly coupled RTN source discussed in Sec.~\ref{sec:RTN} (with reconstructed first spectrum shown in Fig.~\ref{FigRTN2}(c))  could be described well only when taking into account all the cumulants (i.e.~using an exact solution).

Starting from formulas relating $\chi_{k}$ to noise cumulants $C_{k}(t_1,\ldots,t_{k})$ given in Sec.~\ref{sec:gaussianizationNS}, we can generalize the derivation from Sec.~\ref{sec:periodic_ddns} of the ``spectroscopic limit'' of $\chi_{2}$ to the case of higher-order $\chi_{k}$.  
For stationary noise, i.e. a noise for which $C_{k}(t_0 +t_1,\ldots t_0+t_k) = C_{k}(t_1,\ldots t_k)$ for any $t_0$, in the spectroscopic regime of $T\gg \tau_c$ we obtain
\begin{eqnarray}
&& \frac{\chi_{k}(T)}{T} \underset{T\gg \tau_c}{\longrightarrow}  \frac{1}{2^{k-1}}\sideset{}{'}\sum_{m_1\ldots m_k} \left(\prod_{i=1}^k c_{m_i\omega_p}\right)\nonumber\\
&&\times S_{k-1}\big[ m_1\omega_p,(m_1+m_2)\omega_p, \ldots ,\sum_{i=1}^{k-1}m_i \omega_p\big]\,\, , \label{eq:polyspec_formula}
\end{eqnarray}
in which $\sum'$ indicates that only terms for which $\sum_{i=1}^k m_i=0$ should be included in the sum, $c_{m\omega_p}$ are the Fourier series coefficients of the filter function $f_T(t)$ with characteristic frequency $\omega_p$ (see Eq.~(\ref{eq:c})).
This is the generalization of spectroscopic formula (\ref{eq:spectro_formula}) extended to include multi-dimensional spectra of the noise \cite{Norris_PRL16}.

In section \ref{sec:spectrum_reconstruction} we have demonstrated that the spectral density of Gaussian noise (i.e. $S_{1}(\omega)$ in nomenclature of polyspectra) can be reconstructed by solving a set of linear equations obtained form measurements of qubit's coherence function $W^{(s)}(T_s)$ in a series of experiments run with a set of appropriately chosen pulse sequences $f_{T_s}^{(s)}$. Building on this approach, we consider the linear relations between measured $W^{(s)}(T_s)$ and the polyspectra picked out by the multidimensional frequency comb truncated to cut-off $m_c$:
\begin{eqnarray}
&&\frac{\log W^{(s)}(T_s)}{T_s} \approx \sum_{k=2}^K (-i)^k \frac{\chi_k^{(s)}(T_s)}{T_s}\approx\nonumber\\
&& \sum_{k=2}^K \frac{(-i)^k}{2^{k-1}} \!\!\!\sideset{}{'}\sum^{m_c}_{m_1\ldots m_k}\!\!\left(\prod_{i=1}^k c^{(s)}_{m_i \omega_s}\right)\!\!S_{k-1}[\boldsymbol{\omega}(m_1\ldots m_k)]\,,
\end{eqnarray}
thus forming a system of equations with known coefficients $\prod_{i=1}^k c^{(s)}_{m_i \omega_s}$ (excluding those with $\sum_i m_i \neq 0$) for a set of unknowns $S_{k-1}[\boldsymbol{\omega}(m_1\ldots m_k)]$ with $k$ ranging from $2$ to assumed cumulant series cut-off $K$.

An immediate difficulty with the above approach is that even with only the first several cumulants retained (small $K$), and a modest comb cut-off $m_c$, the high dimensionality of the frequency domains necessitates a large number of sequences for a successful reconstruction of the polyspectra. Norris {\it et al.} have proposed in \cite{Norris_PRL16}
to resolve this issue by employing repeated base sequences composed of combinations of randomly drawn Concatenated DD (CDD) sequences, which include unbalanced sequences, capable of reconstructing zero frequency components -- the latter becoming increasingly important necessary for reconstruction of higher-order polyspectra. Nevertheless, the practical implementation will be likely limited by the exponential scaling of the needed number of measurements with the reconstructed polyspectrum order. Moreover, the conversion matrices that connect the measured signals with the polyspectra tend to become ill-conditioned as their size increases, resulting in numerical instabilities. We emphasize that, at the least, this method should be valuable in detecting and quantifying the leading non-Gaussian contributions by enabling the reconstruction of the noise bi- and tri-spectrum ($S_{2}$ and $S_{3}$, respectively).

\subsection{Noise spectroscopy with multiple qubits}  \label{sec:multiple}
The continuing advances in control of multiple qubits, including creation of qubit entanglement in systems in which single-qubit DD-based noise spectroscopy was succesfully implemented (see e.g.~\cite{Shulman_Science12,Brunner_PRL11} for entanglement of QD-based spin qubits, \cite{Dolde_NP13} for entanglement of NV centers, and \cite{Steffen_Science06,DiCarlo_Nature10} for entanglement of superconducting qubits), show that noise spectroscopy with multiple qubits is within experimental reach. It is well known that the multi-qubit decoherence, and consequently entanglement dynamics \cite{Aolita_RPP15}, can depend qualitatively on the presence and degree of correlations between the noises affecting distinct qubits \cite{Gordon_PRL06,Gordon_JPB11,DArrigo_NJP08,Zhou_QIP10,Benedetti_PRA13,Szankowski_QIP15}, and when all the qubits are experiencing the same (common) noise, decoherence-free subspaces, hosting entangled states immune to such noise, appear \cite{Duan_PRA98,Lidar_ACP14}. It is also well understood that dynamical decoupling suppresses decoherence and disentanglement of multi-qubit states \cite{Gordon_JPB11,Jiang_PRA11,Pan_JPB11,Wang_PRB11,LoFranco_PRB14,Orieux_SR15,Paz_NJP16}, even if the unitary operations are applied to single qubits only \cite{Gordon_JPB11,DArrigo_AP14,Bragar_PRB15,Paz_NJP16}. However, the DD-based spectroscopic characterization of cross-correlations of multiple noises has begun to be explored only quite recently \cite{Szankowski_PRA16,Paz_NJP16,Paz_PRA17}.  Such a characterization can not only give interesting information about the relevant environments, but it also can have important implications for theory of fault-tolerance of quantum computers, since the latter relies on assumptions about the strength of cross-correlations between errors experienced by nearby qubits, see \cite{Norris_PRL16} and references therein.

Let us also note that while using entangled qubits was argued to improve estimation of environmental noise parameters \cite{Rossi_PRA15}, the methods discussed below do not require entanglement, only the presence of coherence between mutliple-qubit states. The latter can be nonzero for separable states \cite{Szankowski_PRA16}, although its value is maximal for entangled states.

\subsubsection{Noise cross-correlation spectroscopy with two qubits.}  \label{sec:cross}
The standard single-qubit noise spectroscopy method has an inherent flaw: even if the noise coupled to the qubit originates from distinct sources it is always registered as a single spectrum. For example, suppose that the environment probed by the qubit is actually composed of two independent parts, $A$ and $B$, each emitting its own noise $b_A$ and $b_B$, respectively ($\hat b_A$ and $\hat b_B$ in the quantum case). A single qubit would couple to the total noise $b_z = b_A + b_B$ (or $\hat b_z = \hat b_A+ \hat b_B$ in quantum case), without any means to resolve the two constituents. Unless the resulting spectrum has very characteristic features that would suggest its composite nature, it might be outright impossible to notice that in reality there are two distinct sources behind the signal. In \cite{Szankowski_PRA16} it was proposed that this weakness can be overcome by employing noise spectroscopy with a two-qubit probe instead of single-qubit one.

The probe is assumed to comprise two noninteracting qubits with a Hamiltonian $\hat H_{Q} = \sum_{\alpha=1,2} (\Omega^{(\alpha)}/2)\hat\sigma_z^{(\alpha)}$ being a straightforward extension of single-qubit case. Here $\hat\sigma^{(\alpha)}_i$ is the Pauli operator of the $\alpha$-th qubit. In addition, the two qubits are independently controlled by pulse sequences, which crucially can generate distinct filter functions, $f_T^{(\alpha)}(t)$ for each of them. The qubits coupling to a pure phase noise is defined as
\begin{equation}
\hat V_{QE} = \sum_{\alpha=1,2}\frac{1}{2}b_z^{(\alpha)}(t)\hat\sigma_z^{(\alpha)}
	= \sum_{\alpha=1,2}\frac{1}{2}v^{(\alpha)}\xi^{(\alpha)}(t)\hat\sigma_z^{(\alpha)}\,,
\end{equation}
where $b_z^{(\alpha)}(t)$ is the noise affecting qubit $\alpha$. This model covers a wide variety of possible scenarios, falling into three categories:
\begin{enumerate}
\item Each qubit is coupled to distinct source of noise. In reference to the example described above, the case when $b_z^{(1)} = b_A$ and $b_z^{(2)}=b_B$ would be included into this category.
\item The qubits are coupled to the same source of noise, but with different couplings. For example, if the qubits are located at different positions and the source-qubit coupling law depends on the relative distance. In such a case $b_z^{(\alpha)}(t) = v^{(\alpha)}\xi(t)$ with $v^{(1)}\neq v^{(2)}$ and $\xi(t)$ representing the noise emitted by the single source.\label{enum:same_noise}
\item The qubits are coupled to the same source but the noise at one qubit is retarded with respect to the noise at the other. For example, if the signal emitted by the source has a finite effective propagation speed $c$ then $\xi^{(2)}(t) = \xi^{(1)}(t - d/c)$ where $d$ is the distance between the qubits.
Another important example of retardation is encountered for qubits coupled to a precessing magnetic moment, as in Sec.~\ref{sec:mu}. The noises $\mu_{x}(t)$ and $\mu_{y}(t)$ appearing in Eq.~(\ref{eq:WmuCdef}) are in fact delayed one with respect to another by $\pi/(2\omega_{0})$, where $\omega_{0}$ is the magnetic moment precession frequency, see Eqs.~(\ref{eq:mux}) and (\ref{eq:muy}).
\end{enumerate}
Of course, these categories are not mutually exclusive and in general the noises $b_z^{(\alpha)}$ can arise as a combination of all above types.

The relationship between the noises $b_z^{(\alpha)}$ is captured by the properties of the cross-correlation function
\begin{equation}
C_{12}(t-t') = \langle b_z^{(1)}(t) b_z^{(2)}(t') \rangle
\end{equation}
and its Fourier transform, the cross-spectrum
\begin{equation}
S_{12}(\omega) = \int_{-\infty}^\infty e^{-i \omega t}C_{12}(t)\mathrm{d}t\,.
\end{equation}
Conversely, the spectroscopy of the cross-spectrum $S_{12}(\omega)$ can provide insight into the structure of the environment as a source of noise in greater detail then it is possible with a single qubit.

The simplest nontrivial case is found when the noise at each qubit is a sum of noises emitted by distinct and {\it independent} sources, e.g. $b_z^{(\alpha)}(t) = v_A^{(\alpha)} \xi_A(t) + v_B^{(\alpha)} \xi_B(t)$. Even though the bare noises $\xi_A$ and $\xi_B$ are assumed to be independent (i.e. their cross-correlation vanishes $\langle \xi_A(t)\xi_B(t') \rangle = \langle \xi_A(t)\rangle \langle \xi_B(t') \rangle = 0$
they are cross-correlated at the qubits level, $C_{12}(t-t') = v_A^{(1)}v_A^{(2)}\langle \xi_A(t)\xi_A(t')\rangle + v_B^{(1)}v_B^{(2)}\langle \xi_B(t)\xi_B(t')\rangle$. Since the cross-correlation function is a sum of auto-correlations of classical noises, the corresponding cross-spectrum is real, $S_{12} = v_A^{(1)}v_A^{(2)}S_A + v_B^{(1)}v_B^{(2)}S_{B}=S_{12}^*$. Comparing the cross-spectrum with the self-spectra of the noises affecting each qubit, $S_{\alpha \alpha} = (v_A^{(\alpha)})^2 S_A + (v_B^{(\alpha)})^2 S_B$, allows us to recognize the presence of two noise sources --- a task that cannot be accomplished with single-qubit spectroscopy.

When the noise sources are not independent, for example due to their mutual interaction, the cross-correlation of the bare noises does not vanish. Moreover, since $A$ and $B$ interact, there is a {\it causal relation} between the bare noises -- the value of $\xi_A$ depends on the value of $\xi_B$ at previous times and {\it vice versa}. This means that the cross-correlation function between $\xi_A$ and $\xi_B$ does not have to be symmetric, which implies the same the for cross-correlation between  noises experienced by the qubits, $C_{12}(t-t') \neq C_{12}(t'-t)$, see \cite{Szankowski_PRA16} for an example of a model of interacting sources.
The Fourier transform of a non-symmetric function yields a complex valued cross-spectrum, in contrast to the purely real $S_{12}$ for non-interacting sources. Therefore, the examination of the cross-spectrum can not only reveal the presence of multiple noise sources, but also contains signatures of interactions among them.

Causal relations between qubit noises arise not only due to sources interactions. The third category listed above -- noise retardation -- is counted as such relation as well. Consequently, the cross-spectrum between retarded signals is likewise a complex valued function, but with a very specific form. Taking the example discussed previously, where the retardation between noises is caused by a finite signal propagation, such that $\xi^{(2)}(t) = \xi^{(1)}(t - d/c)$, the cross-spectrum is given by $S_{12}(\omega)
=(v^{(2)}/v^{(1)})e^{i(d/c)\omega}S_{11}(\omega)=(v^{(1)}/v^{(2)})e^{i(d/c)\omega}S_{22}(\omega)$, where $S_{\alpha\alpha}$ is the (purely real) self-spectrum of the noise sensed by qubit $\alpha$. Again, comparison between the self-spectra and the cross-spectrum allows to detect the additional phase factor characteristic of this type of noise relation. Additionally, provided the distance $d$ between the qubits is known, the two-qubit probe can be used to determine the signal propagation speed $c$.

\subsubsection{Reconstruction of the cross-spectrum.}\label{sec:recon_cross}

Analogously to the single-qubit case (see Sec. \ref{sec:PD}) pure phase noise affects only off-diagonal elements of the two-qubit density matrix in the product basis of $\hat\sigma^{(\alpha)}_z$ eigenstates,
\begin{eqnarray}
&&\rho_{\sigma_1\sigma_2,\sigma_1'\sigma_2'}(T)= \langle \sigma_1|\langle\sigma_2|\langle\hat\rho(T)\rangle|\sigma_1'\rangle|\sigma_2'\rangle\nonumber\\
&&=\rho_{\sigma_1\sigma_2,\sigma_1'\sigma_2'}(0)\langle e^{-\frac{i}{2}\sum_{\alpha}(\sigma_\alpha-\sigma_\alpha')\int_0^T f_T^{(\alpha)}(t)b_z^{(\alpha)}(t)\mathrm{d}t}\rangle
\end{eqnarray}
where $\sigma_\alpha,\sigma_\alpha'=\pm$ and $f_T^{(\alpha)}(t)$ is the filter function of the pulse sequence applied to the $\alpha$th qubit. Measuring the elements $\rho_{\sigma_1\sigma_2,{-\sigma_1}\sigma_2}(T)$ and $\rho_{\sigma_1\sigma_2,\sigma_1{-\sigma_2}}(T)$ eliminates contributions from $b_z^{(2)}$ and $b_z^{(1)}$, respectively, yielding self-spectra of the qubit noises via means of standard single-qubit spectroscopy. In order to gain access to the cross-spectrum we must measure either $\rho_{{+}{-},{-}{+}}(T)$ or $\rho_{{+}{+},{-}{-}}(T)$. The initial value of the former matrix element is maximal for the entangled Bell states $|\Psi_{\pm}\rangle=(|{+}\rangle|{-}\rangle\pm|{-}\rangle|{+}\rangle)/\sqrt 2$, while the latter is maximal for the equally entangled GHZ states $|\Phi_\pm\rangle = (|{+}\rangle|{+}\rangle\pm|{-}\rangle|{-}\rangle)/\sqrt 2$.

Under the Gaussian noise assumption, the coherence functions $W(T)$ (see Sec. \ref{sec:WPD}) associated with $\Psi/\Phi$ -- type coherences read
\begin{equation}
W_{\Psi/\Phi}(T) =e^{ -\chi_{11}(T) - \chi_{22}(T) \pm 2 \chi_{12}(T)}\, ,
\end{equation}
Where $\chi_{\alpha\alpha}(T)$ is the local attenuation factor derived from noise sensed by the $\alpha$th qubit, $b_z^{(\alpha)}$, given by the single-qubit formula, Eq.~(\ref{eq:chi}). The attenuation factor related to the cross-spectrum is found as \cite{Szankowski_PRA16}:
\begin{eqnarray}
\chi_{12}(T) &=&\frac{1}{2}\int_{-\infty}^\infty \frac{\mathrm{d}\omega}{2\pi}\tilde{f}^{(1)}_T(\omega)\tilde{f}_T^{(2)}(-\omega)S_{12}(\omega)\nonumber\\
&\approx&T\sum_m \mathrm{Re}\{c_{m\omega_p}^{(1)}(c_{m\omega_p}^{(2)})^*\}\mathrm{Re}\{S_{12}(m\omega_p)\}\nonumber\\
&&- T\sum_m\mathrm{Im}\{c_{m\omega_p}^{(1)} (c_{m\omega_p}^{(2)})^*\}\mathrm{Im}\{S_{12}(m\omega_p)\}\,,
\end{eqnarray}
where the approximate result is valid in the spectroscopic regime and $c_{m\omega_p}^{(\alpha)}$ are the Fourier series coefficients of $f_T^{(\alpha)}(t)$ (see Sec.~\ref{sec:periodic_ddns}). Assuming the local attenuation factors, $\chi_{\alpha\alpha}$, are known (for example, from measurements of $\rho_{\sigma_1\sigma_2,{-\sigma_1}\sigma_2}(T)$ and $\rho_{\sigma_1\sigma_2,\sigma_1{-\sigma_2}}(T)$) the measurement of $W_{\Psi/\Phi}$ allows to reconstruct the cross-spectrum using a procedure similar to the one described in Sec.~\ref{sec:periodic_ddns}. We note that an analogous formula for two-qubit decoherence was derived in \cite{Gordon_JPB11} for two qubits driven by ac fields with periodically modulated amplitude or phase.

The flexibility in choosing independent pulse sequences for each qubit allows to target the specific parts of the cross-spectrum. Applying identical sequences enables the reconstruction of the real part of $S_{12}$, since $c_{m\omega_p}^{(1)} = c_{m\omega_p}^{(2)}$ and $c_{m\omega_p}^{(1)}(c_{m\omega_p}^{(2)})^* = |c_{m\omega_p}^{(1)}|^2 = |c_{m\omega_p}^{(2)}|^2$ are real, and the term proportional to $\mathrm{Im}\,S_{12}$ vanishes. The access to the imaginary part of the cross-spectrum is gained by exploiting the relation between the Fourier series coefficients of equivalent sequences, described in Sec.~\ref{sec:periodic_ddns}. In particular, in \cite{Szankowski_PRA16} it was proposed to use $f_T^{(1)}=f_T^{(\mathrm{PDD})}$ and $f_T^{(2)}=f_T^{(\mathrm{CP})}$, both having the same characteristic frequency $\omega_p \! =\! \pi/\tau$ (where $\tau$ is the interpulse time), so that the PDD sequence consists of $n$ pulses (with $n$ being odd), and the CP sequence has $n+1$ pulses. Then the coefficients of the first sequence are purely imaginary, $c_{(2m+1)\omega_p}^{(1)}=2/(i\pi (2m+1))$, and those of the second are purely real, $c_{(2m+1)\omega_p}^{(2)}= e^{i \frac{\pi}{2} (2m+1)} c_{(2m+1)\omega_p}^{(1)}=(-1)^m 2/(\pi(2m+1))$. In this case the only part of $\chi_{12}(T)$ that survives is the one proportional to $\mathrm{Im}\,S_{12}$, allowing the reconstruction of the imaginary part of cross-spectrum. Such a reconstruction was performed in \cite{Szankowski_PRA16} using a model system consisting of two interacting TLSes (i.e.~the noise consisted of two causally cross-correlated RTN signals), inspired by recent reports \cite{Burin_PRB15,Muller_PRB15,Lisenfeld_NC15,Burnett_NC14} of signatures of interaction between TLSes affecting superconducting qubits.

\subsubsection{Multiqubit spectroscopy of classical and quantum noise.}
In a pair of recent works \cite{Paz_NJP16,Paz_PRA17} a broad generalization of the single- and two-qubit noise spectroscopy methods was presented. It encompassed the use of multiple qubits for sensing not only classical, but also quantum noise spectra, associated with a commutator of the relevant bath operators instead of an anticommutator appearing in Eq.~(\ref{eq:AQ}). The latter was made possible by considering a more general class of DD operations, involving SWAP operations applied to pairs of qubits. Utilizing composite DD sequences with specific timing symmetries, Paz-Silva {\it et al} proposed protocols that are able to generate frequency combs suitable for quantum noise reconstruction \cite{Paz_PRA17}, similarly to the way in which the CP/PDD pair of sequences described in the previous section gave access the imaginary part of the cross-spectrum of classical noise. Note that with both classical and quantum noise reconstructed in some frequency range, one gains access to temperature of the environment \cite{Paz_PRA17}, since the ratio of the Fourier transform of symmetric and antisymmetric combination of bath operators at frequency $\omega$ is proportional to $\coth \beta \omega/2$, where $\beta$ is the inverse temperature of the bath.

A different extension of the DD-based noise spectroscopy method introduced in \cite{Paz_PRA17} was to consider a more general form of pure dephasing Hamiltonians, given by $\ket{+}\bra{+}\hat{b}_{+} + \ket{-}\bra{-}\hat{b}_{-}$. Note that above we focused on the commonly encountered case of $\hat{b}_{-}=-\hat{b}_{+} = -\hat{b}_{z}/2$ leading to Eq.~(\ref{eq:Hpd}), but for quantum-dot cased excitonic qubits interacting with phonons \cite{Roszak_PRA06,Krzywda_SR16} and for NV center qubits employing $m=0$ and $m=1$ levels of spin-one center, one has $\hat{b}_{-}=0$. In the latter case, for a bath of noninteracting bosons (i.e.~a quantum Gaussian environment) the coherence under a DD sequence acquires nontrivial phase dynamics related to the quantum part of the noise \cite{Paz_PRA17}.

\section{Summary and outlook}
The dynamical-decoupling spectroscopy methods described in this review are now widely used to characterize the environments affecting the coherence of various kinds of qubits. This means that single qubits and, hopefully soon, multi-qubit registers are being used as sensitive probes of their nanoscale environments, allowing to gain new insights into the physics of nanoscale materials, single molecules, and devices. Characterization of the correlation of noises experienced by multiple qubits is also expected to be crucial for realistic implementations of quantum error correction protocols.

We hope that we have explained the theoretical basis of these methods, and that we have elucidated the reasons for their success. The latter was not {\em a priori} obvious, because the widespread applicability of the classical Gaussian noise model in describing the environmental influence on qubits was rather unexpected. Some (but likely not all) of the reasons for this success have been discussed here: operating the qubit in the regime where its coupling to the environment is characterized by pure dephasing, the suppression of non-Gaussian features of the noise filtered by dynamical decoupling, and the emergence of Gaussianity for environments consisting of multiple noninteracting (or weakly interacting) subsystems.

Possible directions of future research include full elucidation of the role of quantum dynamics of the environment (including practical refinements of quantum noise spectroscopy protocols \cite{Paz_PRA17}), finding control protocols that will enable discrimination between classical non-Gaussian noise and quantum non-Gaussian noise, a more careful analysis of the influence of imperfect pulses (see e.g.~\cite{Loretz_PRX15} for a recent example of spectroscopic artifacts introduced by a sequence of realistic pulses), and investigating the relation between creating nonclassical qubit-environment correlations (that can be quite easily quantified for a qubit subjected to pure dephasing \cite{Roszak_PRA15}) and the ability to and the ability to treat the quantum environment as a source of classical noise.

\begin{acknowledgements}
We would like to thank Friedemann Reinhard and Lorenza Viola for helpful comments, and Leigh Norris for  enlightening explanations concerning the range of frequencies in which the noise spectrum can be reconstructed.
The work was supported by the Polish National Science Centre (NCN) Grants No. DEC-2012/07/B/ST3/03616 and DEC-2015/19/B/ST3/03152, and by the National Science Foundation Grant No. DMR 1207298.
\end{acknowledgements}

\appendix

\section{Pure dephasing due to quantum Gaussian noise} \label{app:quantum}
The evolution operator for classical noise (\ref{eq:evo_op}) is generalized to quantum noise with the addition of the time ordering symbol
\begin{eqnarray}
&&\hat U(T,0) =\mathcal T e^{-i \frac{1}{2}\hat\sigma_z \int_0^T dt f_T(t) \hat b_z(t)}\nonumber\\
&&=\sum_{k=0}^\infty \left(-i \frac{1}{2}\hat\sigma_z\right)^k \int_0^T dt_1 \int_0^{t_1}dt_2\ldots \int_0^{t_{k-1}}dt_k\nonumber\\
&&\times f_T(t_1)\hat b_z(t_1)\,f_T(t_2)\hat b_z(t_2)\ldots f_T(t_k)\hat b_z(t_k)
\end{eqnarray}
where $\hat b_z(t) = e^{i \hat H_E t}\hat b_z e^{-i \hat H_E t}$ is the interaction picture of the noise operator with respect to the free Hamiltonian of the environment $\hat H_E$. The coherence factor is then given by
\begin{equation}
W(T) = \frac{\langle{+} |\mathrm{Tr}_E\{ \hat U(T,0)\hat\rho(0)\hat U(T,0)^\dagger\} |{-}\rangle}{\langle{+}| \mathrm{Tr}_E\{\hat \rho(0)\} |{-}\rangle}\,,\label{apdx:W}
\end{equation}
with $\mathrm{Tr}_E$ denoting partial trace over environmental degrees of freedom. Here the initial density matrix is assumed to be of the product form $\hat\rho(0) = \hat\rho_Q(0)\hat \rho_E(0)$, and the environment part is assumed to be stationary, i.e. $[ \hat H_E ,\hat \rho_E(0) ] =0$. Using Eq.~(\ref{eq:Wchi}), the coherence can be expressed in terms of the attenuation factor, $\chi(T)$, which is second order in the noise operator $\hat b_z$ for Gaussian noise. Our goal below is to show that the Gaussian attenuation factor is given by
\begin{eqnarray}
\chi(T) &=& \frac{1}{2}\int_0^T \!\!\!\! dt_1dt_2 f_T(t_1)f_T(t_2)C^\mathrm{qm}(t_1-t_2)\nonumber\\
 &=&\frac{1}{2}\int_0^T \!\!\!\! dt_1dt_2 f_T(t_1)f_T(t_2)\nonumber\\
&&\times \frac{1}{2}\mathrm{Tr}_E[ \{ \hat b_z(t_1),\hat b_z(t_2)\} \hat\rho_E(0)]\,,\label{apdx:chi}
\end{eqnarray}
where $\{ \hat A , \hat B \} = \hat A\hat B + \hat B\hat A$.

The numerator of Eq.~(\ref{apdx:W}) up to the second order in $\hat b_z(t)$ reads:
\begin{eqnarray}
&&\left\langle + \left|\mathrm{Tr}_E\left\{ \hat U(T,0)\hat\rho(0)\hat U(T,0)^\dagger \right\} \right|- \right\rangle =\nonumber\\
&& \mathrm{Tr}_E\left\{\left\langle + \left|\left(1-i\frac{\hat\sigma_z }{2}\int_0^T dt_1 f_T(t_1)\hat b_z(t_1) \right.\right.\right.\right. \nonumber\\
&&\left.\left.\left.\left. \!\!\!\! -\left(\frac{\hat\sigma_z}{2}\right)^2 \!\! \int_0^T \!\!\! dt_1\int_0^{t_1}\!\!\! dt_2\, f_T(t_1)\hat b_z(t_1)\,f_T(t_2)\hat b_z(t_2)\right) \times \right. \right. \right. \nonumber\\
&& \left.\left.\left. \!\!\! \hat \rho_Q(0) \hat \rho_E(0)\!\! \left( 1 \!+i \frac{\hat\sigma_z}{2}\int_0^T dt_1'\,f_T(t_1') \hat b_z(t_1')-\!\! \left(\frac{\hat\sigma_z}{2}\right)^2 \!\! \times \right. \right.\right.\right. \nonumber\\
&&\left.\left.\left.\left. \int_0^T \!\!\! dt_1'\int_0^{t_1'}\!\!\! dt_2'\,f_T(t_2')\hat b_z(t_2')\, f_T(t_1')\hat b_z(t_1')\right) \right| - \right\rangle \right\}=\nonumber\\
&&\langle {+}|\hat \rho_Q(0)|{-}\rangle \times \nonumber\\
&& \!\!\! \left( 1\!-i\frac{1}{2} \int_0^T \!\!\! dt_1\, f_T(t_1)\mathrm{Tr}_E\left\{ \hat b_z(t_1)\hat \rho_E(0)\!+\hat\rho_E(0)\hat b_z(t_1)\right\} \right. \nonumber\\
&&\left. -\frac{1}{4}\int_0^T \!\!\! dt_1 \!\!  \int_0^T \!\!\! dt_1'\, f_T(t_1)f_T(t_1')\mathrm{Tr}_E\left\{\hat b_z(t_1)\hat\rho_E(0)\hat b_z(t_1') \right\} \right.  \nonumber\\
&&\left. -\frac{1}{4}\int_0^T dt_1\int_0^{t_1}dt_2\, f_T(t_1)f_T(t_2) \times \right. \nonumber\\
&& \left. \mathrm{Tr}_E\left\{\hat b_z(t_1)\hat b_z(t_2)\hat\rho_E(0)+\hat\rho_E(0)\hat b_z(t_2)\hat b_z(t_1)\right\}\right),
\end{eqnarray}
where we used relations $\langle {+}|\hat\sigma_z =\langle {+}|$ and $\hat\sigma_z |{-}\rangle = -|{-}\rangle$. Utilizing the cyclic property of the trace we have
\begin{eqnarray}
&&\langle {+}|\hat \rho_Q(0)|{-}\rangle\Big( 1 -i \int_0^T dt_1\, f_T(t_1)\langle \hat b_z(t_1) \rangle_E\nonumber\\
&&-\frac{1}{4}\int_0^T \!\! dt_1 \int_0^T \!\! dt_1'\,f_T(t_1)f_T(t_1')\langle \hat b_z(t_1')\hat b_z(t_1)\rangle_E\nonumber\\
&&-\frac{1}{4}\int_0^T \!\!\!\! dt_1\!\!\int_0^{t_1}\!\!\!\! dt_2\,f_T(t_1)f_T(t_2)\langle \{\hat b_z(t_1), \hat b_z(t_2)\} \rangle_E \Big)
\end{eqnarray}
where $\langle \ldots \rangle_E = \mathrm{Tr}_E\{ \ldots \hat \rho_E(0)\}$. Using the relation $\int_0^T dt_1\int_0^T dt_2 = \int_0^T dt_1\int_0^{t_1}dt_2 + \int_0^T d{t_2}\int_0^{t_2}dt_1$ and the symmetry of the above integrals with respect to interchanging the integration variables we arrive at the result
\begin{eqnarray}
&&\langle {+}|\hat \rho_Q(0)|{-}\rangle\Big( 1 -i \int_0^T dt_1\, f_T(t_1)\langle \hat b_z(t_1) \rangle_E\nonumber\\
&&\!\! -\frac{1}{2}\int_0^T \!\!\! dt_1 \!\! \int_0^T \!\!\! dt_2 f_T(t_1)f_T(t_2) \frac{1}{2}\langle \{ \hat b_z(t_1),\hat b_z(t_2)\} \rangle_E \Big).
\end{eqnarray}
Without loss of generality, we assume $\langle \hat b_z(t) \rangle_E =0$ to get {
\bea
W(T) &=& 1 - \frac{1}{2} \int_0^T \!\! dt_1 \int_0^T \!\! dt_2 f_T(t_1)f_T(t_2) \times \nonumber \\ &&C^\mathrm{qm}(t_1-t_2) + O(\hat b_z^3) \,\, ,
\eea
in which  $C^{\rm qm}$ is defined by Eq.~(\ref{eq:AQ}).
On the other hand, according to Gaussian assumption
\begin{equation}
W(T) = e^{-\chi(T)} = 1  -\chi(T) + O(\hat b_z^3)\,.
\end{equation}
Comparing the terms of second order we arrive at the final result (\ref{apdx:chi}).

\section{The exact result for ${\bf W^{(1)}(T)}$}\label{app:W_1}
Here we calculate the single spin contribution to the coherence function (\ref{eq:full_coh_spins}):
\beq
W^{(1)}(T) = \mathrm{Tr}\left[ \hat{U}_{+}(T)\hat{U}^{\dagger}_{-}(T)   \right ] \,\, ,  \label{apdx:W1}
\eeq
where $\hat{U}_{\pm}(T)$ are the evolution operators of environmental spin conditioned on the initial state of the qubit being $|\pm\rangle$. For a pulse sequence defined by pulse times $t_k$ we have
\begin{eqnarray}
\hat{U}_{\pm}(T) & = & e^{-i\hat{H}_{\mp}(T-t_{n})} e^{-i\hat{H}_{\pm}(t_{n}-t_{n-1})} \ldots \nonumber\\
& & \times e^{-i\hat{H}_{\mp}(t_{2}-t_1)}  e^{-i\hat{H}_{\pm}t_{1}}
\end{eqnarray}
for odd number of pulses $n$, while for even $n$, the leftmost exponential should be evaluated with $\hat{H}_{\pm}$ instead. The conditional Hamiltonians $\hat{H}_{\pm}$ are defined as
\beq
\hat{H}_{\pm} = \omega_{0}\hat{J}_{z} \pm \frac{1}{2} (A_{\perp}\hat{J}_{x} + A_{\parallel}\hat{J}_{z})   \,\, .
\eeq
When $t_k = k \tau$ and $n\in \mathrm{odd}$, i.e. a PDD sequence with characteristic frequency $\omega_p = \pi/\tau$, we can write
\beq
\hat{U}_{\pm}(T) = [\hat{u}_{\pm}(2\tau)]^{(n+1)/2} \,\, ,
\eeq
where
\begin{eqnarray}
\hat{u}_{\pm}(2\tau) & = & e^{-i\hat{H}_{\mp}\tau}e^{-i\hat{H}_{\pm}\tau} \equiv e^{-i\gamma_{\mp}\mathbf{n}_{\mp}\cdot\mathbf{\hat{J}}} e^{-i\gamma_{\pm}\mathbf{n}_{\pm}\cdot\mathbf{\hat{J}}} \nonumber \\
& \equiv & e^{-i\phi_{\pm}\mathbf{m}_{\pm}\cdot\mathbf{\hat{J}}} \,\, ,  \label{apdx:mdef}
\end{eqnarray}
with
\begin{eqnarray}
\mathbf{n}_{\pm} & = & \frac{1}{\sqrt{1\pm \epsilon_\parallel + \frac{\epsilon^2_\perp}{4}+ \frac{\epsilon^2_\parallel}{4}}}
	\left(\begin{array}{c} \pm \epsilon_{\perp}/2 \\ 0 \\  1 \pm \epsilon_{\parallel}/2\\ \end{array}\right)\nonumber\\
& \approx & \left(\begin{array}{c}\pm \frac{\epsilon_{\perp}}{2} - \frac{\epsilon_\perp \epsilon_\parallel}{4}\\0\\1-\frac{\epsilon^2_\perp}{8}\\ \end{array}\right) + O(\epsilon^3) \,\, ,
\end{eqnarray}
and
\bea
\gamma_{\pm} &=&
\omega_0 \tau \sqrt{1\pm \epsilon_\parallel + \frac{\epsilon^2_\perp}{4}+ \frac{\epsilon^2_\parallel}{4}} \nonumber \\ & \approx & \omega_0 \tau \left( 1\pm \frac{\epsilon_\parallel}{2}+\frac{\epsilon^2_\perp}{8} \right) + O(\epsilon^3)\,\, .
\eea
The quantities $\mathbf{m}_{\pm}$ and $\phi_{\pm}$ defined by Eq.~(\ref{apdx:mdef}) correspond to the axis and the angle of the effective rotation, respectively. For a PDD sequence they can be obtained by solving the following non-linear equations:
\begin{eqnarray}
\label{apdx:effrotPDD}
\cos\frac{\phi_\pm}{2} &=& \cos\frac{\gamma_+}{2}\cos\frac{\gamma_{-}}{2} \nonumber\\
&& - (\mathbf{n}_{+} \cdot \mathbf{n}_{-})\sin\frac{\gamma_+}{2}\sin\frac{\gamma_{-}}{2}  \,,\label{apdx:rel_1}\\
\mathbf{m}_\pm\,\sin\frac{\phi_\pm}{2}  &=& \mathbf{n}_{+} \sin\frac{\gamma_+}{2}\cos\frac{\gamma_{-}}{2}+ \mathbf{n}_{-}\sin\frac{\gamma_{-}}{2}\cos\frac{\gamma_+}{2} \nonumber \\
&& \pm ( \mathbf{n}_{-}\times\mathbf{n}_{+} )\sin\frac{\gamma_{+}}{2}\sin\frac{\gamma_{-}}{2}\label{apdx:rel_2}
\end{eqnarray}
from which we get that $\phi_{+}=\phi_-\equiv\phi$. The conditional evolutions $\hat{U}_{\pm}(T)$ amount to $\frac{n+1}{2}$ repetitions of the $\hat{u}_{\pm}$ rotations, and their product appearing in Eq.~(\ref{apdx:W1}) is a rotation of spin $J$ about axis $\mathbf{m}_{\pm}$ by angle $\frac{n+1}{2}\phi_{\pm}$.
This leads to the following exact expression for the coherence of the qubit probe, coupled to a single environmental spin:
\begin{equation}
W^{(1)}=1-\sin^2\left(\frac{(n+1)}{4}\,\phi\right)(1-\mathbf{m}_+\cdot\mathbf{m}_-) \,\, .  \label{apdx:W1phi}
\end{equation}
Using the relations (\ref{apdx:rel_1}) and (\ref{apdx:rel_2}) we obtain an approximate rotation angle $\phi = 2\epsilon_{\perp}+O(\epsilon^3)$ and $\big(\mathbf{m}_+\cdot\mathbf{m}_{-}\big)= -1+O(\epsilon^2)$. Substituting these results to Eq.~(\ref{apdx:W1phi}) we get for $n \! \ll \! \epsilon^{-3}$:
\begin{equation}
W^{(1)}\left(T=\frac{(n+1)\pi}{\omega_0}\right)\approx\cos\left[(n+1)\,\epsilon_{\perp}\right]  \,\, . \label{apdx:W1cos}
\end{equation}


%

\end{document}